\documentclass[11pt,a4paper]{article} 
\usepackage{jheppub}
\pdfoutput=1
\usepackage{graphicx}
\usepackage{amsmath}
\usepackage{amssymb}
\usepackage{subcaption}
\usepackage{bbm} 
\usepackage{placeins}
\usepackage{hhline} 
\usepackage{cleveref} 
\usepackage{tikz-feynman}
\usepackage{tikz-cd}

\newcommand{\msbar}{\overline{\mathrm{MS}}}

\newcommand{\bare}{\mathrm{bare}}

\newcommand{\cA}{{\mathcal A}}
\newcommand{\cM}{{\mathcal M}}

\newcommand{\cO}{{\mathcal O}}
\newcommand{\cL}{{\mathcal L}}

\newcommand{\cT}{{\mathcal T}}

\newcommand{\cTcut}{\cT_{\mathrm{cut}}}

\newcommand{\Idiv}{I_{\mathrm{div.}}}

\title{Soft functions for generalised angularity event shapes at NNLO}

\author[a]{Emmet P. Byrne,}
\author[b]{Jonathan R. Gaunt,}
\affiliation[a]{Department of Physics and Astronomy, The University of Manchester, Manchester M13 9PL, UK}
\affiliation[b]{Department of Physics, University of Cyprus, Nicosia 1678, Cyprus}

\emailAdd{emmet.byrne@manchester.ac.uk}
\emailAdd{gaunt.jonathan@ucy.ac.cy}

\abstract{
We consider a class of dijet event shapes which tend to angularity in the forward limit. The factorisation formula for this class of event shapes differs from angularity only in the soft function. The NNLO soft function of an event shape in this class is therefore the only missing ingredient required for NNLL$'$ resummation. We describe a modular approach to the calculation of the soft function for any event shape in this class, at NLO and NNLO. Much of the structure of the soft function can be written in terms of a single rapidity integral. The remainder can be computed as a three-dimensional numerical integral for which we provide a code. To demonstrate the flexibility of our approach, we compute the NNLO soft function for three new families of event shapes in this class. Potential applications include the investigation of hadronisation effects at future $e^+ e^-$ colliders, or through the re-analysis of archived data.
}


\begin{document}

\vspace{1cm}

\maketitle

\section{Introduction}
\label{sec:intro}

Event shapes in electron-positron annihilation have long been a testing ground for QCD, as well as our understanding of it. In particular, event shapes serve as a rather clean environment for the measurement of the fundamental strong coupling constant $\alpha_s$ \cite{Abbate:2012jh, Hoang:2014wka, Abbate:2010xh, Aglietti:2025jdj}. 

One important complication in extractions of $\alpha_s$ via event shapes is that these observables are affected by significant non-perturbative power corrections. For a large class of event shapes (including thrust, C-parameter, and the event shapes we will consider below), the leading correction to the event shape distribution in the dijet region beyond the peak is a simple shift~\cite{Dokshitzer:1997ew}. Neglecting the effect of final-state hadron masses, the size of the shift is given by $c_e \Omega_1/Q$, with $Q$ the energy of the collision, $\Omega_1$ a universal non-perturbative parameter, and $c_e$ a perturbatively calculable coefficient
that depends on the form of the event shape \cite{Webber:1994cp, Dokshitzer:1995qm, Dokshitzer:1995zt, Dokshitzer:1998pt, Korchemsky:1998ev,  Korchemsky:1999kt, Gardi:2001ny, Gardi:2002bg, Lee:2006fn, Lee:2006nr, Becher:2013iya}. There have been many further studies into event-shape power corrections, including the impact of hadron masses \cite{Salam:2001bd,Mateu:2012nk}, and the study of power corrections in the three-jet region \cite{Luisoni:2020efy,
Caola:2021kzt,
Caola:2022vea,Nason:2023asn}. Extractions of $\alpha_s$ from event shapes typically involve a simultaneous fit of $\alpha_s(M_Z)$ and $\Omega_1$, and it is well-known that there is a degeneracy between these two fit parameters (a decrease in $\alpha_s(M_Z)$ can be compensated by an increase in $\Omega_1$).

One way to break this degeneracy is to consider data at different $Q$ values, combining data from different colliders (since the non-perturbative shift scales as $1/Q$). An alternative approach is to consider several event shape observables with different $c_e$ values. This approach was used in ref.~\cite{Bell:2018gce}, which studies the angularity event shape \cite{Berger:2003iw,Berger:2003pk}. The angularity event shape has an adjustable parameter $a$ which effectively determines the strength of the radiation veto in the soft region with respect to that in the collinear region; for $a=0$, angularity reduces to thrust~\cite{Brandt:1964sa,Farhi:1977sg} while for $a=1$ it reproduces broadening~\cite{Rakow:1981qn,Catani:1992jc,Dokshitzer:1998kz}. The coefficient $c_e$ for angularity scales as $1/(1-a)$, and thus one may use measurements at different $a$ values to break the degeneracy. In ref.~\cite{Bennett:2025jli}, we studied the C-angularity event shape (originally mentioned in ref.~\cite{Lee:2006nr}), which is an angularity-like generalization of C-parameter (as opposed to thrust). Considering different values of $a$ for this observable represents a further avenue for investigating the power corrections.

Note that for (C-)angularity, changing $a$ changes the shape of the observable everywhere, both in the (central) soft and (forward) collinear regions. This means that when we change $a$ we not only change the non-perturbative corrections, but also the perturbative resummation part, since the anomalous dimensions change, as well as the soft and jet functions. An alternative, potentially cleaner, approach to altering the sensitivity to $\Omega_1$ is to consider a number of event shape variables that differ in shape only in the soft wide-angle region. These will have different sensitivities to $\Omega_1$, but essentially the same perturbative resummation structure (the only difference being the soft function).

This is the approach we pursue in this paper. We consider a general class of additive event shapes, which converge to angularity sufficiently quickly in the collinear limit, but whose behaviour in the soft wide-angle region is left arbitrary\footnote{Subject to basic constraints to maintain infrared safety and avoid non-global logarithms -- the event shape should be insensitive to soft particles with $p_T \to 0$, and the contribution from a particle with given $p_T$ to the event shape should not vary suddenly with rapidity.} -- we refer to these as generalised angularity event shapes. In this paper we construct an efficient algorithm for the numerical computation of the soft function for generalised angularities up to the two loop order (or NNLO). Utilising other known theoretical results from the angularity case, this enables theoretical predictions for this broader class of observables at the same NNLL$'$ precision as angularity, and allows them to be used in precision studies of the non-perturbative corrections and extractions of $\alpha_s$. 
These new event shapes could be studied at a future $e^+ e^-$ collider \cite{FCC:2018evy, CEPCStudyGroup:2018ghi, CLICdp:2018cto, ILCInternationalDevelopmentTeam:2022izu} or with archived $e^+e^-$ data (in a similar spirit to refs.~\cite{Bossi:2025nux, Zhang:2025nlf, Badea:2025wzd,Electron-PositronAlliance:2025hze}). Aside from phenomenological motivations, experimental considerations (e.g. detector sensitivity, data quality) may also inform the precise choice of generalised angularity observables to study. Analogous event shapes can also be defined and studied in the context of deep inelastic scattering (DIS), in a similar way to how DIS angularity is defined in ref.~\cite{Zhu:2021xjn}\footnote{For some alternative recent approaches to determining $\alpha_S$ and $\Omega_1$ via event shapes in DIS, see e.g. refs.~\cite{Dotson:2026ttc, Boughezal:2026dvu, Boughezal:2026iva}.}.

For the computation of the two-loop soft function, we follow the same broad strategy as has previously been used in refs.~\cite{Jouttenus:2011wh,Kasemets:2015uus,Banfi:2014sua,Gangal:2016kuo,Bauer:2020npd,Abreu:2022sdc, Abreu:2022zgo, Bennett:2025jli, Buonocore:2026yai, Abreu:2026FutureJV} -- i.e. we compute it as a correction to a soft function with a simplified measurement, where the latter is more straightforward to compute (semi-)analytically, and the correction has a simpler divergence structure than the full computation, enabling a numerical computation. In fact, the strategy for this calculation closely follows our previous computation of the two-loop soft function for C-angularity \cite{Bennett:2025jli}, and we are able to directly exploit results from that paper here. Refining this strategy, we can write the majority of the structure of the soft function in terms of, at worst, simple one-dimensional numerical integrals. The remaining terms can be computed as a three-fold numerical integral over the two-parton real emission amplitudes. We provide an explicit implementation of this algorithm in Mathematica as an ancillary file.

For the event shapes we consider here, the NNLO soft functions could alternatively be computed using the program~\texttt{SoftSERVE}~\cite{Bell:2018vaa,Bell:2018oqa,Bell:2020yzz}. It is worth distinguishing our methodology and motivation. Our goal was to develop a highly efficient method of calculating soft functions for a specific class of event shapes. Certain features of this class of observables (being azimuthally symmetric and having constrained collinear behaviour) have allowed us to maintain analytic control over much of the calculation, and the remaining numerical integrals are fast to evaluate. Another goal was to develop a method of calculation that can be extended to higher orders -- it is readily possible to extend the strategy discussed here to compute the soft function at N$^3$LO, which is required for N$^3$LL$'$ resummation (and, for the case $a=0$, is the last missing piece).

To illustrate the flexibility of this approach, we propose three new families of event shapes that generalise angularity, and we compute the soft functions for these event shapes up to NNLO. These three examples depend on one, two and three parameters, respectively, in addition to $a$. The first of these, which we refer to as $L_p$ angularity, is a generalization of the angularity and C-angularity event shapes, with each of these being recovered from $L_p$ angularity for a particular choice of the additional parameter. The other two have plateau-like weightings for  radiation in the central soft region, with one family being better suited to a lower plateau, and the other being better suited to a higher one. These three families of event shape are chosen because they map out a wide variety of possible shapes in the wide-angle soft region, and thus sensitivities to non-perturbative effects (that is, values of $c_e$).

This paper is organised as follows. In \cref{sec:setup} we specify the class of generalised angularities we consider in this paper, and we introduce three families of observables that belong to this class of event shapes. In \cref{sec:properties} we outline some basic properties of these event shapes, compute their soft function at NLO, and perform a preliminary investigation of their leading hadron-mass effects. In \cref{sec:NNLO} we describe the computation of the NNLO soft function for any member of the class of event shapes considered. Finally, we give our conclusions in \cref{sec:conc}.

\section{Generalised angularity event shapes}
\label{sec:setup}

Two-jet event shapes, $\tau_e$, are distributions which are dominated by two almost back-to-back collimated sets of particles in the limit $\tau_e\to 0$. Additive event shapes in $e^+ e^- \to X$ are given by 
\begin{equation}
    \tau_e(X) = \frac{1}{Q}\sum_{i \in X} \cT_e(k_i)\,,
    \label{eq:taue}
\end{equation}
where $Q$ is the centre-of-mass energy, and $i$ is a sum over particles in the hadronic final state $X$. We focus on a broad class of event shapes for which the summand takes the form
\begin{equation}
    \cT_e(k_i)=f_e(\eta_i)|\vec{p}_{i\perp}|\,,
\end{equation}
where $\eta_i$ and $\vec{p}_{i\perp}$ are the rapidity and transverse momentum of particle $i$, with respect to the thrust axis of the event\footnote{Other axes may be considered, for example the broadening axis~\cite{Larkoski:2014uqa} or the winner-take-all axis~\cite{Bertolini:2013iqa}.}. The classic event shapes thrust,  C-parameter and broadening are given by a choice of\footnote{The C-parameter definition given here differs from  the standard Lorentz-invariant event shape of ref.~\cite{Ellis:1980wv,Gardi:2003iv} but the definitions coincide in the dijet limit, up to a factor of 6~\cite{Salam:2001bd}.} 
\begin{equation}
    f_\tau(\eta)=e^{-|\eta|}\,, \qquad 
    f_C(\eta)=\frac{1}{2 \cosh \eta}\,, \qquad 
    f_b(\eta)=1\,,
    \label{eq:f_classic}
\end{equation}
respectively. 
The above definitions assume massless hadrons. For a physical observable one must specify a hadron mass scheme (see e.g. refs.~\cite{Salam:2001bd,Mateu:2012nk} and \cref{eq:mass_schemes}). Our perturbative calculations for the soft function are insensitive to this scheme so, with the exception of \cref{sec:hadron}, we take hadron masses to be zero. 

\begin{figure}
    \centering
    \includegraphics[width=1\linewidth]{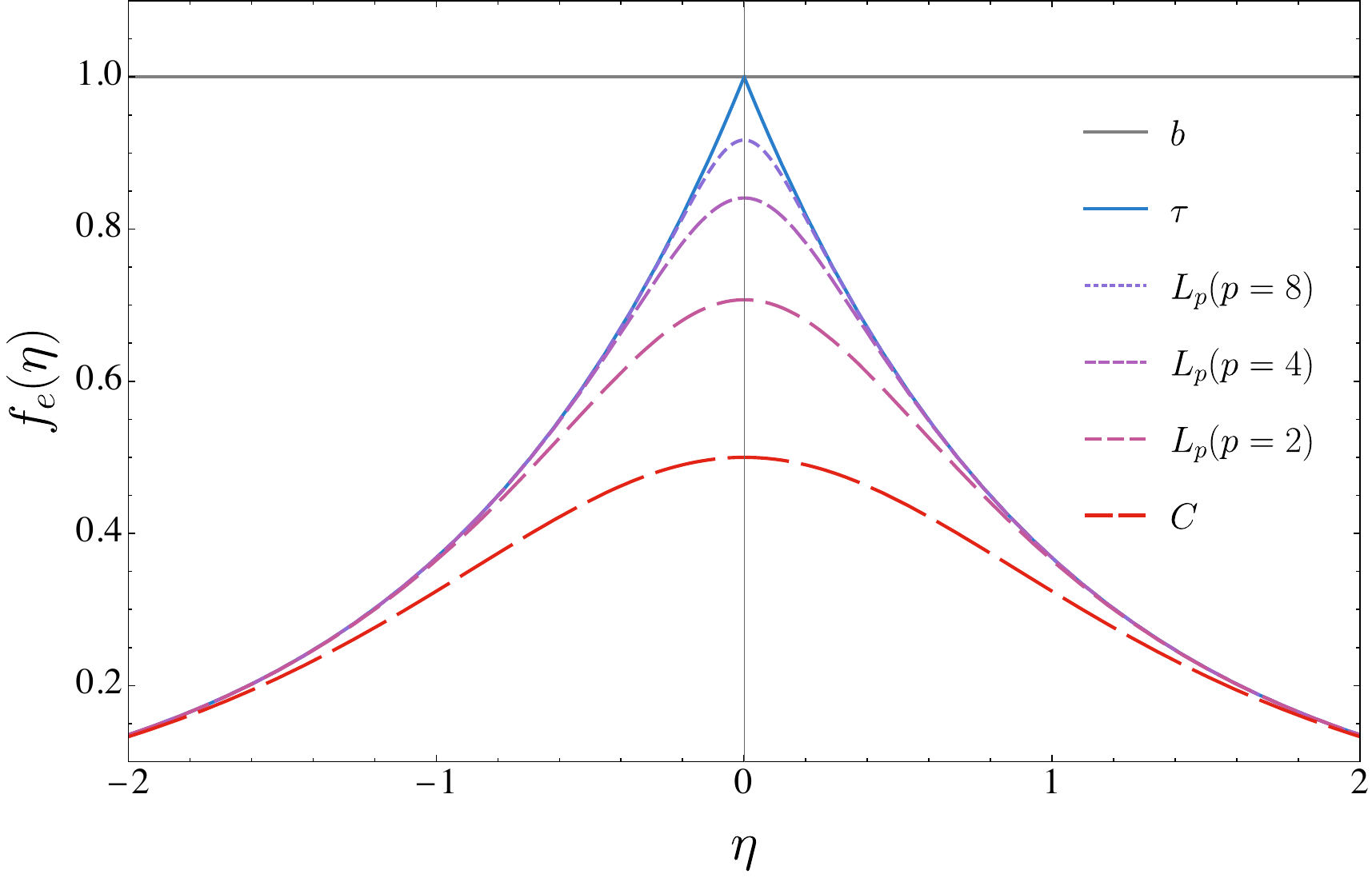}
\caption{A plot comparing the rapidity weighting of $L_p$ with the classic event shapes thrust ($e=\tau$) and C-parameter ($e=C$), for several choices of $p$. $L_p$ recovers thrust and C-parameter in the limits $p\to \infty$ and $p \to 1$ respectively.}
\label{fig:Lp}
\end{figure}

We note that there is a natural interpolation between the first two event shapes in \cref{eq:f_classic}. We will refer to this as the $L_p$ event shape, which is defined via the following rapidity weighting:
\begin{equation}
f_{L_p}(\eta) = (2\cosh p\eta)^{-\frac{1}{p}}\,.
\label{eq:fLp}
\end{equation}
For $p=1$, $f_{L_p}$ becomes $f_C$ while for $p\to \infty$, $f_{L_p}$ tends to $f_\tau$.
It is instructive to move to light-cone coordinates, where the corresponding single particle event shapes are 
\begin{subequations}
\label{eq:Te_LC}
\begin{align}
\cT_{\tau}(k_i) &= \frac{k_i^+ k_i^-}{\max(k_i^+, k_i^-)}\,,\\
\cT_{C}(k_i) &= \frac{k_i^+ k_i^-}{k_i^+ +k_i^-}\,,\\
\cT_{L_p}(k_i) &= \frac{k_i^+ k_i^-}{\left[ (k_i^+)^p + (k_i^-)^p \right]^{\frac{1}{p}}}\,.
\label{eq:TLp_LC}
\qquad
\end{align}
\end{subequations}
We define the lightcone coordinates using the (SCET) conventions given in Appendix \ref{sec:kin}. The name \textit{$L_p$} refers to the geometric structure of the denominator in \cref{eq:TLp_LC}. If we group the light-cone components of a given emission into a two-dimensional positive vector $\vec{k}_{i,\text{lc}} \equiv (k_i^+, k_i^-)$, this denominator is precisely the mathematical $L_p$ norm of $\vec{k}_{i,\text{lc}}$ (or \textit{Minkowski distance}). Certain choices correspond to well-known standard distance metrics:
\begin{itemize}
    \item For $p=2$, the norm is the familiar \textit{Euclidean} distance $\sqrt{(k_i^+)^2 + (k_i^-)^2}$.
    \item For $p=1$ (corresponding to C-parameter), the norm becomes the simple sum $k_i^+ + k_i^-$. This is called the \textit{Manhattan} or \textit{taxicab} distance (analogous to a taxi constrained to navigate a grid of city blocks without diagonal movement).
    \item In the limit $p \to \infty$ (corresponding to thrust), the norm is strictly dominated by the largest component, $\max(k_i^+, k_i^-)$, and this is referred to as the \textit{Chebyshev} distance.
\end{itemize}
Thus, the measurement in \cref{eq:TLp_LC} can be thought of as a measurement of the transverse momentum squared of the particle, divided by a geometric `length' in light-cone space. In \cref{fig:Lp} we plot the rapidity weighting $f_{L_p}$ for $p\in \{2,4,8\}$, as well as the functions in \cref{eq:f_classic}.

In the large rapidity limit, the $L_p$ event shape for any $p>0$ coincides with thrust and C-parameter,
\begin{equation}
    \lim_{|\eta| \to \infty } \ln f_{L_p}(\eta)= -|\eta|\,.
\end{equation}
Thus, the $L_p$ event shapes can be viewed as a one-parameter family of event shapes which alter the sensitivity to soft, wide angle emissions while leaving the sensitivity to forward emissions fixed. This can be contrasted with angularity~\cite{Berger:2003iw,Berger:2003pk}, 
\begin{equation}
    f_{\tau,a}(\eta)=e^{-(1-a)|\eta|}\,,
\end{equation}
which alters the large-$|\eta|$ behavior while fixing the sensitivity at $\eta=0$. 

Here we wish to explore event shapes with an even more general shape, such that the rapidity weighting function in the forward region can have any shape from the angularity event class (rather than being restricted to only the thrust shape with $a=0$), and the behaviour in the soft region can be modified (as for the $L_p$ event shape). To a limited extent, this wish can be fulfilled by comparing the standard angularity to the \emph{C-angularity} event shape~\cite{Lee:2006nr, Bennett:2025jli}
\begin{equation}
    f_{C,a}=(2 \cosh \eta )^{-(1-a)}\,,
\end{equation}
since the two event shapes coincide in the forward region (for equal $a$) but differ in the central region. However, this allows us only two possible shapes in the central region -- we would like to extend this to a continuum of possible shapes.

Thus, we introduce the \textit{generalised angularities} that are the focus of this paper. We require that the rapidity weighting function $f_{e,a}(\eta)$ for an event shape in this class should approach that of standard angularity sufficiently quickly in the forward limit; in particular we must have\footnote{
To see this at one loop consider \cref{eq:1Lfullmasy}, which must be free of divergences for \cref{eq:fac} to hold. If $\delta \le 0$ then we encounter divergences associated with the $\eta \to \infty$ limit. For the explicit observables we consider in this paper, the correction terms are actually exponentially suppressed in the forward limit as opposed to being a power of $|\eta|$, so this requirement is satisfied.}
\begin{equation}
    \lim_{|\eta| \to \infty} \ln f_{e,a}(\eta) = - (1-a) |\eta| + \mathcal{O}\left( \frac{1}{|\eta|^{1+\delta}}\right)\,,\qquad \delta > 0\,.
\end{equation}
We also require $f_{e,a}(\eta)$ to be finite everywhere (to ensure that the observable is infrared safe) and impose that there should not be any large discontinuities in $f_{e,a}(\eta)$, to avoid non-global logarithms. Other than this, we do not put any restrictions on $f_{e,a}(\eta)$ in the central region, allowing a multitude of possible shapes. 

Then, for $a<1$, the observable factorises in the $\tau_{e,a} \ll 1$ limit according to the SCET$_{\textrm{I}}$ form~\cite{Bauer:2008dt,Hornig:2009vb}:
\begin{align}
\begin{split}
    \frac{1}{\sigma^{(0)}}\frac{\mathrm{d}\sigma}{\mathrm{d} \tau_{e,a}}= \frac{H(Q,\mu)}{Q^2}
    \int& 
    \mathrm{d} \cT_S \,
    \mathrm{d} \cT_1 \,
    \mathrm{d} \cT_2 \,
    \delta(Q \,\tau_{e,a}-\cT_1-\cT_2-\cT_S)\\
    &
    \times S_{e,a}(\cT_S,\mu)
    \,
    J_a(Q, \cT_1,\mu)
    \,
    J_a(Q, \cT_2,\mu)\,.
    \label{eq:fac}
\end{split}
\end{align}
The quantities $J_a$ are the angularity jet functions, which are already known to both one \cite{Hornig:2009vb} and two loops \cite{Bell:2018gce,Bell:2021dpb}. The hard function $H$ is known up to four loops \cite{Lee:2022nhh}. This means that the only missing piece for NNLL$'$ resummation of the event shape $\tau_{e,a}$ is the computation of the soft function up to two loops, which is the focus of this paper. The soft function is the only part that depends on the non-trivial observable shape near $\eta =0$. In the DIS case, one of the jet functions is replaced by an angularity beam function \cite{Zhu:2021xjn} -- this is known at the two-loop order required for NNLL$'$ precision \cite{Bell:2024lwy}.

The strategy we will introduce below to compute the NLO and NNLO soft function works for any $f_{e,a}$ that approaches angularity suitably quickly in the forward limit, but we will base our discussion around three novel families of observables. 


The first family of event shapes we introduce is a two-parameter family that is a straightforward generalization of the $L_p$ event shape from \cref{eq:fLp}, and contains all the event shapes mentioned thus far. We call this $L_p$\emph{-angularity}, $\tau_{L_p,a}$ which has rapidity weighting:
\begin{equation}
f_{L_p,a}(\eta;p) = \left(f_{L_p}(\eta)\right)^{1-a} =(2\cosh p\eta)^{-\frac{(1-a)}{p}}\,.
\end{equation}
It is easy to see that this reduces to C-angularity when $p=1$, standard angularity in the limit $p\to \infty$, and the $L_p$ event shape for $a=0$. Recalling \cref{fig:Lp}, one observes that changing $p$ alters two aspects of the shape. First, decreasing $p$ makes the size of the quasi-flat region near $\eta \sim 0$ larger. More generally, one can see this directly by considering the expansion of $f_{L_p,a}$ near $\eta = 0$:
\begin{equation}
    f_{L_p,a} = 2^{\frac{a-1}{p}}
    \left(1 + \frac{(a-1) }{2} p \eta^2
    \right) + \mathcal{O}(\eta^4)\,.
\end{equation}
Thus, the size of the quasi-flat region goes roughly as $1/\sqrt{p}$. Second, the transition point $\eta_t$ where the asymptotic forward behaviour takes over is also set by $p$, going roughly as $\eta_t \sim 1/p$. For $p \ll 1$ the transition point migrates into the forward region, such that we should avoid this region for the applicability of our SCET$_{\mathrm{I}}$ factorization approach.

\begin{figure}
\begin{subfigure}{.32\textwidth}
    \centering
    \includegraphics[width=1\linewidth]{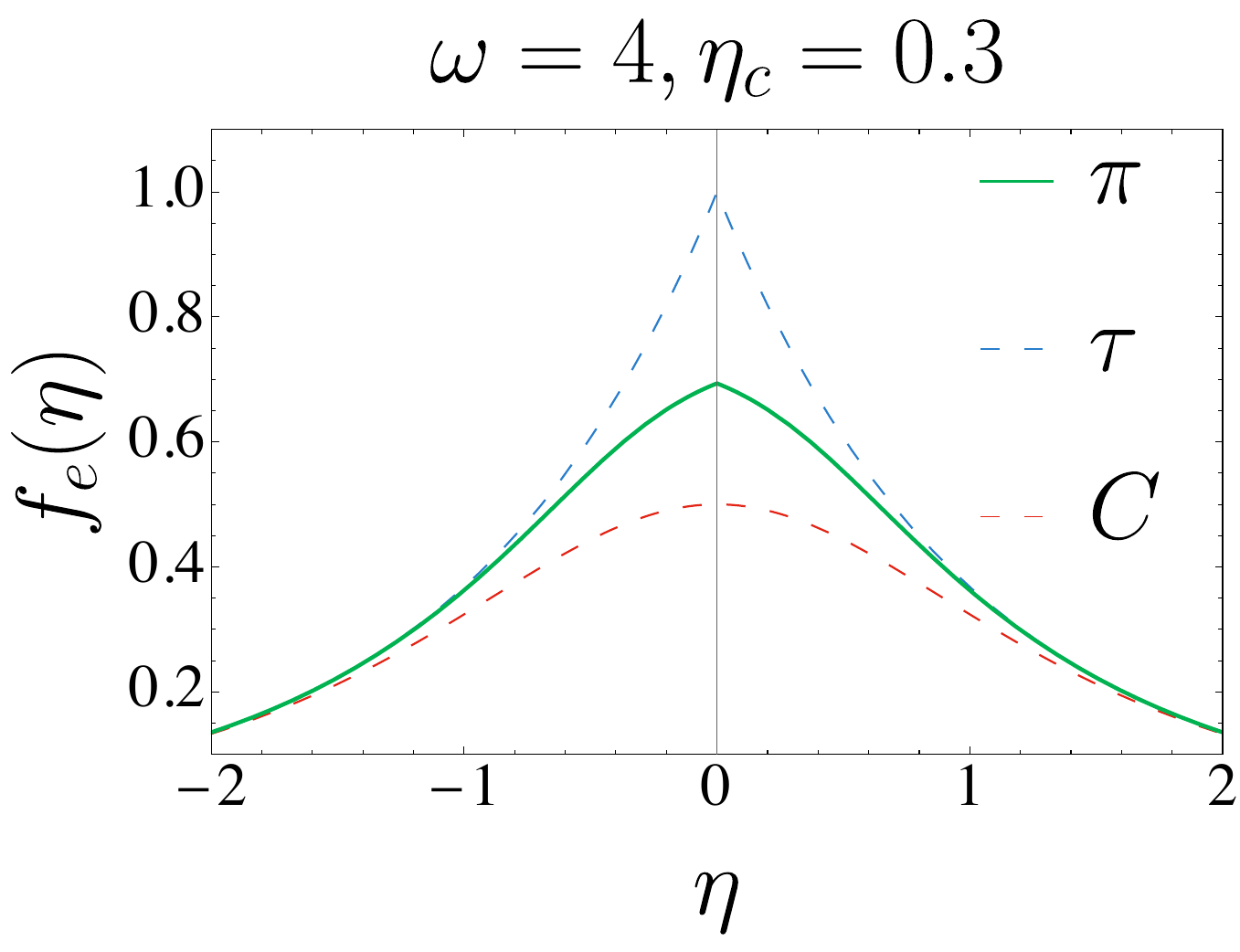}
    \label{fig:mp11}
\end{subfigure}
\begin{subfigure}{.32\textwidth}
    \centering
    \includegraphics[width=1\linewidth]{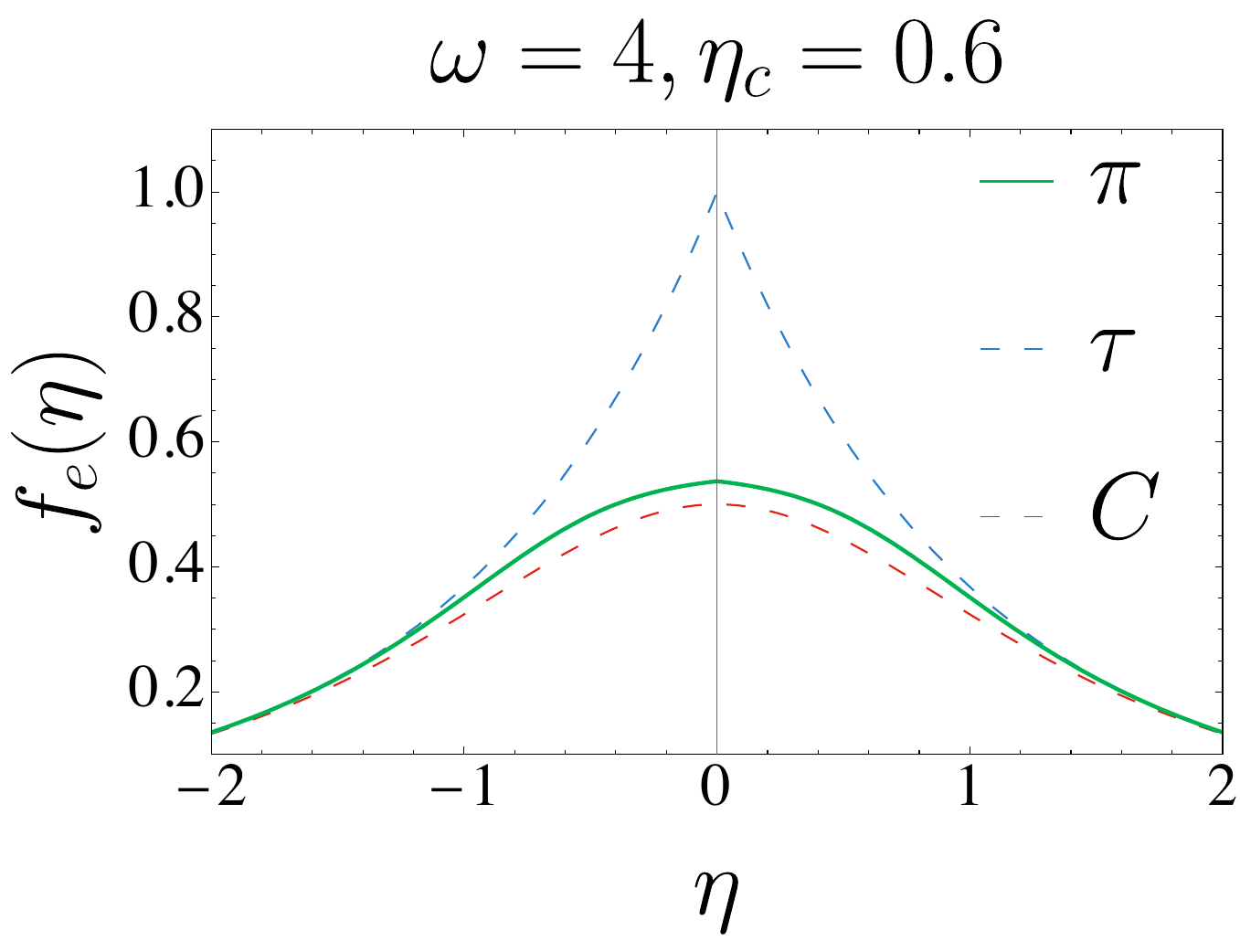}
    \label{fig:mp12}
\end{subfigure}
\begin{subfigure}{.32\textwidth}
    \centering
    \includegraphics[width=1\linewidth]{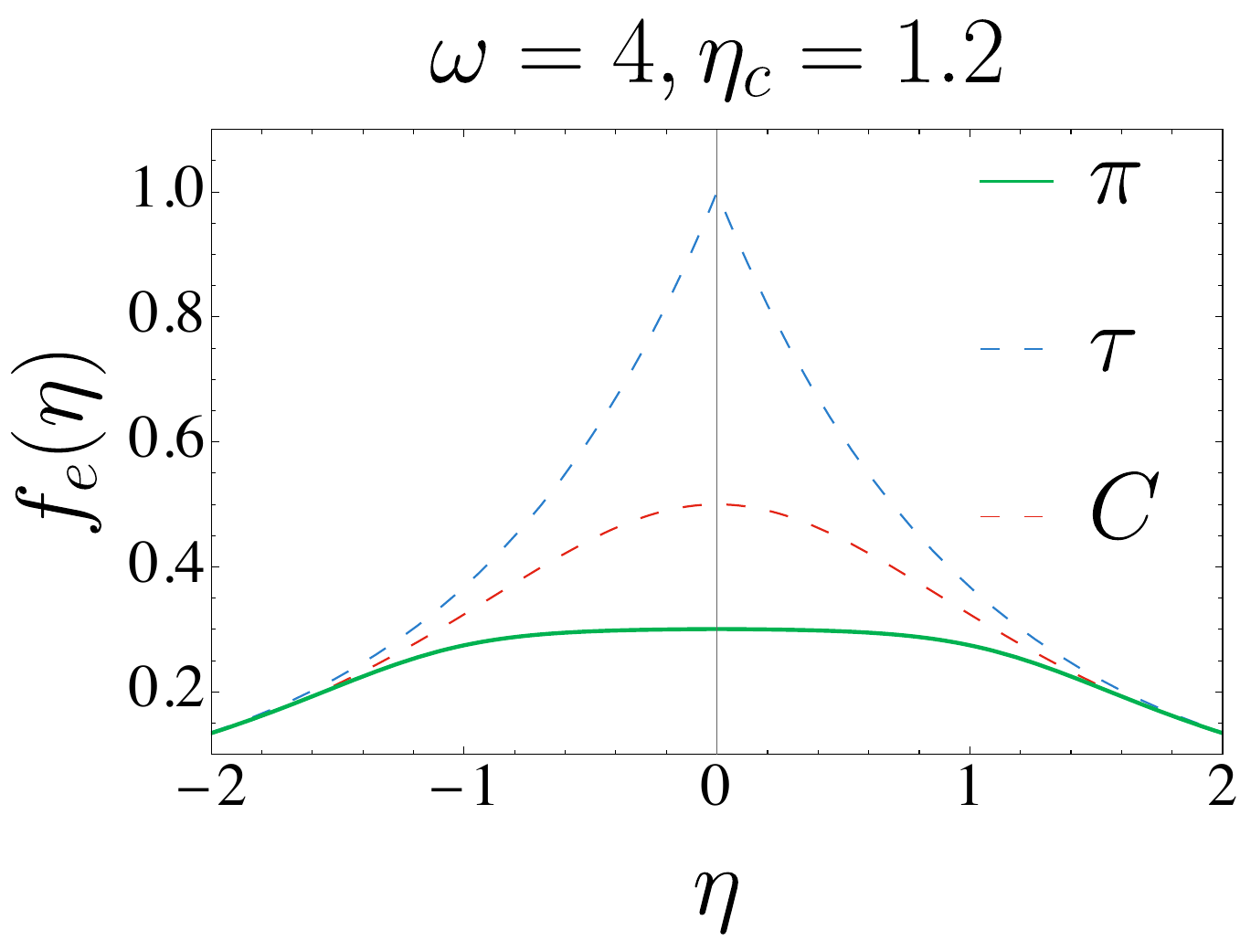}
    \label{fig:mp13}
\end{subfigure}
\\
\begin{subfigure}{.32\textwidth}
    \centering
    \includegraphics[width=1\linewidth]{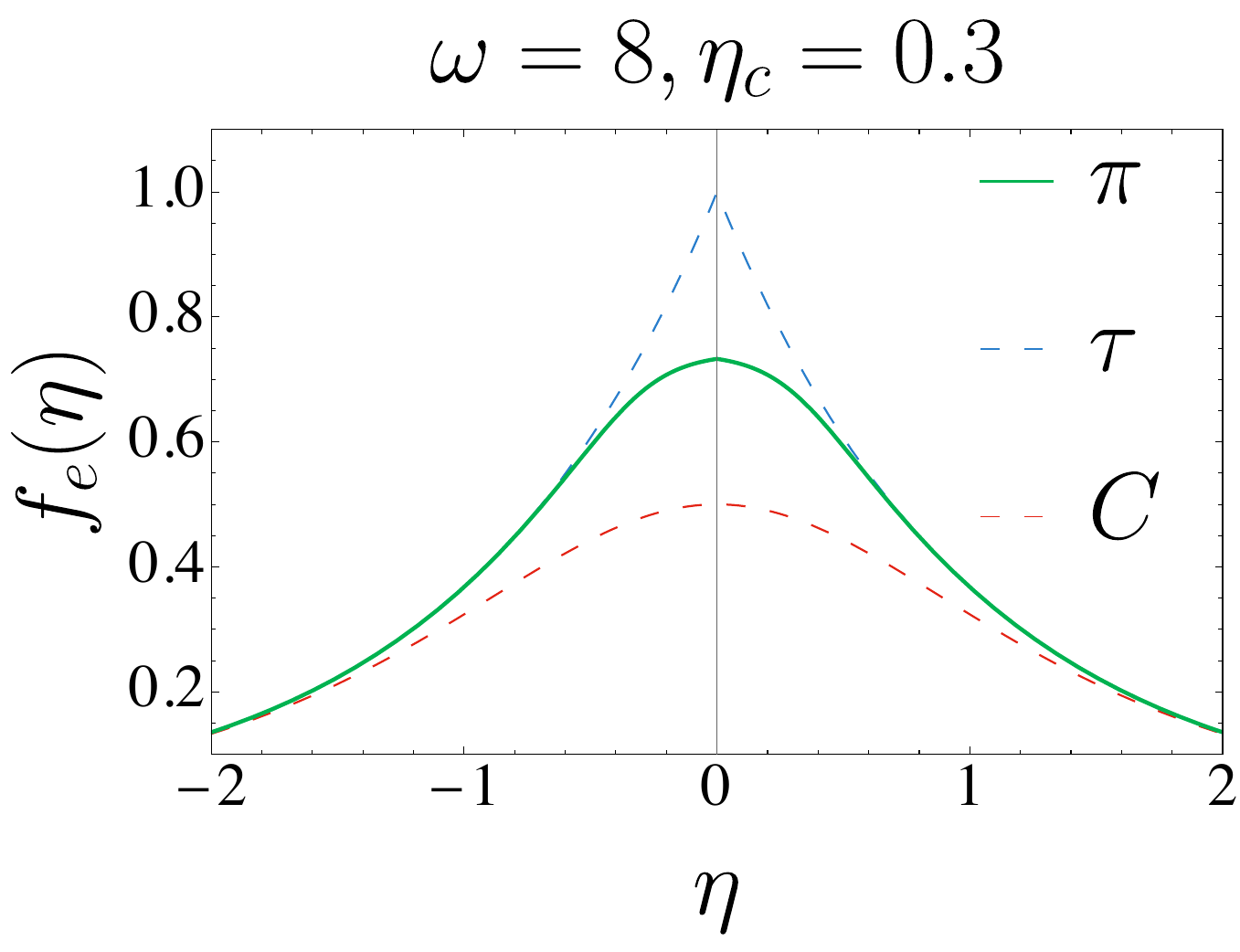}
    \label{fig:mp21}
\end{subfigure}
\begin{subfigure}{.32\textwidth}
    \centering
    \includegraphics[width=1\linewidth]{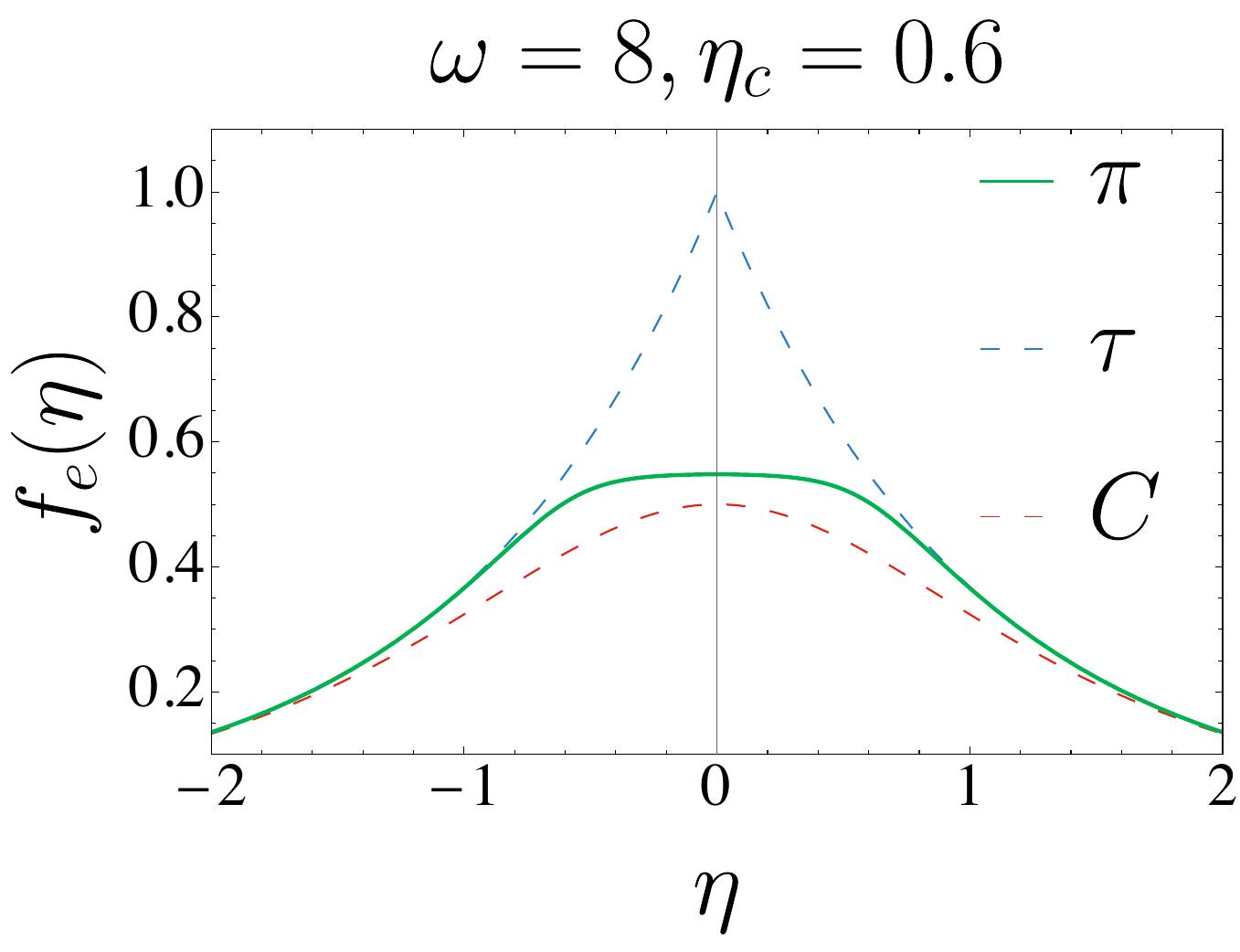}
    \label{fig:mp22}
\end{subfigure}
\begin{subfigure}{.32\textwidth}
    \centering
    \includegraphics[width=1\linewidth]{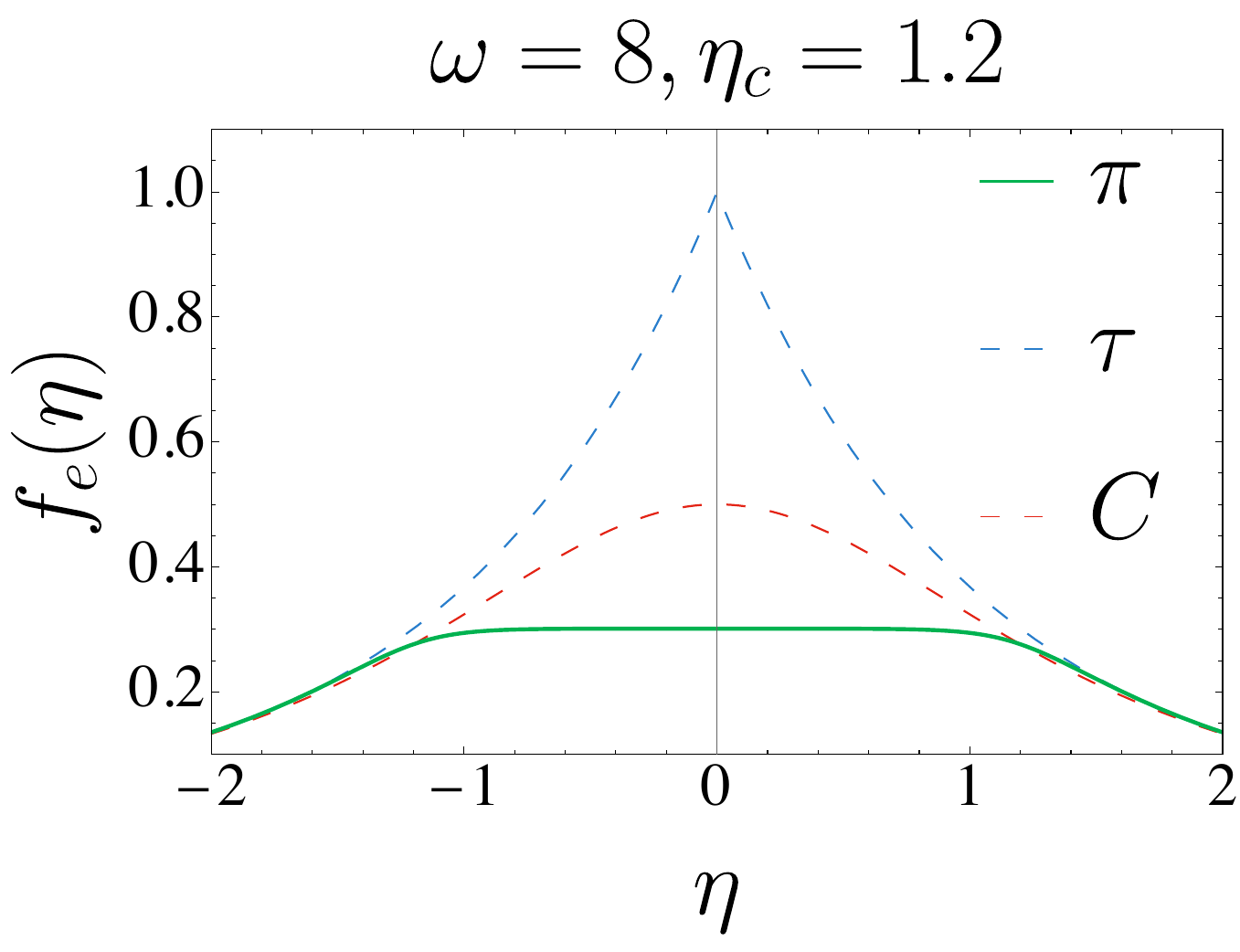}
    \label{fig:mp23}
\end{subfigure}
\\
\begin{subfigure}{.32\textwidth}
    \centering
    \includegraphics[width=1\linewidth]{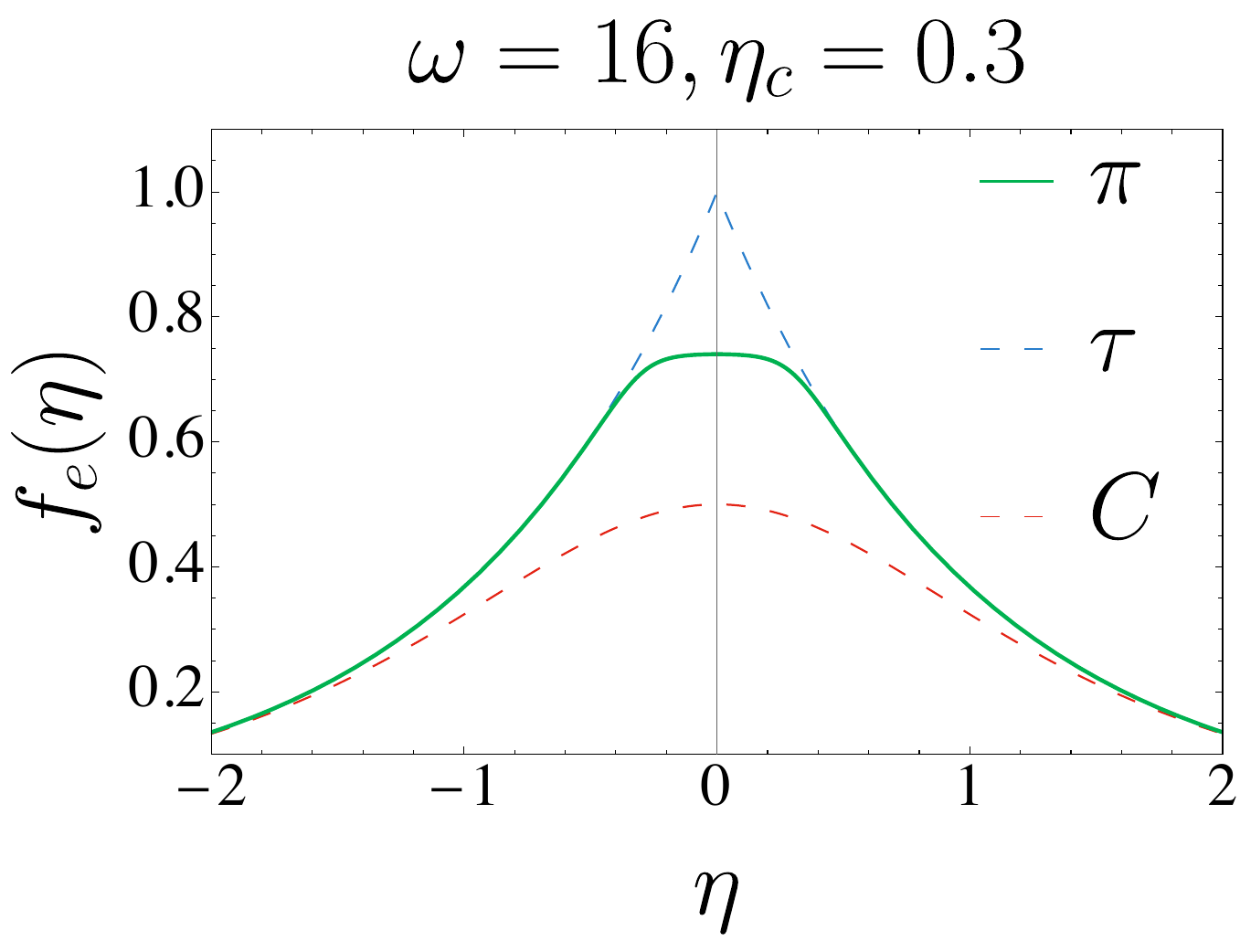}
    \label{fig:mp31}
\end{subfigure}
\begin{subfigure}{.32\textwidth}
    \centering
    \includegraphics[width=1\linewidth]{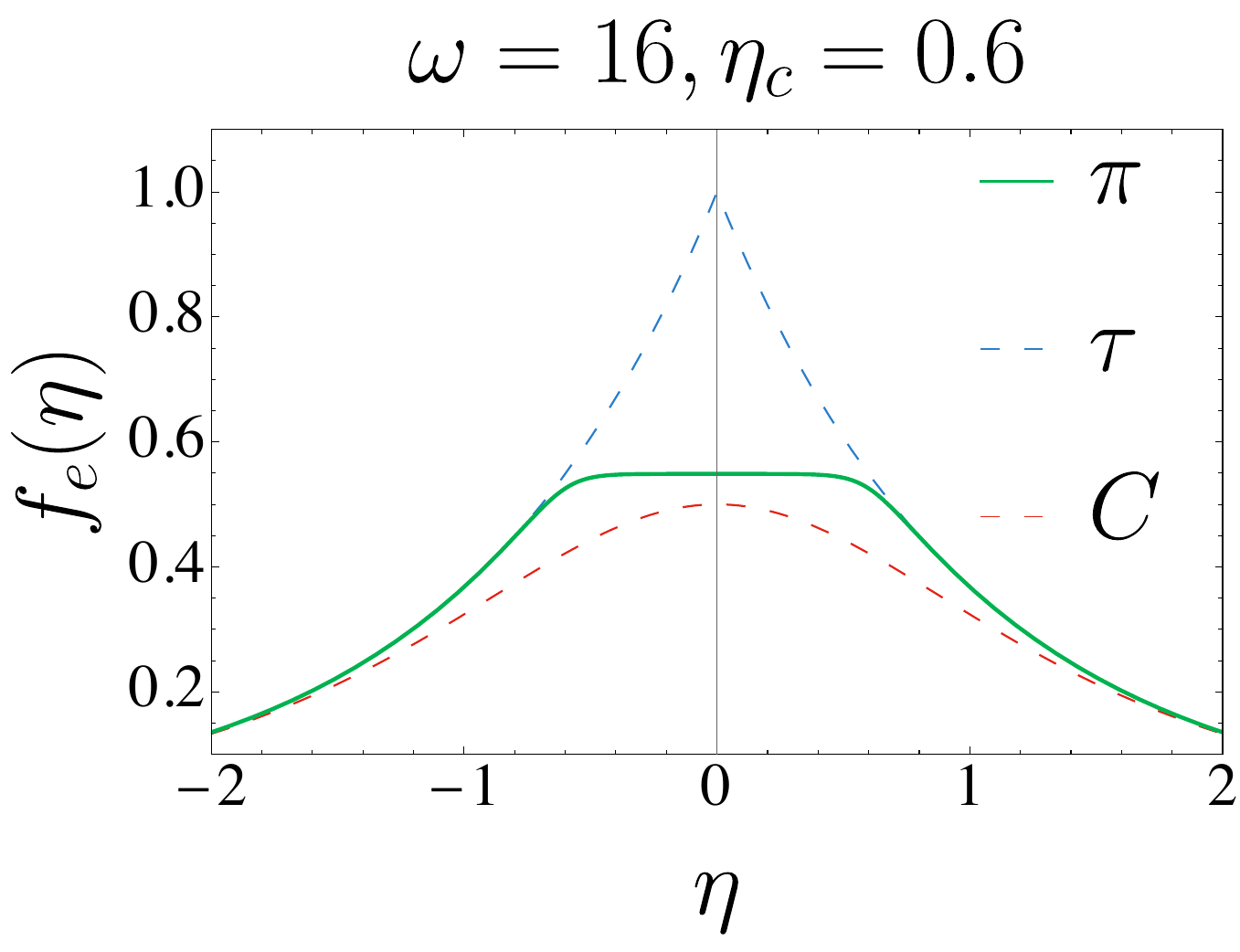}
    \label{fig:mp32}
\end{subfigure}
\begin{subfigure}{.32\textwidth}
    \centering
    \includegraphics[width=1\linewidth]{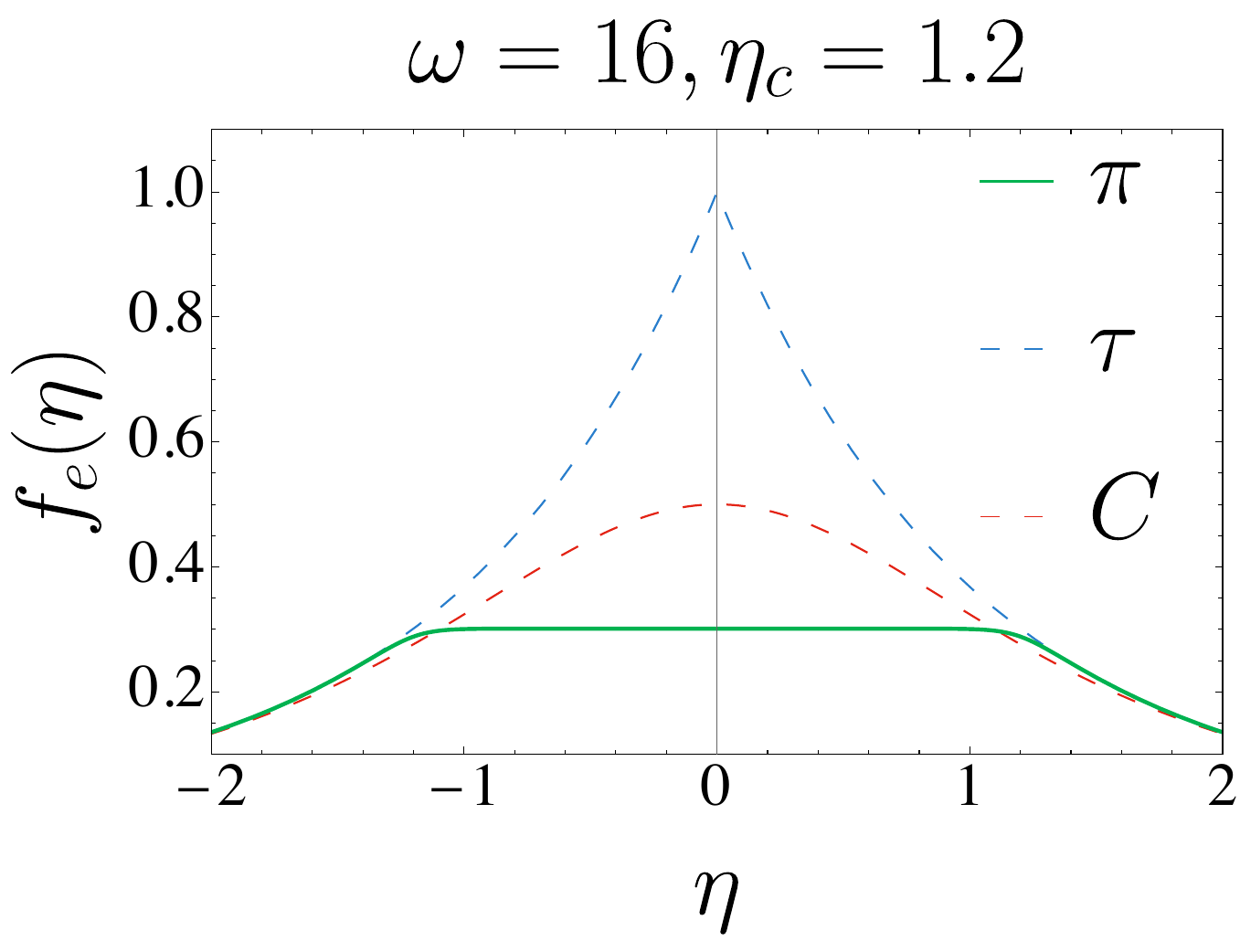}
    \label{fig:mp33}
\end{subfigure}
\caption{The rapidity weighting of the multiplicative plateau event shape, $\pi(\eta_c,\omega)$, for $\eta_c \in \{0.3,0.6,1.2\}$ and $\omega \in \{4,8,16\}$. The rapidity weighting of thrust and C-parameter are plotted for reference. Comparing panels from left to right one sees that increasing $\eta_c$ (for fixed $\omega$) leads to a wider plateau and reduced height at $\eta=0$. Comparing panels from top to bottom one sees that increasing $\omega$ (for fixed $\eta_c$) leads to  a more abrupt transition from the plateau region to the thrust-like region at large $|\eta|$.}
\label{fig:mp}
\end{figure}

We see that for $L_p$ angularity, it is not possible to have a significant flat region in $f_e$ near the origin of width $\sim 1$ in $\eta$ without setting $p$ to be very small and disrupting the behaviour of the observable in the collinear region. Let us now introduce an alternative observable where one can arrange this, which we'll call the \textit{multiplicative plateau angularity}, $\tau_{\pi,a}$. The rapidity weighting function for this observable is:
\begin{align}
    f_{\pi,a}(\eta;\omega,\eta_c) &=  \left( 
    e^{\omega |\eta|} 
    +
    e^{\omega \eta_c}\right)^{-\frac{(1-a)}{\omega}}\,,\label{eq:fmp}
    \\ \nonumber
    &= e^{-(1-a)|\eta|}\sigma(\eta;\omega,\eta_c)^{\frac{(1-a)}{\omega}}\,,
\end{align}
where $\sigma(\eta; \omega,\eta_c)$ is a standard logistic step function: 
\begin{equation}
    \sigma(\eta; \omega,\eta_c) = \frac{1}{1 + e^{-\omega(\vert{}\eta\vert{} - \eta_c)}}\,.
\end{equation}
In \cref{fig:mp} we have plotted \cref{eq:fmp} for several choices of $\omega$ and $\eta_c$, and $a=0$.
The function $f_{\pi,a}$ has a flat plateau of width $\sim \eta_c$ and height $\sim e^{-(1-a)\eta_c}$ for $\eta \sim 0$, and transitions to the asymptotic angularity behaviour over a rapidity range $\sim 1/\omega$ around $|\eta| = \eta_c$. The latter property means that one should not pick values of $\omega$ that are too small (i.e. $\omega \ll 1$), since then the transition will be smeared all the way into the collinear region. Further, if $\omega$ is sufficiently small, the transition region also extends all the way down to $\eta =0$ and causes $f_e$ to develop a kink there. One may see the latter property in the top-left panel of \cref{fig:mp}, or, more generally, by considering the expansion of $f_{\pi,a}$ for small $\eta$:
\begin{equation}
    f_{\pi,a}(\eta) = \left(1 + e^{\omega \eta_c}\right)^{-\frac{(1-a)}{\omega}} \left( 1 - \frac{(1-a) |\eta|}{1+e^{\omega \eta_c}}  + \mathcal{O}(|\eta|^2) \right)\,.
\end{equation}
Unless $\omega \eta_c \gg 1$, the coefficient of the $|\eta|^1$ cusp term is non-negligible. This particular behaviour at $\eta \sim 0$ is not intrinsically a problem -- thrust shares the same behaviour. 

For \cref{eq:fac} to apply, $\eta_c$ must fall in the soft region $\eta_c \sim \mathcal{O}(1)$, and be much smaller than the rapidity of the collinear modes $\eta_J \equiv \ln(1/\sqrt{\tau})$. If $\eta_c \sim \eta_J$ then the transition falls in the collinear region, and one would have to use the broadening soft function together with jet functions that depend on $\eta_c$ and $\omega$. This scenario is conceptually somewhat similar to the case of a jet veto with a jet rapidity cut, where for $\eta_{\textrm{cut}} \sim \ln(Q/p_{T,\textrm{cut}})$ one has a factorization formula with TMD soft functions and $\eta_{\textrm{cut}}$-dependent beam functions \cite{Michel:2018hui}. If $1 \ll \eta_c \ll \eta_J$ then a SCET$_{+}$ factorisation \cite{Procura:2014cba, Procura:2018zpn} should be appropriate, where the soft function is refactorised into a broadening soft function and two collinear-soft functions that can see the `knee' in the observable. Finally, if $\eta_c \gg \eta_J$, the transition is so far forward that it is no longer relevant, and the appropriate factorisation is just that of the broadening case.

\begin{figure}
\begin{subfigure}{.32\textwidth}
    \centering
    \includegraphics[width=1\linewidth]{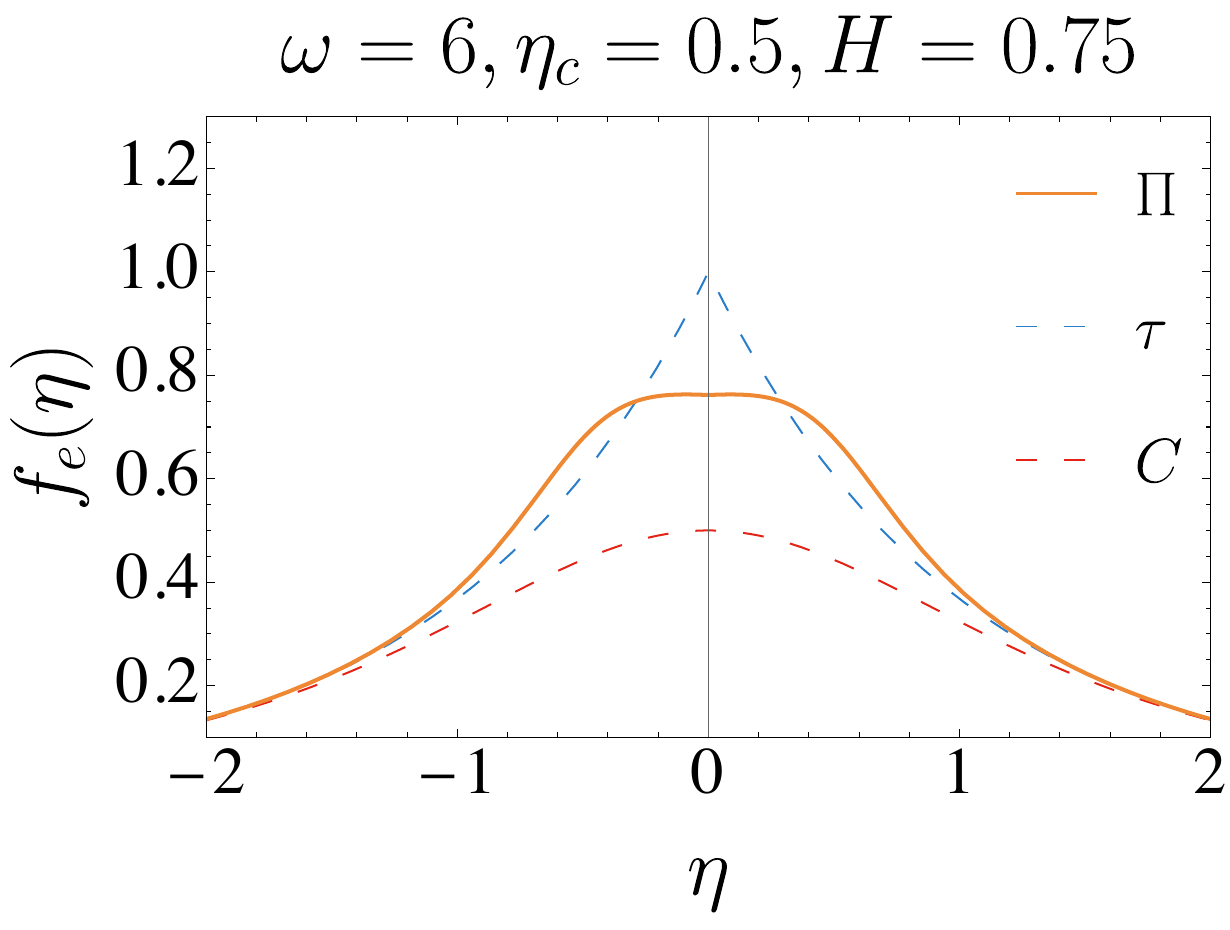}
    \label{fig:ap11}
\end{subfigure}
\begin{subfigure}{.32\textwidth}
    \centering
    \includegraphics[width=1\linewidth]{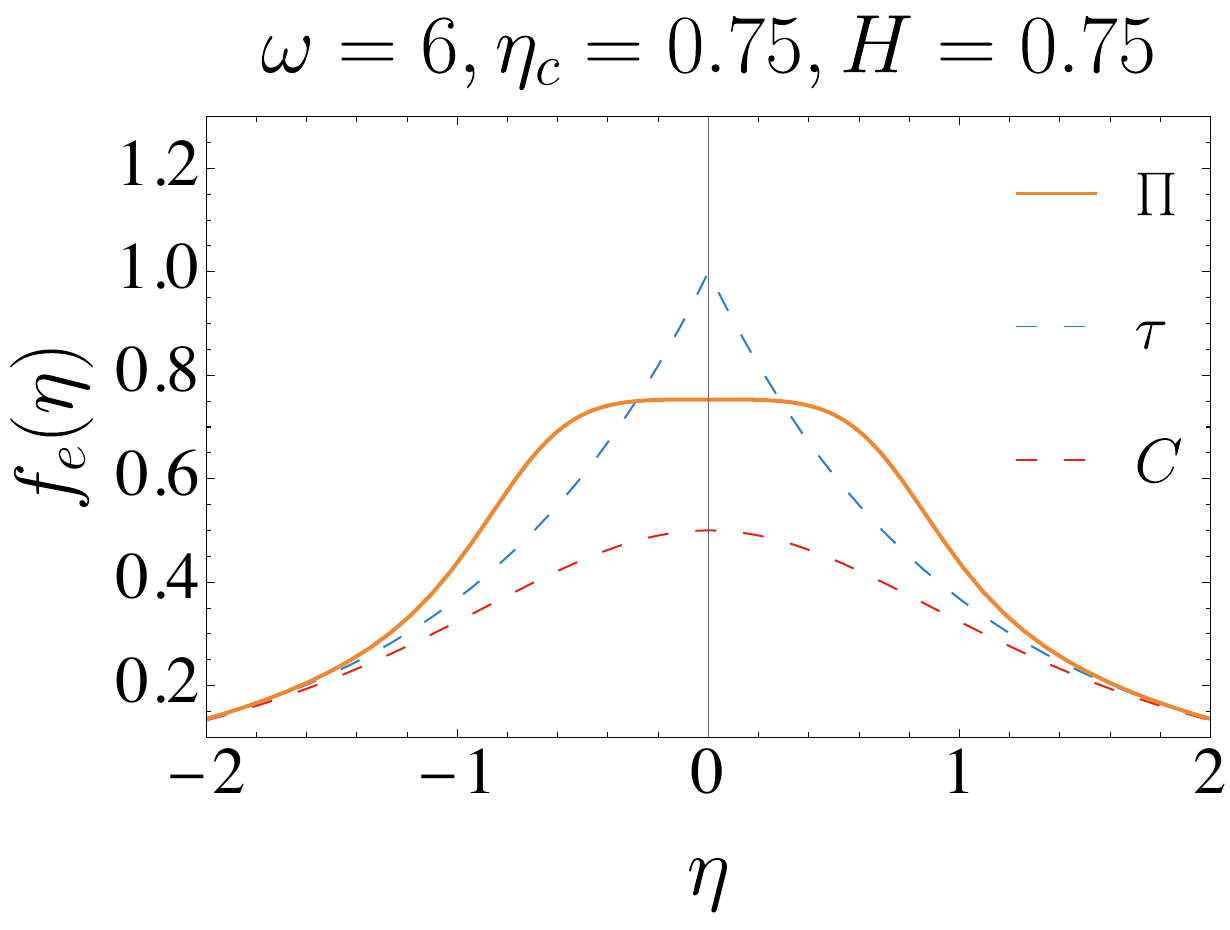}
    \label{fig:ap12}
\end{subfigure}
\begin{subfigure}{.32\textwidth}
    \centering
    \includegraphics[width=1\linewidth]{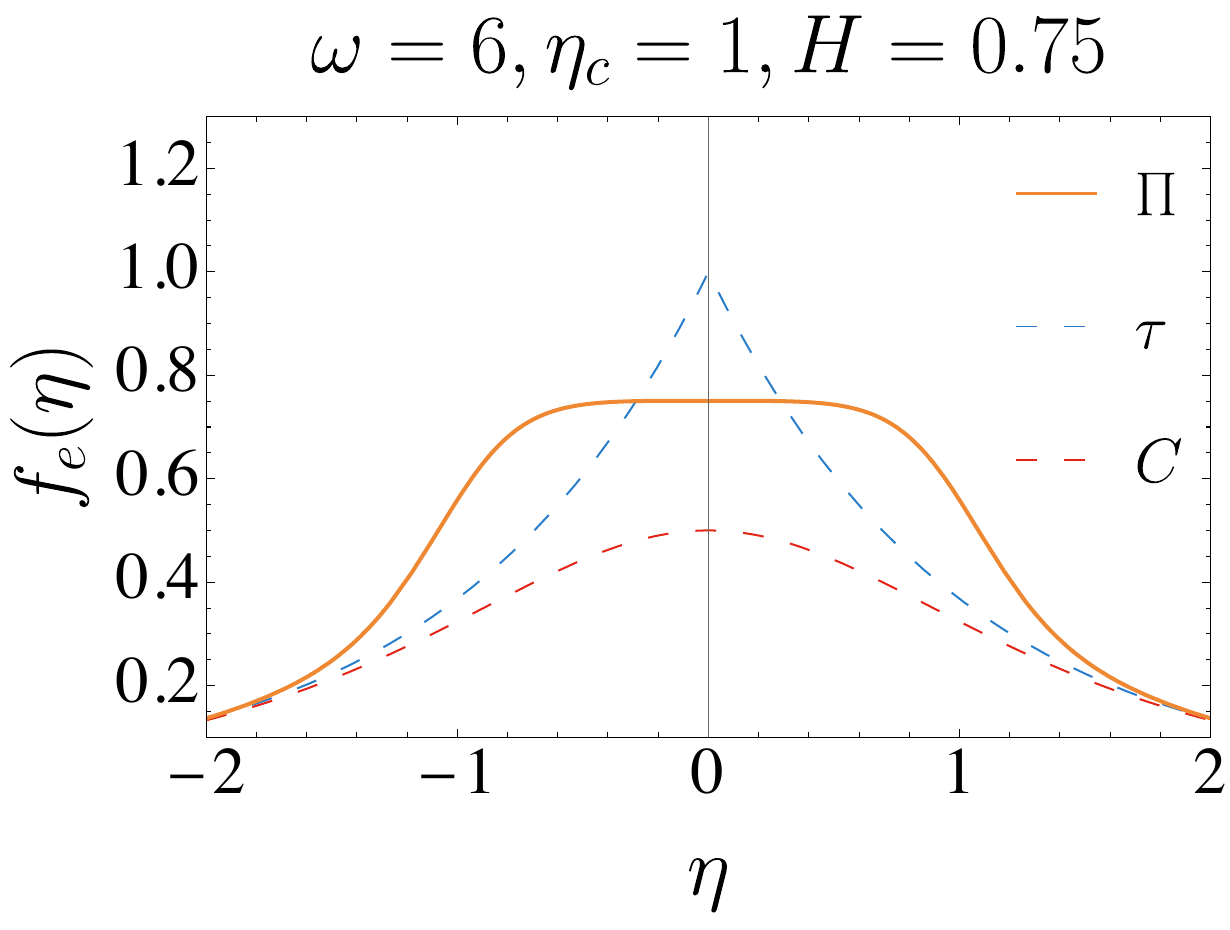}
    \label{fig:ap13}
\end{subfigure}
\\
\begin{subfigure}{.32\textwidth}
    \centering
    \includegraphics[width=1\linewidth]{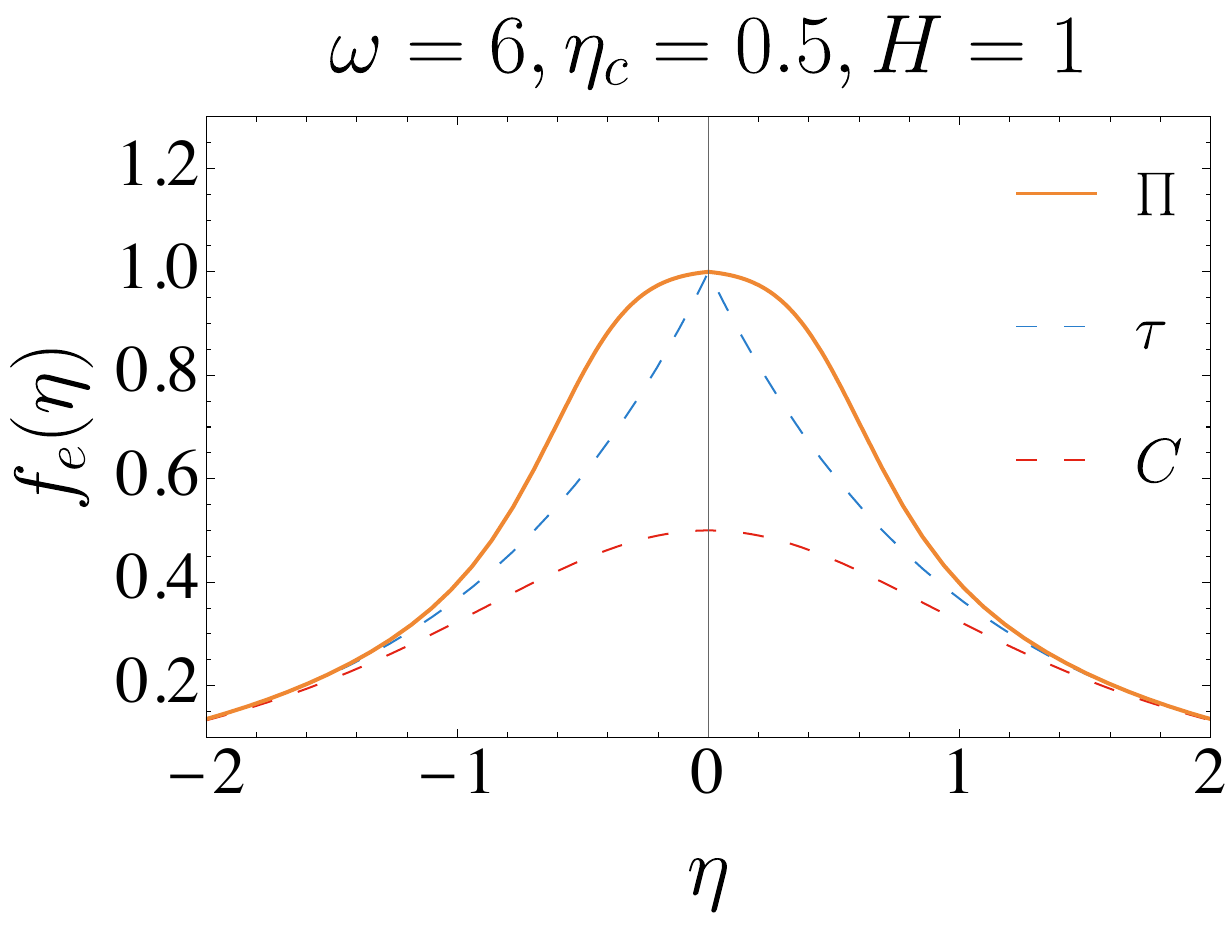}
    \label{fig:ap21}
\end{subfigure}
\begin{subfigure}{.32\textwidth}
    \centering
    \includegraphics[width=1\linewidth]{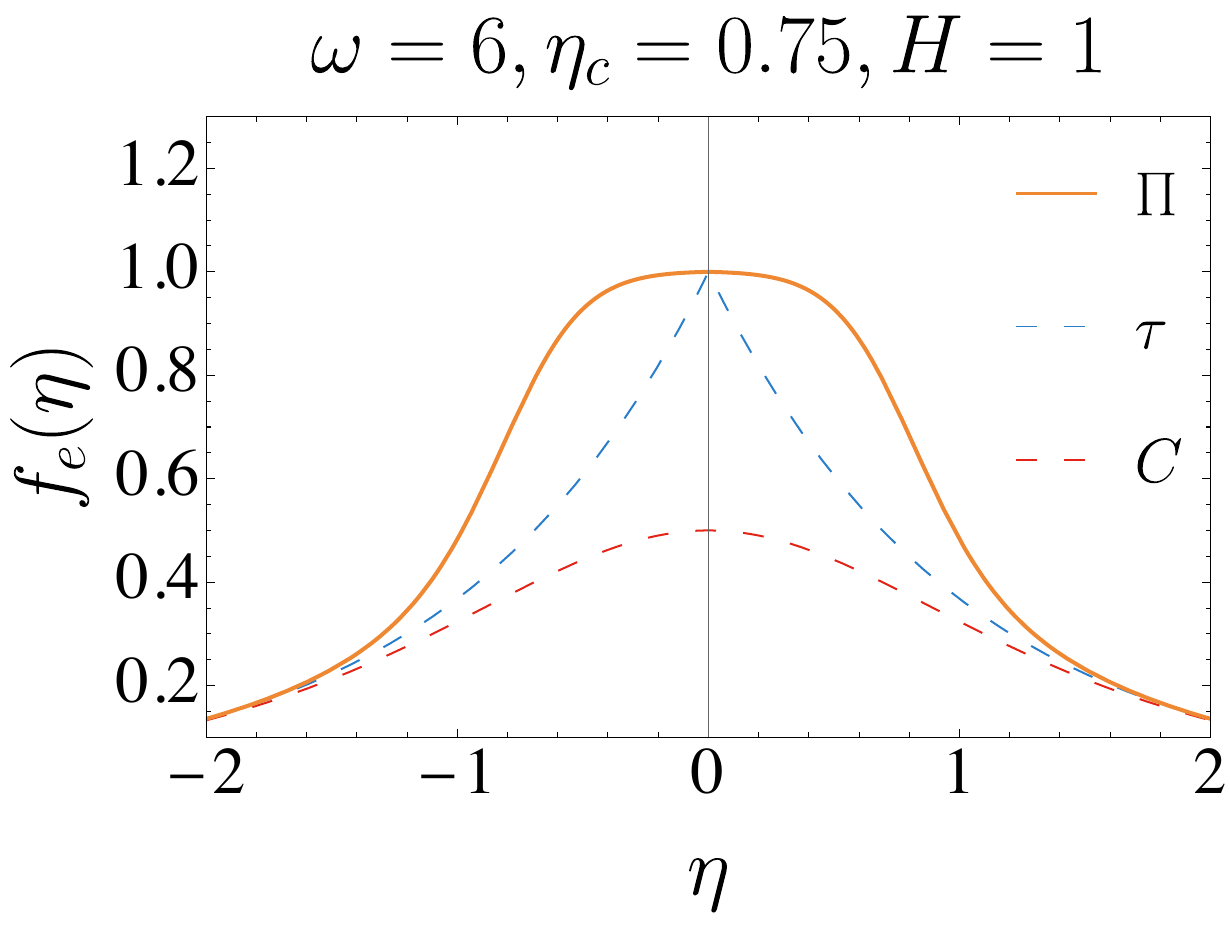}
    \label{fig:ap22}
\end{subfigure}
\begin{subfigure}{.32\textwidth}
    \centering
    \includegraphics[width=1\linewidth]{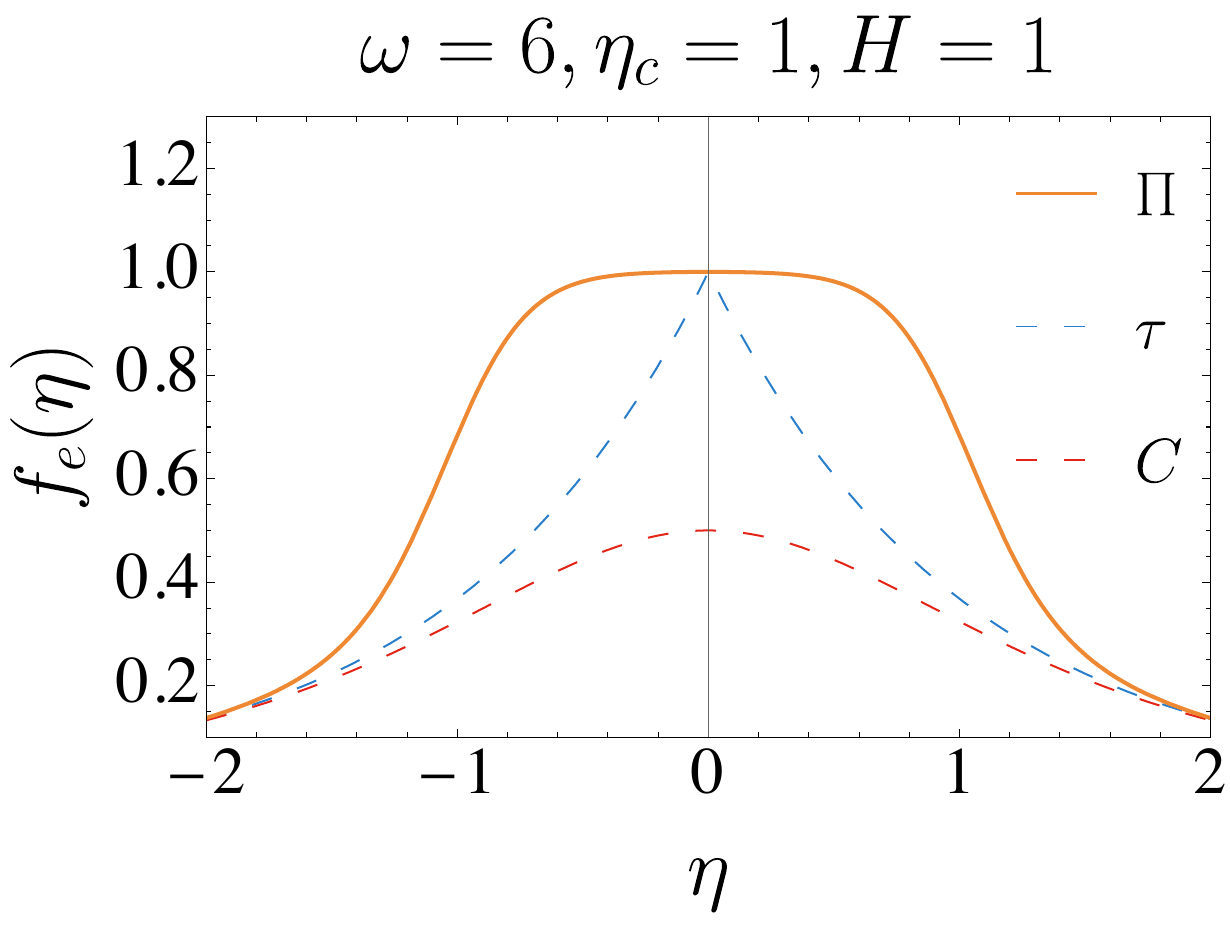}
    \label{fig:ap23}
\end{subfigure}
\\
\begin{subfigure}{.32\textwidth}
    \centering
    \includegraphics[width=1\linewidth]{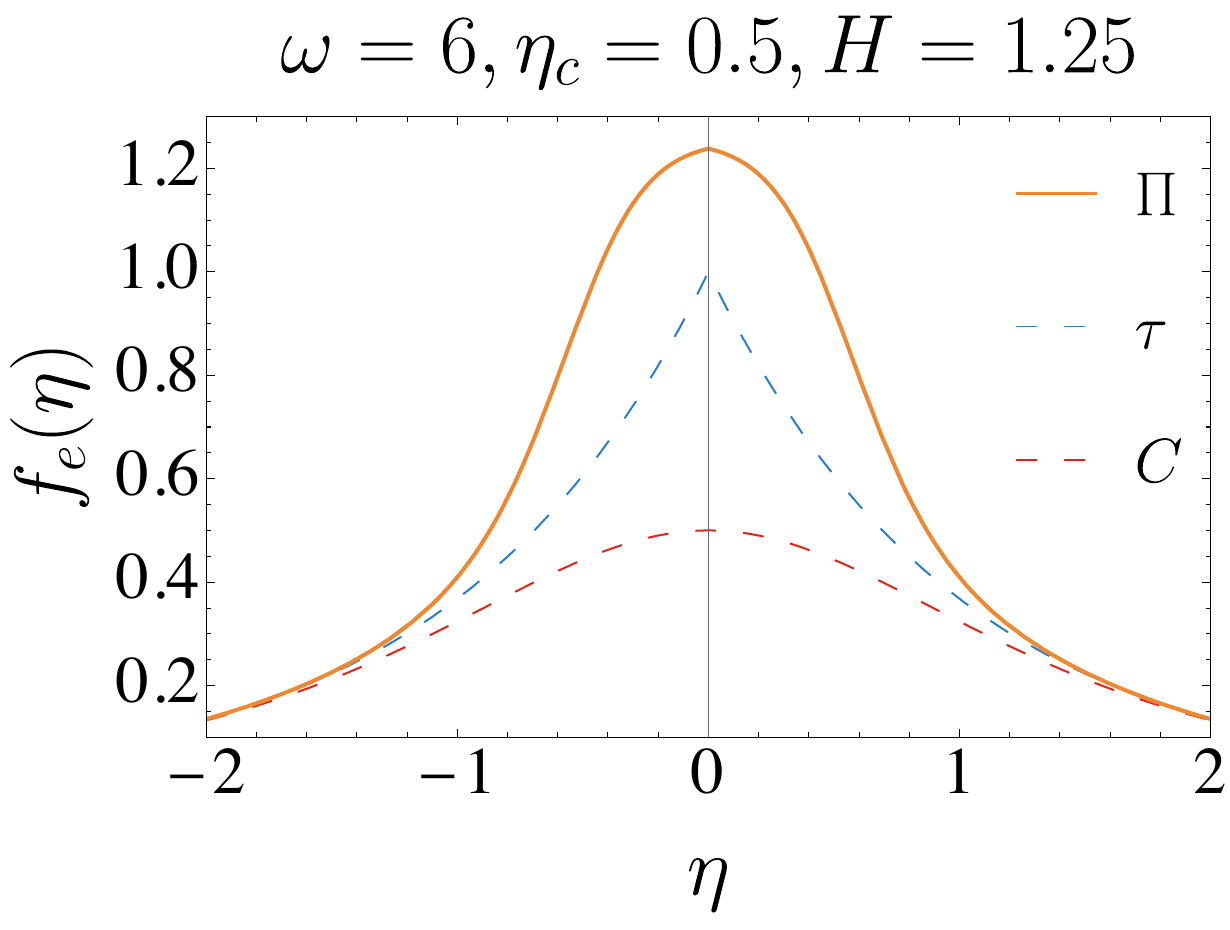}
    \label{fig:ap31}
\end{subfigure}
\begin{subfigure}{.32\textwidth}
    \centering
    \includegraphics[width=1\linewidth]{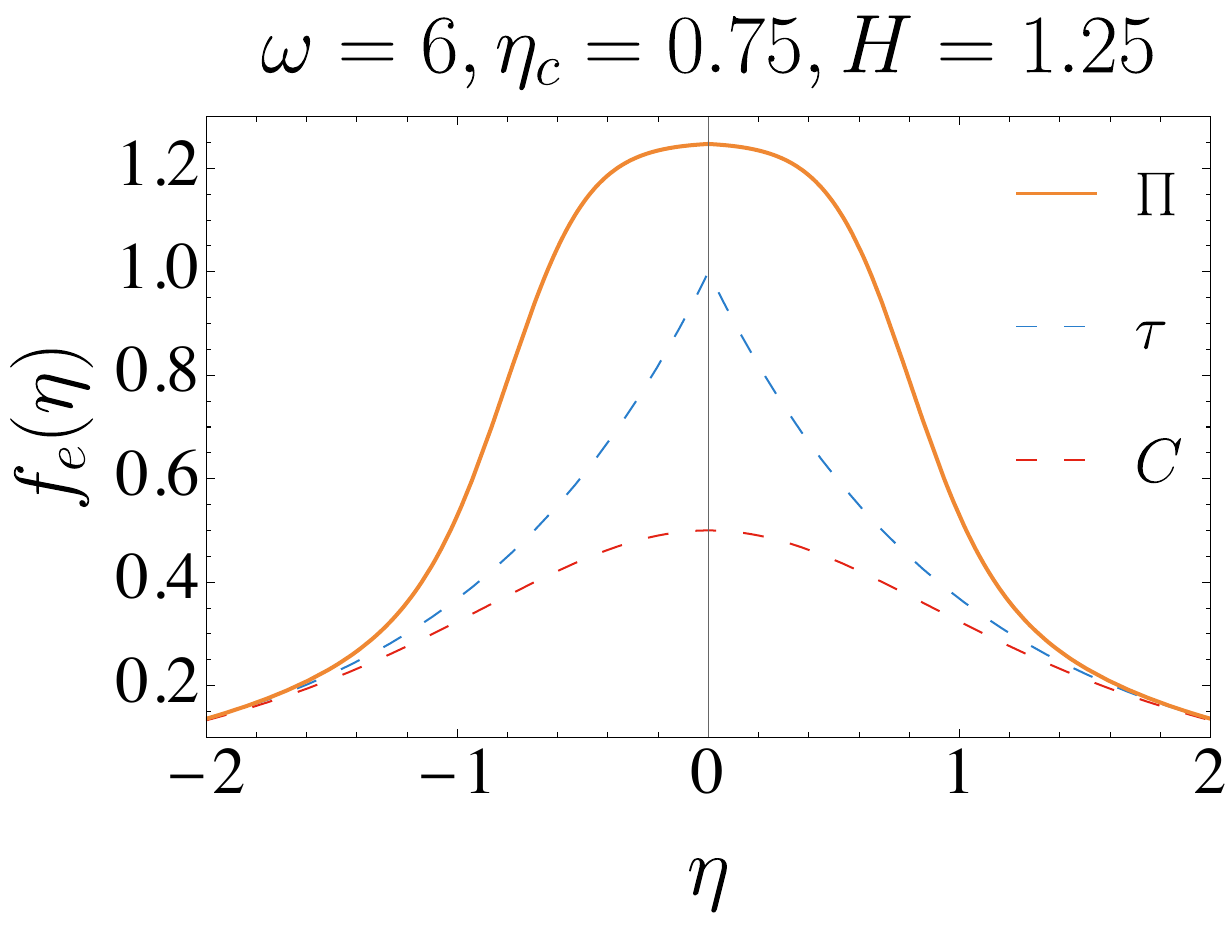}
    \label{fig:ap32}
\end{subfigure}
\begin{subfigure}{.32\textwidth}
    \centering
    \includegraphics[width=1\linewidth]{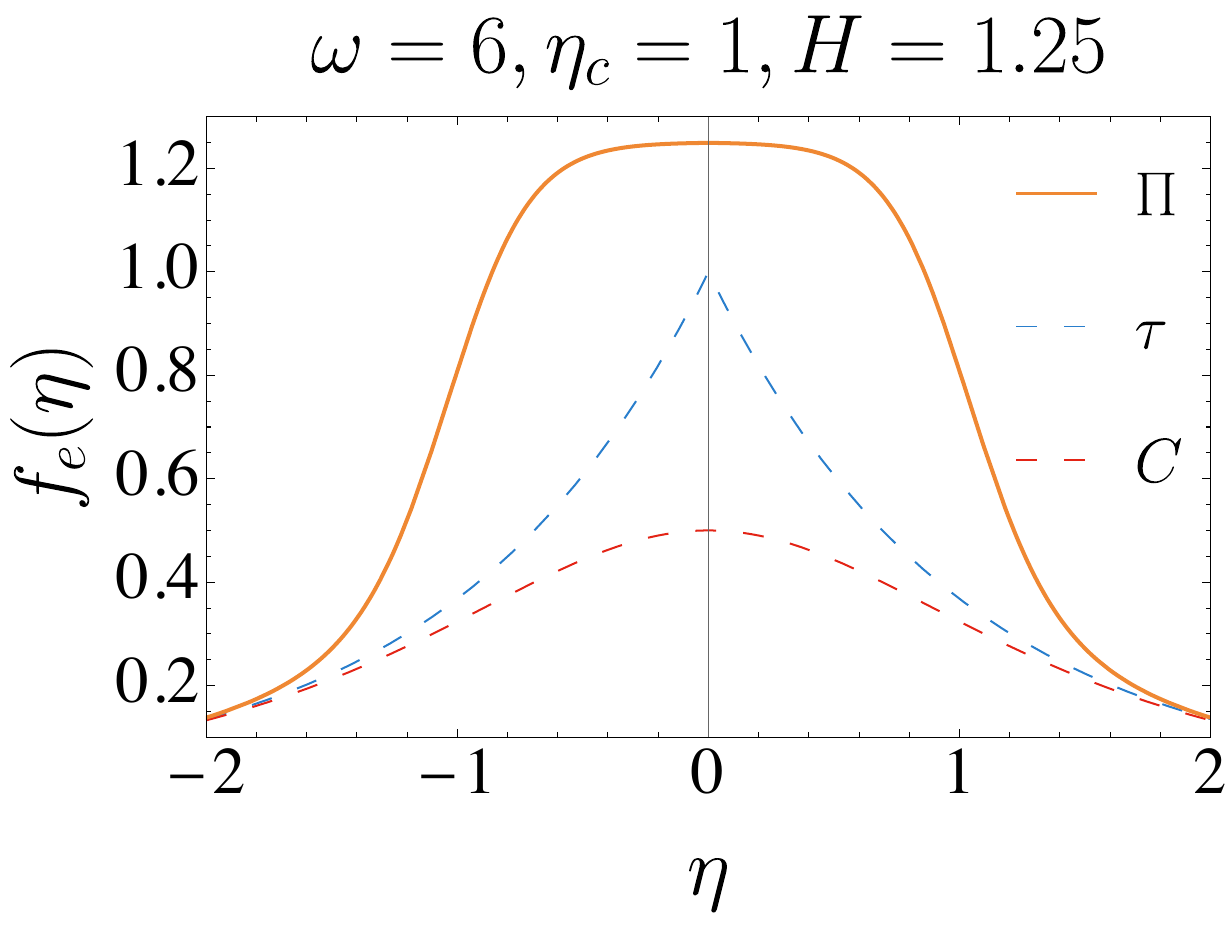}
    \label{fig:ap33}
\end{subfigure}
\caption{The rapidity weighting of the additive plateau event shape, $\Pi(\eta_c,\omega,H)$, for $\eta_c \in \{0.5,0.75,1\}$ and $H \in \{0.75,1,1.25\}$. The rapidity weighting of thrust and C-parameter are plotted for reference. Comparing panels from left to right one sees that increasing $\eta_c$ (for fixed $\omega,H$) leads to a wider plateau. Comparing panels from top to bottom one sees that increasing $H$ (for fixed $\omega,\eta_c$) leads to a higher plateau.}
\label{fig:AddPlateau}
\end{figure}

\begin{figure}
\begin{subfigure}{.49\textwidth}
    \centering
    \includegraphics[width=1\linewidth]{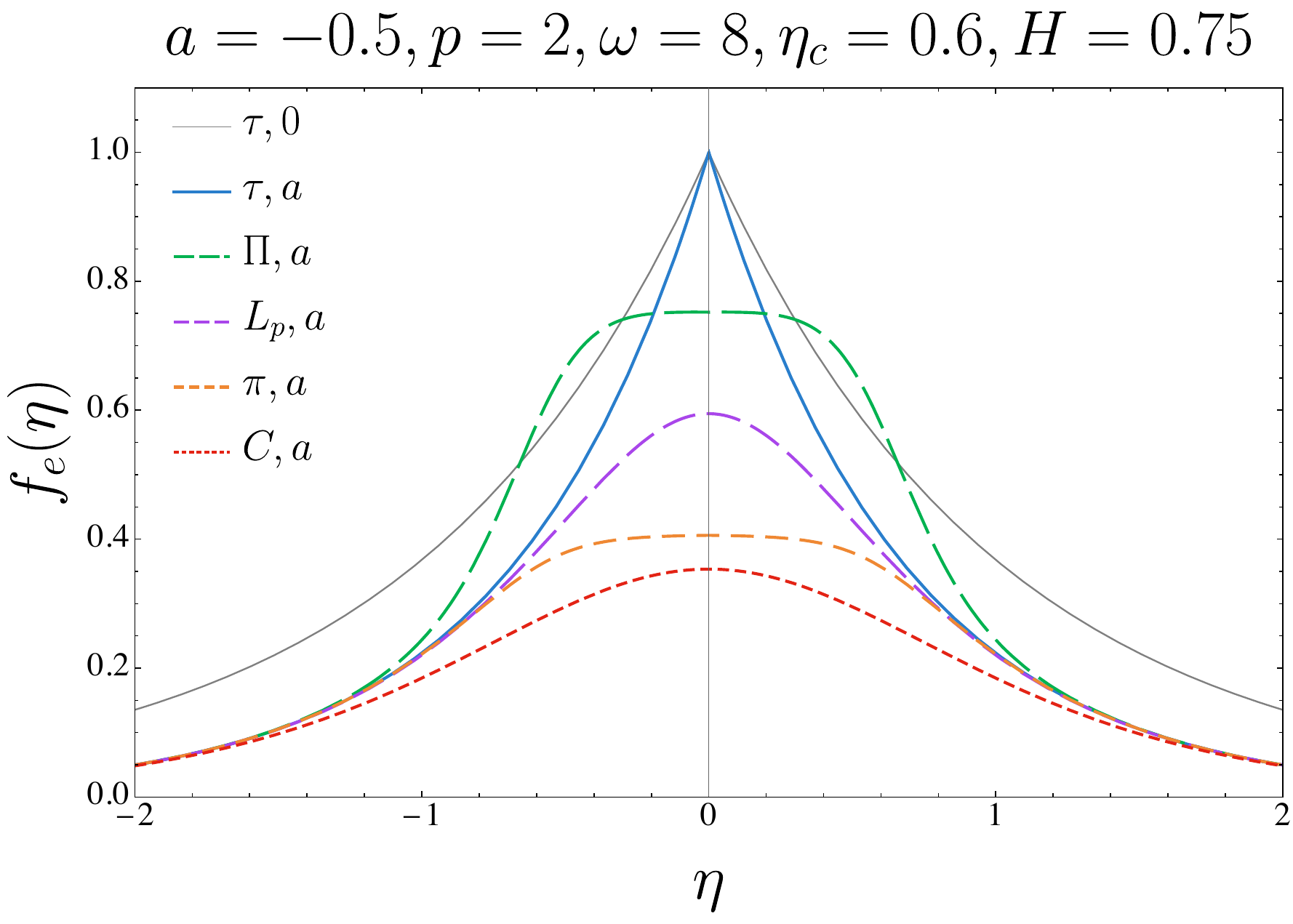}
    \caption{}
    \label{fig:fe_am05}
\end{subfigure}
\begin{subfigure}{.49\textwidth}
    \centering
    \includegraphics[width=1\linewidth]{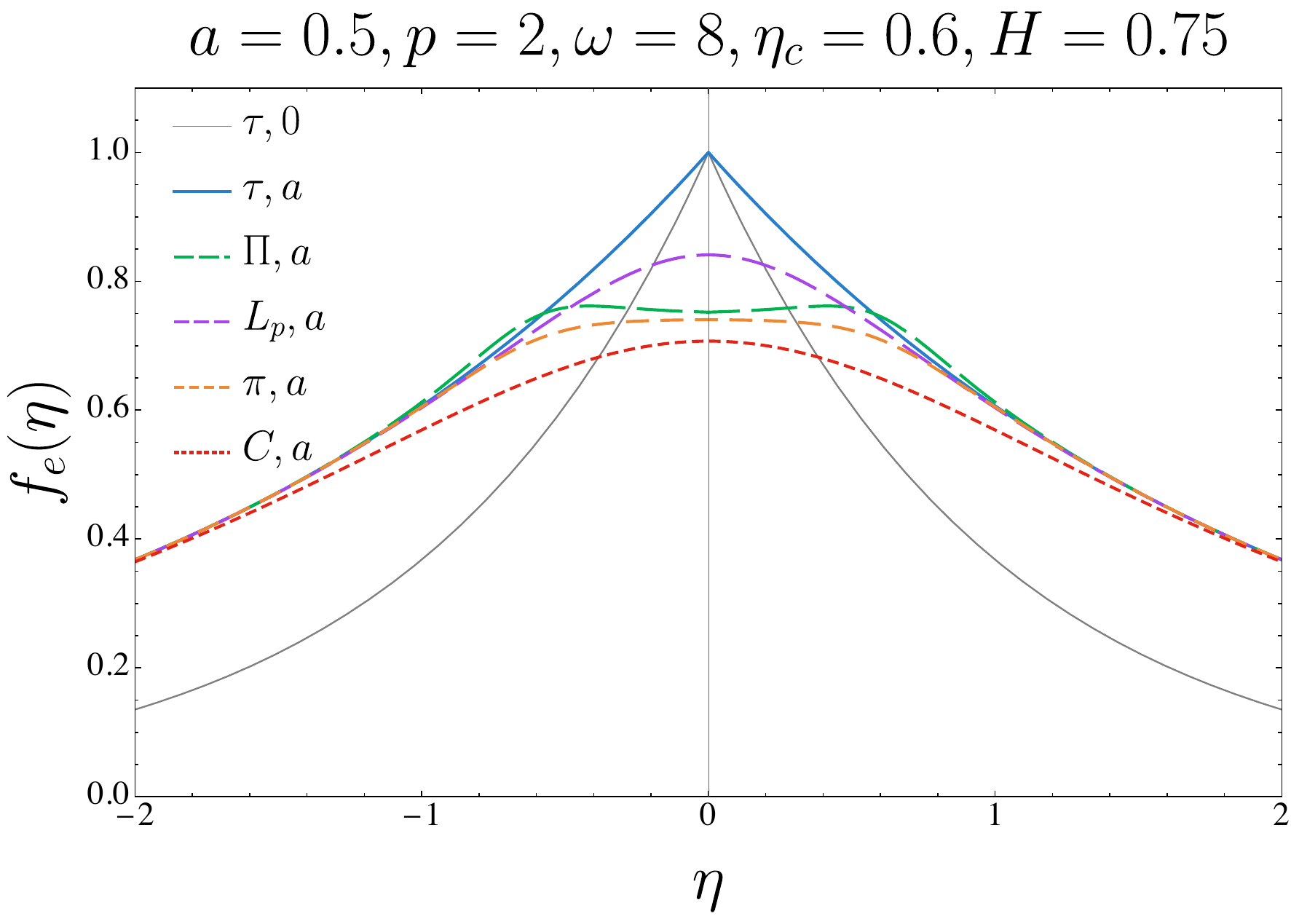}
    \caption{}
    \label{fig:fe_a05}
\end{subfigure}
\caption{One representative of each family of generalised angularity in \cref{eq:gen_ang} for (a) $a=-0.5$ and (b) $a=+0.5$. The rapidity weighting of thrust is also plotted in gray for reference.}
\label{fig:vary_a}
\end{figure}

It is easy to see that the rapidity weighting functions for $L_p$-angularity and the multiplicative plateau angularity lie below that of standard angularity for all $\eta$, approaching it from below as $|\eta|$ increases. One might wish to use an alternative plateau observable where part (or all) of the plateau can extend above the angularity curve, and the weighting function then tends to the angularity one from above as $|\eta| \to \infty$. We can achieve this using an additive matching procedure between the $\eta \to 0$ plateau behaviour and the $\eta \to \infty$ decay:
\begin{equation}
    f_{\Pi,a}(\eta;\omega,\eta_c,H) = H \Big( 1 - \sigma(\eta; \omega,\eta_c) \Big) + e^{-(1-a) \vert{}\eta\vert{}} \sigma(\eta; \omega,\eta_c)\,.
\end{equation}
Here $H$ is a further parameter that dictates the height of the plateau. Let us call this observable the \textit{additive plateau angularity}. Similar considerations apply for  $\omega$ and $\eta_c$ here as for the multiplicative angularity case; for this case we additionally note that we must have $\omega > (1-a)$ in order that the asymptotic behaviour is controlled by $a$ rather than $\omega$. The height $H$ should be chosen to be $\sim 1$. Values of $H \gg 1$ will yield a very aggressive veto in the soft region which risks extending into the non-perturbative region and/or inducing non-global logarithms. Further, for the rapidity weighting function to be monotonically decreasing as $|\eta|$ increases, $H$ cannot be too small. In particular it must satisfy:
\begin{equation}
    H \ge 1 - \frac{1-a}{\omega} \left( 1 + e^{-\omega \eta_c} \right)
\end{equation}
There is no intrinsic issue with non-monotonically decreasing functions of $|\eta|$ -- an example of such a shape is depicted in \cref{fig:fe_a05}.

To conclude this section, we summarise the various families of generalised angularity event shapes discussed above, and the relationships among them, in the following diagram:
\begin{equation}
\begin{tikzcd}
	{\color[HTML]{00B250}\pi,a} &&&&&&& \\
	&&& 
    {\color[HTML]{297ECC}\tau,a} 
    && 
    {\color[HTML]{A946EA}L_p,a} 
    &&
    {\color[HTML]{E32315}C,a} \\
	{\color[HTML]{EF8833}\Pi,a}
	\arrow[
    "{\eta_c=0,\omega\to\infty}",
    sloped,
    from=1-1, to=2-4
    ]
	\arrow["{p\to\infty}", from=2-6, to=2-4]
	\arrow["{p=1}"', from=2-6, to=2-8]
	\arrow[
        "{\eta_c=0,\omega\to\infty,H=1}"',
    sloped,
    from=3-1, to=2-4
]
\end{tikzcd}
\label{eq:gen_ang}
\end{equation}
\Cref{fig:Lp,fig:mp,fig:AddPlateau} all plot $f_e$ for $a=0$. To illustrate the dependence on $a$, in \cref{fig:vary_a} we compare the rapidity weighting for representatives from each of these five generalised angularities for $a=\pm \tfrac{1}{2}$.

\section{Properties of the soft function} \label{sec:properties}
In this section we establish some basic properties of the soft function for any member of the class of generalised angularities. When referring to a generic member of this class, we shall simply use the subscript $e$, suppressing the subscript $a$ for notational simplicity. We use standard dimensional regularisation in $4-2 \epsilon$ dimensions for bare quantities, and perform renormalization using the $\msbar$ scheme.

The ultraviolet renormalisation of the soft function is given by a convolution
\begin{equation}
    S_{e}^{\bare}(\mathcal{T}) = Z_{e}(\mathcal{T},\mu) \otimes_{\cT} S_{e}(\mathcal{T},\mu)\,,
    \label{eq:ren}
\end{equation}
where we use the short-hand notation for a convolution
\begin{equation}
    f(\cT) \otimes_\cT g(\cT) = \int_0^\cT \mathrm{d} \cT^\prime f(\cT-\cT^\prime) g(\cT^\prime)\,.
\end{equation}
The renormalised soft function satisfies the all-order renormalisation group equation (RGE)
\begin{equation}
    \mu \frac{\mathrm{d}}{\mathrm{d} \mu}S_{e}(\mathcal{T},\mu) = 
\left[
\Gamma_e \, \mathcal{L}_0(\mathcal{T},\mu) 
+\hat{\gamma}_e \, \delta(\mathcal{T}) \right]  \otimes_{\mathcal{T}}S_e(\mathcal{T},\mu)\,,
\label{eq:SRGE}
\end{equation}
where we have decomposed the anomalous dimension into a cusp ($\Gamma_e$) and non-cusp ($\hat{\gamma}_e$) term. Here we use the two-argument plus distributions of ref.~\cite{Ebert:2016gcn}:
\begin{equation}
       \cL_n(k,\mu) = \dfrac{1}{\mu}\cL_n\left(\dfrac{k}{\mu}\right)
\end{equation}
where
\begin{align}
\cL_n(x)=\left[\frac{\theta(x) \ln^n x}{x}\right]_{+}=\lim _{\epsilon \rightarrow 0} \frac{\mathrm{d}}{\mathrm{d} x}\left[\theta(x-\epsilon) \frac{\ln^{n+1} x}{n+1}\right]\,.
\end{align}
We expand all quantities as a series in $\alpha_s(\mu)/(4\pi)$, with $\alpha_s(\mu)$ the renormalised coupling. We use a superscript $(n)$ to refer to the term proportional to $[\alpha_s(\mu)/(4\pi)]^n$ in this series, except for anomalous dimensions, where the superscript $n$ refers to the term proportional to $[\alpha_s(\mu)/(4\pi)]^{n+1}$ (since these begin only at $\mathcal{O}(\alpha_s^1)$). Expanding \cref{eq:SRGE} in the strong coupling, and solving up to two loops, we obtain,
\begin{subequations}
\begin{align}
    S_e^{(0)}&=\delta(\cT)\,,\\
    S_e^{(1)}&=-\Gamma^0_e \, \cL_1(\cT,\mu) 
    -\hat{\gamma}^0_e \, \cL_0(\cT,\mu) 
    +S_{e,\delta}^{(1)} \,  \delta(\cT)\,,
    \label{eq:SRGE1}\\
\begin{split}
    S_e^{(2)}&=\frac{1}{2}(\Gamma^0_e)^2 \, \cL_3(\cT,\mu) 
    +
    (\beta_0 +\frac{3}{2} \hat{\gamma}^0_e ) \Gamma^0_e \, \cL_2(\cT,\mu)\\
    &+
    \left[
    \hat{\gamma}^0_e (2 \beta_0 + \hat{\gamma}^0_e) 
    - S_{e,\delta}^{(1)} \, \Gamma^0_e 
    - \Gamma^1_e 
    - (\Gamma^0_e)^2 \zeta_2
    \right] \, \cL_1(\cT,\mu) \\
    &+
    \left[
    \Gamma^0_e(-\hat{\gamma}^0_e \zeta_2 +\Gamma^0_e \zeta_3)
    -\hat{\gamma}^1_e
- S_{e,\delta}^{(1)} (2 \beta_0 + \hat{\gamma}^0_e)\ 
     \right] \, \cL_0(\cT,\mu)
     +S_{e,\delta}^{(2)} \,  \delta(\cT)\,.
     \label{eq:SRGE2}
    \end{split}
\end{align}
\end{subequations}
where $\beta_0 = (11 C_A-4 n_f T_F)/3$ and we have used the leading-order result $S_{e,\delta}^{(0)}=1$. From \cref{eq:SRGE1} we conclude that the one-loop soft function is fully determined by the one-loop anomalous dimension coefficients $\Gamma^0_e$ and $\hat{\gamma}^0_e$,
and the one-loop boundary coefficient $S_{e,\delta}^{(1)}$. In turn, the two-loop soft function is fully determined by \smash{$\Gamma^1_e,\hat{\gamma}^1_e$} and \smash{$S_{e,\delta}^{(2)}$}.

We outline our strategy by investigating the one-loop soft function. As noted in ref.~\cite{Hoang:2014wka}, the one-loop soft function can be written in the form
\begin{align}
    S_{e}^{\mathrm{bare}(1)}(\mathcal{T}) &=
    \frac{8 C_R \, e^{\epsilon \gamma_E}}{\Gamma(1-\epsilon)}
    \,
    \int \frac{\mathrm{d} k^+ \mathrm{d} k^- }{k^+  k^-}
\left(\frac{\mu^2}{k^+  k^-}\right)^{\epsilon} \delta\left(\mathcal{T}_{e}(k^+, k^-)-\mathcal{T}\right)\,,
\label{eq:SNLOint}
    \\
    &=\frac{8 C_R \, e^{\epsilon \gamma_E}}{\Gamma(1-\epsilon)} F_e(\epsilon) \frac{1}{\mu}\left(\frac{\mathcal{T}}{\mu} \right)^{-1-2\epsilon}\,,
    \label{eq:SNLO}
\end{align}
where $C_R$ is the quadratic casimir of the representation $R$ of the final-state Wilson lines. In all plots and numerical evaluations we take $C_R=C_F$. The only dependence on the event-shape weight, $f_e$, is contained in the integral
\begin{equation}
F_e = \int_{-\infty}^{\infty} \mathrm{d} \eta \, \big( f_e(\eta) \big)^{2 \epsilon}\,.
\label{eq:Fe}
\end{equation}
We wish to find expressions for the anomalous dimension and boundary coefficients in terms of this integrated shape function. To this end, it will be useful to expand $F_e$ as a Laurent expansion in $\epsilon$, 
\begin{equation}
    F_e = \sum_{i=-1}^{\infty} \epsilon^i \, F_e^{[i]} \,.
    \label{eq:Fei}
\end{equation}
The pole in $\epsilon$ arises from the large-rapidity tails of the integral, and can be computed by integrating over the forward and backward asymptotic limits,
\begin{equation}
    \frac{F_e^{[-1]}}{\epsilon}=
    \int_{-\infty}^{0} \mathrm{d} \eta \left[
    \lim_{\eta \to -\infty}\big( f_e(\eta) \big)^{2 \epsilon}
    \right]+
    \int_{0}^{\infty} \mathrm{d} \eta \left[
    \lim_{\eta \to \infty}\big( f_e(\eta) \big)^{2 \epsilon}
    \right]\,.
    \label{eq:Fem1}
\end{equation}
We can then deduce that $F_e^{[0]}=0$ for all event shapes $e$. To see this, consider the difference
\begin{equation}
F_e - \frac{F_e^{[-1]}}{\epsilon}
=
2\int_{0}^{\infty} \mathrm{d} \eta \, \left( f_e(\eta)^{2 \epsilon} -\left[
    \lim_{\eta \to \infty}\big( f_e(\eta) \big)^{2 \epsilon}
    \right]\right)\,. \label{eq:1Lfullmasy}
\end{equation}
Expanding in $\epsilon$ we note that the $\cO(\epsilon^0)$ term always vanishes.

For many of the event shapes reviewed in \cref{sec:setup} the integral \cref{eq:Fe} can be performed to all orders in $\epsilon$. In particular, the integrated event shapes for angularity and C-angularity are
\begin{align}
F_{\tau,a}(a) &= \frac{1}{(1-a)\epsilon} \,, \qquad
\qquad
F_{C,a}(a) = \frac{\Gamma(\epsilon(1-a))^2}{2\Gamma(2\epsilon(1-a))} \,.
\label{eq:Fea}
\end{align}
We note that $F_{\tau,a}^{[-1]}=F_{C,a}^{[-1]}$, in line with our knowledge that these two observables coincide in the $\eta \to \infty$ limit. Taking $a\to 0$ we recover the integrated event shapes for thrust and C-parameter respectively~\cite{Hoang:2014wka}
\begin{align}
F_{\tau} = F_{\tau,a}(0) = \frac{1}{\epsilon} \,, \qquad \qquad
F_{C} = F_{C,a}(0)=\frac{\Gamma(\epsilon)^2}{2\Gamma(2\epsilon)} \,.
\label{eq:Fe0}
\end{align}
We can also perform the integral \cref{eq:Fe} for the two-parameter event-shape $L_p$-angularity,
\begin{align}
F_{L_p,a} (p,a)&=\frac{\Gamma\left(\epsilon\left(\frac{1-a}{p}\right)\right)^2}{2 \, p \,\Gamma\left(2\epsilon\left(\frac{1-a}{p}\right)\right)}
\,,
\qquad
F_{L_p,a}(p,0) =F_{L_p}(p)=\frac{\Gamma\left(\frac{\epsilon}{p}\right)^2}{2 \, p \,\Gamma\left(\frac{2\epsilon}{p}\right)}\,,
\label{eq:FLpa}
\end{align}
which in the limits $p \to \infty$ or $p \to 1$ correctly reproduces the expressions in \cref{eq:Fea}. The corresponding result for the multiplicative plateau angularity is:
\begin{equation}
    F_{\pi,a}(\eta_c,\omega,a) = \frac{2e^{-2 (1-a) \epsilon \eta_c}}{\omega} B\left( \frac{1}{1 + e^{-\eta_c \omega}}; \frac{2 (1-a) \epsilon}{\omega}, 0 \right)
    \label{eq:Fpia}
\end{equation}
where $B(z;x,y)$ is the incomplete beta function. Taking the limit of large $\omega$,
\begin{equation}
    \lim_{\omega \to \infty} F_{\pi,a}(\eta_c,\omega,a) = 
    e^{-2 (1-a) \epsilon \eta_c}
    \left(
    \frac{1}{(1-a)\epsilon}+2 \eta_c
    \right)
\end{equation}
we see that further setting $\eta_c=0$ recovers $F_{\tau,a}$.

For the additive plateau angularity we were not able to obtain a closed form expression for general $\epsilon$. However, as we shall see below, to compute the soft function to a given fixed order, $\alpha_s^n$, we only need to consider the truncation of the Laurent expansion of \cref{eq:Fei} to $\mathcal{O}(\epsilon^n)$. For the convenience of the reader, we collect the $\mathcal{O}(\epsilon^1)$ coefficients for the three new families of angularity event shapes discussed above, which are needed for the NLO soft function, where the additive plateau angularity coefficient does happen to have a simple closed-form representation:
\begin{subequations}
\label{eq:F1list}
\begin{align}
F^{[1]}_{L_p,a} &= \frac{(a-1)}{p^2} \zeta_2
\,, \\
F^{[1]}_{\pi,a} &= \frac{4(1-a)}{\omega^2}\mathrm{Li}_2(-e^{\eta_c \omega})
\,, \\
F^{[1]}_{\Pi,a} &=4 \left( \frac{\text{Li}_2(-e^{\omega\eta_c})}{\omega} - \frac{\text{Li}_2(-H e^{\omega\eta_c})}{\omega + a - 1} \right)\,.
\end{align}
\end{subequations}
For the NNLO soft function we will need in addition the $\mathcal{O}(\epsilon^2)$ coefficients:
\begin{subequations}
\label{eq:F2list}
\begin{align}
F^{[2]}_{L_p,a} &= \frac{2(1-a)^2}{p^3} \zeta_3
\,, \\
\begin{split}
F^{[2]}_{\pi,a} &= \frac{8 (1-a)^2}{\omega ^3}
    \Bigg[
    \zeta_3
    -\mathrm{Li}_3\left(-e^{\eta_c \omega }\right)-\text{Li}_3\left(1+e^{\eta_c \omega }\right)
    \\
    &+\frac{1}{2} \ln \left(-e^{\eta_c \omega }\right) \ln ^2\left(1+e^{\eta_c \omega }\right)+\ln \left(1+e^{\eta_c \omega }\right) \text{Li}_2\left(1+e^{\eta_c \omega }\right)\Bigg]\,
\,, 
\end{split}
\\
\begin{split}
    F^{[2]}_{\Pi,a} &= 8(1-a) \left[ \frac{\text{Li}_3(-H e^{\omega\eta_c})}{(\omega + a - 1)^2} - \frac{\text{Li}_3(-e^{\omega\eta_c})}{\omega^2} \right] \\ &\quad + 4 \int_0^\infty \ln^2\left( \frac{1 + H e^{\omega\eta_c} e^{-(\omega-(1-a))\eta}}{1 + e^{\omega\eta_c} e^{-\omega\eta}} \right)  d\eta\,.
\end{split}
\end{align}
\end{subequations}
Here the additive plateau angularity coefficient has been left in terms of an  integral which may be computed numerically. While closed-form coefficients are a nice feature of the $L_p,a$ and $\pi,a$ families of angularities, we emphasise that this is not a requirement for our method of calculation, and $f_e$ can be designed for experimental and phenomenological reasons rather than theoretical ones.

Having studied $F_e$ for several examples of $e$, let us return to the goal of determining the anomalous dimension and boundary coefficient in terms of $F_e$. First, we expand \cref{eq:SNLO} in $\epsilon$, making use of the distributional expansion
\begin{equation}
    \frac{1}{\mu}\left(\frac{\cT}{\mu}\right)^{-1+m\epsilon} = \frac{\delta(\cT)}{m\epsilon}+ \sum_{n=0}^{\infty} \frac{(m\epsilon)^n }{n!}\cL_n(\cT,\mu)\,.
\end{equation}
Next, we expand \cref{eq:ren} to NLO, 
\begin{equation}
    S_{e}^{\bare(1)}(\mathcal{T}) = Z_{e}^{(1)}(\mathcal{T},\mu) + S_{e}^{(1)}(\mathcal{T},\mu)\,,
    \label{eq:ren1}
\end{equation}
and obtain the one-loop renormalised soft function. 
Finally, we compare the resulting coefficients of plus distributions with \cref{eq:SRGE1}, whereby we can determine the anomalous dimension in terms of $F_e$:
\begin{align}
    \Gamma^0_e &= 4 F_e^{[-1]} \Gamma^0_{\mathrm{cusp}} \,,
    \label{eq:Gamma0e}
  \\
    \hat{\gamma}^0_e &= -8 C_R F_e^{[0]} = 0 \,.
    \label{eq:gamma0e}
\end{align}
where $\Gamma^0_{\mathrm{cusp}}=4C_R$ is the usual one-loop cusp anomalous dimension. In a similar fashion, the boundary coefficient is
\begin{equation}
    S^{(1)}_{e,\delta} = 2 C_R \left[ \zeta_2 F_e^{[-1]}-2F_e^{[1]}\right]\,.
    \label{eq:Sdelta1}
\end{equation}
Recalling the coefficients listed in \cref{eq:F1list}, we see that the one-loop soft function $S^{(1)}_{e,\delta}$ for $e=L_{p},a$ becomes very large when $p \ll 1$, whilst for $e=\pi,a$ the soft function becomes large for $\omega \ll 1$ or $\eta_c \gg 1$, and for $e=\Pi,a$ the soft function also becomes large for $H\gg1$ or $\omega \sim 1-a$. This is consistent with the statements that the SCET$_{\mathrm{I}}$ factorisation of \cref{eq:fac} breaks down in these limits, as discussed in the previous section. 

It is worthwhile noting that for certain choices of the parameters, e.g. $\eta_c=0$ and $\omega=\sqrt{2}p$ the coefficients become equal,
$$
F^{[1]}_{L_p,a}(p)
=
F^{[1]}_{\Pi,a}(0,\sqrt{2}p)
$$
and therefore the corresponding soft functions are equal at NLO even though the corresponding rapidity-weighting functions $f_e$ differ.

The goal of \cref{sec:NNLO} is to extend the above procedure to NNLO. Before turning to that calculation, we show how the above discussion relates to the topic of hadronisation.

\subsection{Leading power corrections in the dijet limit}
\label{sec:hadron}

\begin{figure}
\begin{subfigure}{.99\textwidth}
    \hspace{-0.5 em}
    \includegraphics[width=0.87\linewidth]{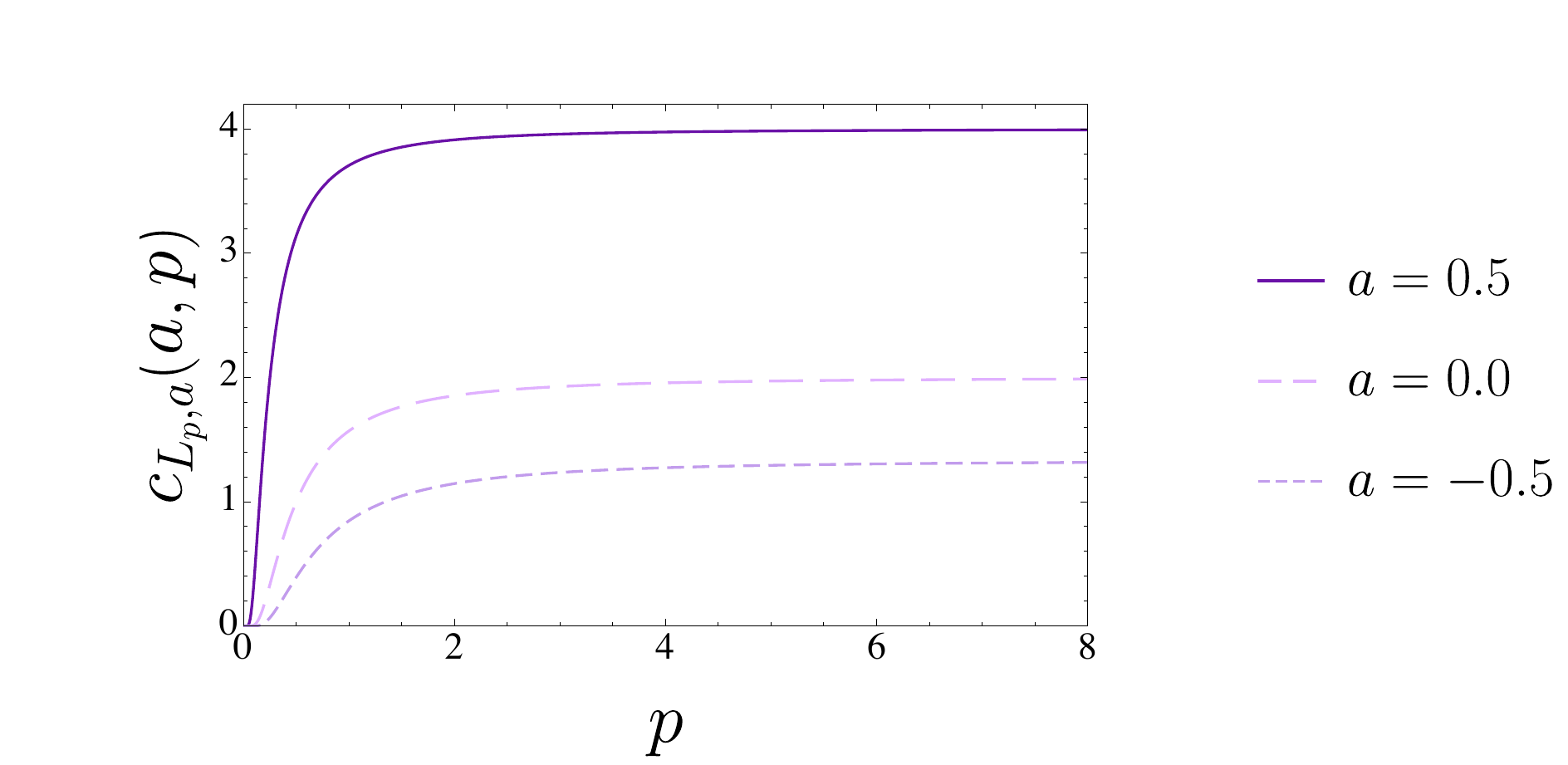}
    \label{fig:cLp}
    \caption{}
\end{subfigure}
\\
\begin{subfigure}{.99\textwidth}
    \hspace{-0.3 em}
    \includegraphics[width=0.97\linewidth]
    {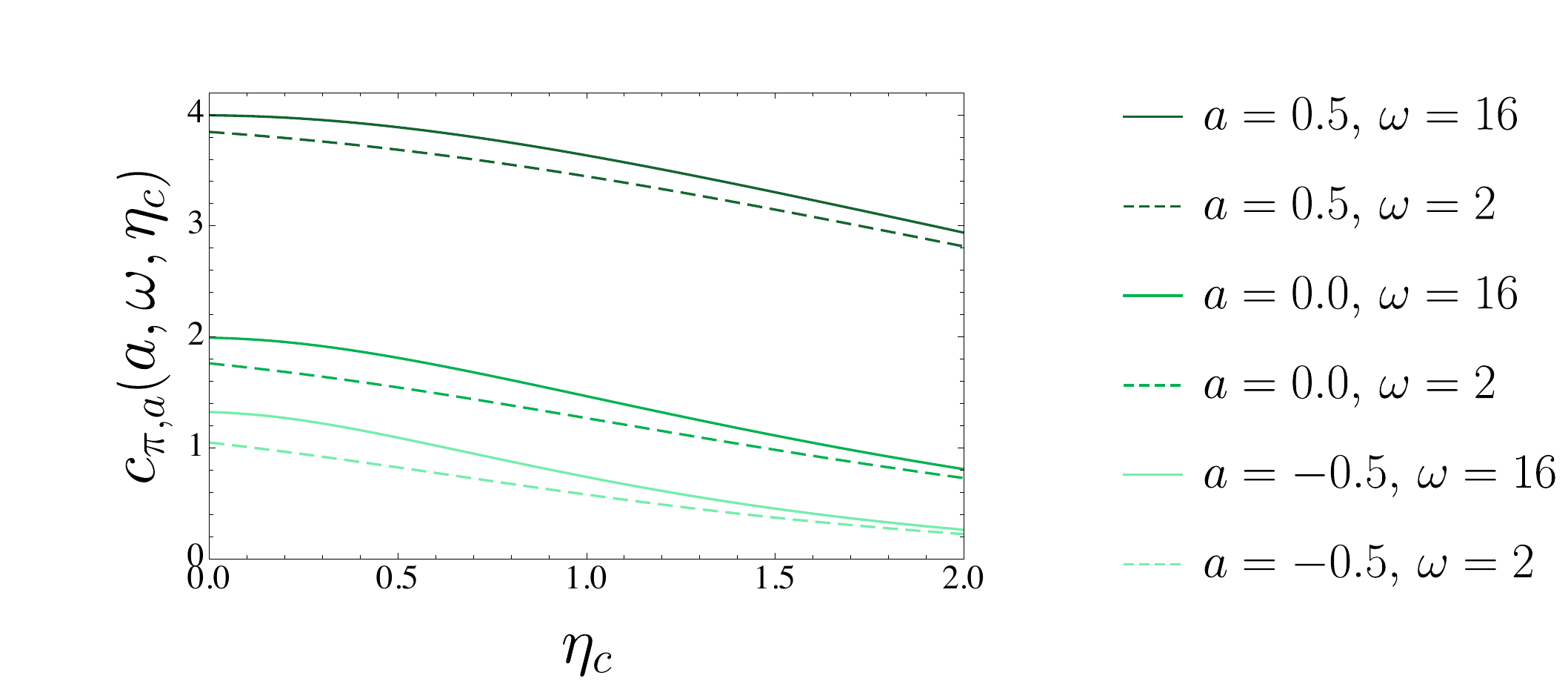}
    \label{fig:cmp}
    \caption{}
\end{subfigure}
 \\
\begin{subfigure}{0.99\textwidth}
    \includegraphics[width=1\linewidth]
    {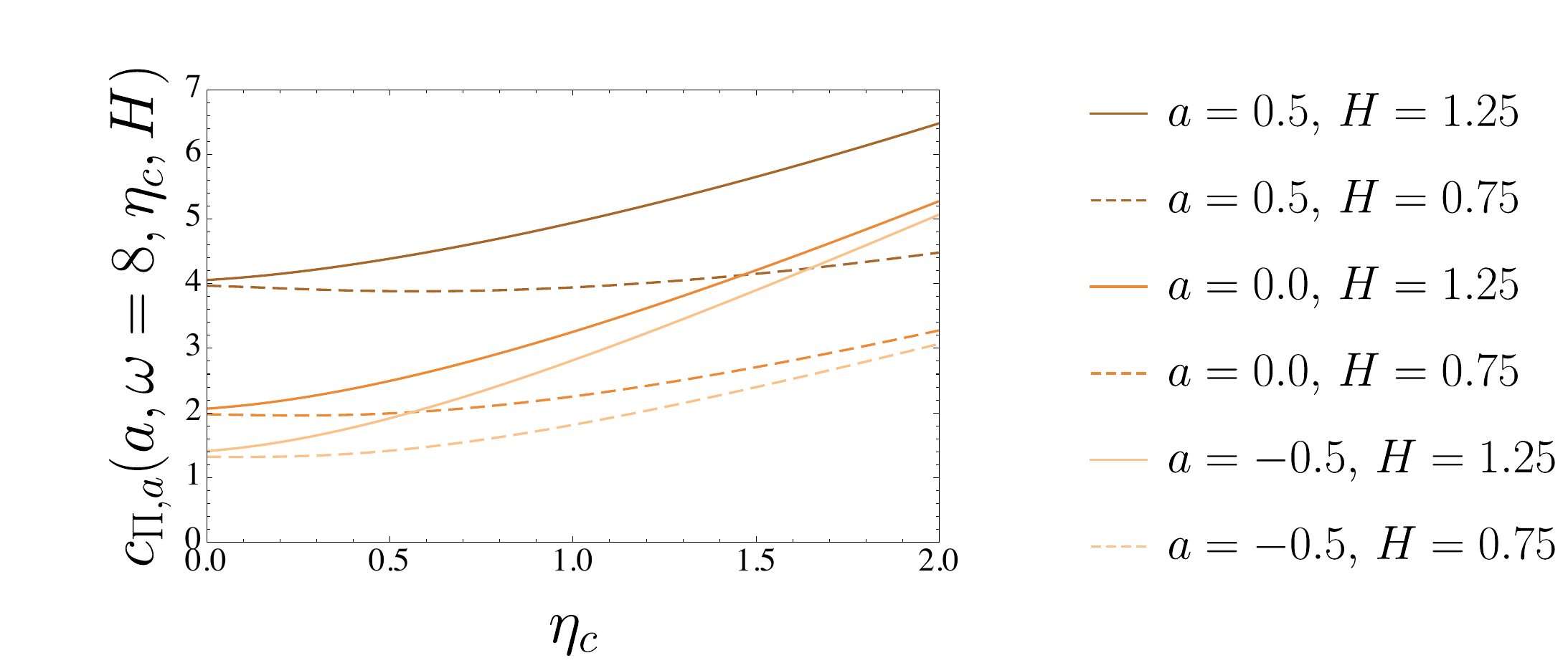}
    \label{fig:cap}
    \caption{}
\end{subfigure}
\caption{The scaling of the leading $1/Q$ power corrections (in the massless limit), $c_e$, for several choices of the parameters $\{a,p,\omega,\eta_c,H\}$. }
\label{fig:ce}
\end{figure}
One nice feature of \cref{eq:Fe} is that it has a direct connection to the leading non-perturbative behaviour in the dijet region, discussed in the introduction. In particular, the calculable coefficient $c_e$ of the leading non-perturbative correction, neglecting the effects of hadron masses, is given by the integral  \cite{Webber:1994cp, Dokshitzer:1995qm, Dokshitzer:1995zt, Dokshitzer:1998pt, Korchemsky:1998ev,  Korchemsky:1999kt, Gardi:2001ny, Gardi:2002bg, Lee:2006fn, Lee:2006nr, Becher:2013iya}
\begin{equation}
    c_e=\int_{-\infty}^{\infty} \mathrm{d} \eta \, f_e(\eta)\,,
    \label{eq:ce}
\end{equation}
which is nothing other than $F_e(\epsilon=\frac{1}{2})$. Thus, we can immediately obtain closed-form expressions for $c_e$, for several families of generalised angularity, from \cref{eq:Fea,eq:FLpa,eq:Fpia}.

To indicate the relative size of power corrections, we plot $c_e$ for the three families of generalised angularity introduced in section \ref{sec:setup}, and several choices of the parameters.  We have chosen to plot $c_e$ in \cref{fig:ce} for a wide range of the parameters, specifically, $p\in (0,8)$ and $\eta_c \in (0,2)$. We remind the reader that for $p \ll 1$ and $\eta_c \gg 1$ the factorisation formula \cref{eq:fac} is invalid, and the perturbative soft function diverges.
For the convenience of the reader, in \cref{tab:ce} we have collected some values of $c_e$ for comparison with \cref{fig:ce}, for angularity and C-angularity. These values were obtained from \cref{eq:Fea}. 

We note that, for fixed $a$, the generalised angularities allow one to explore a greater range of $c_e$ than that of conventional and C-angularity. To be concrete, let us consider $a=0$, and let us restrict the parameter space of $\{p,\eta_c,\omega,H\}$ for our three generalised angularity families by imposing that the one-loop soft function should not be too large (such that the perturbative series breaks down). Here we shall impose the conservative constraint
\begin{equation}
    \left| S_{e,\delta}^{(1)} \right| \leq \frac{3}{2} S_{C,\delta}^{(1)}\,.
\end{equation}
Then we can explore a continuous range of $c_e$ between the extreme values of 
\begin{equation}
    c_\pi(\omega \! = \! 4,\, \eta_c \! = \! 1.1)=1.340\,,
    \qquad
    c_\Pi(\omega \! = \! 9, \, \eta_c \! = \! 1.0, \, H \! = \! 1.6)=3.951,
\end{equation}
which nearly saturate the inequality imposed on $S_{e,\delta}^{(1)}$. The conservative range $c_e \in (1.340,3.951)$ provided by the plateau event shapes should be compared to the two discrete values of 1.571 and 2 provided by C-parameter and thrust respectively (see \cref{tab:ce}).

\begin{table}[htb]
\centering
$
\begin{array}{|c|c|c|}
\hline
c_e
& e=\tau,a &  e= C,a
\\ \hhline{|=|=|=|}
\vphantom{\Big[}
a=-0.5
&    
\frac{4}{3}
& 
0.847213
\\ \hline
\vphantom{\Big[}
a=0.0
& 
2
&   
\frac{\pi}{2}
\\ \hline
\vphantom{\Big[}
a=0.5
& 
4
&     
3.70815
\\ \hline
\end{array}
$
\caption{Some values of $c_e$ for angularity and C-angularity.}
\label{tab:ce}
\end{table}

As shown in ref.~\cite{Salam:2001bd}, the effects of hadron masses can be numerically important. Let us return to \cref{eq:taue} and relax the assumption of massless final state particles. Then one can write the event shape summand as
\begin{equation}
    \cT_e(k_i,m_i)=\sqrt{|\vec{p}_{i\perp}|^2+m_i^2} \, f_e(r_i, y_i),
\end{equation}
where the `transverse velocity' is given by the dimensionless ratio
\begin{equation}
    r_i=\frac{|\vec{p}_{i\perp}|}{\sqrt{|\vec{p}_{i\perp}|^2+m_i^2}}\,.
\end{equation}
The function $f_e(r,y)$ can be any function that reduces to $f_e(\eta)$ in the massless limit. We consider several schemes discussed in refs.~\cite{Salam:2001bd,Mateu:2012nk}:
\begin{subequations}
\label{eq:mass_schemes}
\begin{align}
    \text{P-scheme:}\qquad f_e(r,y)&=r f_e(\eta)\,, \\ 
    \text{E-scheme:}\qquad f_e(r,y)&=\frac{r}{v} f_e(\eta)\,, \\
    \text{R-scheme:}\qquad f_e(r,y)&=r f_e(y)\,, \\
    \text{J-scheme:}\qquad f_e(r,y)&= f_e(y)\,,
\end{align}
\end{subequations}
where the pseudo-rapidity and velocity are, in terms of $r$ and $y$,
\begin{equation}
    \eta=\sinh^{-1}\left( \frac{\sinh y}{r}\right)\,, \qquad v=\frac{\sqrt{r^2+\sinh^2 y }}{\cosh y}\,.
\end{equation}
Here it is worth distinguishing our usage of the phrase `generalised angularity' to the usage in ref.~\cite{Mateu:2012nk}\,, which uses the same term to define a generalisation of the R-scheme, with $r\mapsto r^n$ for $n\ge0$. We suggest that `generalised hadron-mass scheme' might be a more useful term for this generalisation. 
Semantics aside, let us summarise a key result of ref.~\cite{Mateu:2012nk}, which adapts the method of ref.~\cite{Lee:2006nr} to the case of massive hadrons. They find the leading $1/Q$ behaviour is given by
\begin{equation}
    \Omega_1^e = \int_0^1 \mathrm{d} r\left[\int_{-\infty}^{\infty} \mathrm{d} y \, f_e(r,y)\right] \Omega_1(r)\,,
\end{equation}
where $\Omega_1(r)$ is a well-defined non-perturbative matrix element. They write this leading-power correction as
\begin{equation}
    \Omega_1^e = c_e  \, \Omega_{g_e}\,, \qquad \Omega_{g_e}= \int_0^1 \mathrm{d} r \, \Omega_1(r) g_e(r)
\end{equation}
with $c_e$ as in \cref{eq:ce} and
\begin{equation}
    g_e(r)=\frac{1}{c_e} \int_{-\infty}^{\infty} \mathrm{d} y \, f_e(r,y)\,.
\end{equation}
Thus, the scaling relation observed in the massless limit holds only for event shapes with a common $g_e(r)$, in which case they are said to belong to the same `universality class'. 

\begin{figure}
    \centering
    \includegraphics[width=0.8\linewidth]{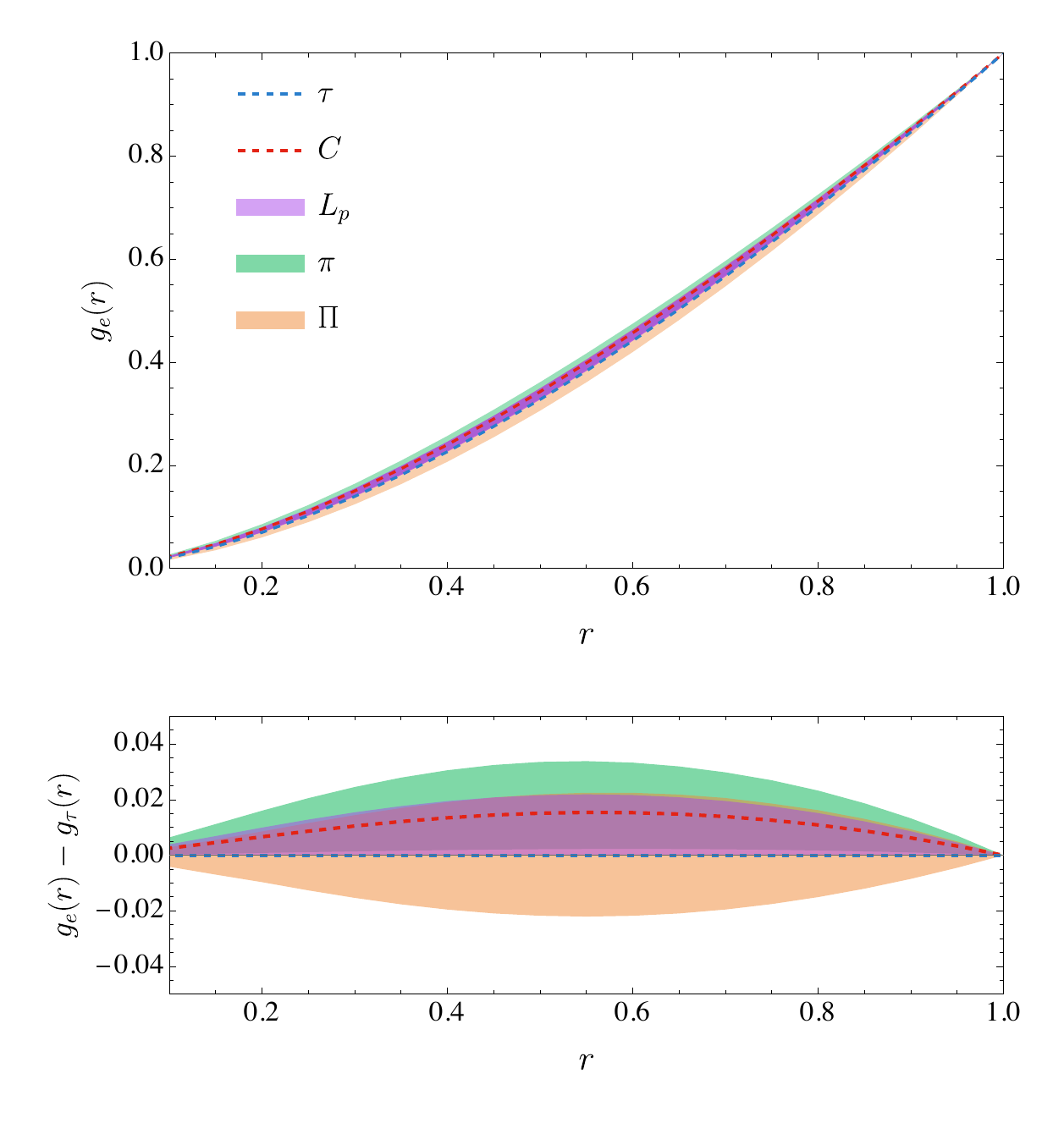}
\caption{The function $g_{e}(r)$ for thrust, C-parameter, and the three families of generalised event shapes introduced in \cref{sec:setup}. All event shapes are in the P-scheme of hadron mass dependence. For $L_p$, the plotted envelope is over the values $p \in \{0.8,32\}$. For $\pi$ the envelope is over the values $\eta_c \in \{0.25,1.25\}$ and $\omega \in \{2,32 \}$, while for $\Pi$ we also sampled over $H \in \{0.5,1.5\}$.}
\label{fig:gr}
\end{figure}

Let us now summarise the behaviour of $g_{e,a}$ for the three families of generalised angularity introduced in \cref{sec:setup}. For the E- and R- scheme, $g_{e,a}(r)=r$, while for the J-scheme, $g_{e,a}(r)=1$, for all $e\in \{ L_p, \pi, \Pi\}$. The most interesting case is the P-scheme, where in general $g_{e,a}(r)$ is a non-trivial integral that we perform numerically. In refs.~\cite{Salam:2001bd,Mateu:2012nk} it was found that thrust and C-parameter lie in different, but approximately the same universality class. As anticipated in ref.~\cite{Salam:2001bd}, the same is true for  $L_p$, $\pi$ and $\Pi$, for any reasonable choice of the parameters $\{p,\omega, \eta_c, H\}$. In \cref{fig:gr} we plot the envelope of $g_e(r)$ in the P-scheme, obtained by varying these parameters over a wide range. While $a=0$ in \cref{fig:gr}, a similar pattern is observed for other values of $a$. In this sense, the modification of event shapes considered in \cref{sec:setup} (which modifies $c_e$ but not $g_e$) is orthogonal to the generalisation of hadron mass schemes considered in ref.~\cite{Mateu:2012nk} (which modifies $g_e$ but not $c_e$).

To conclude, in the P-scheme, we expect the leading-power correction in the dijet limit to approximately obey the classical $c_e$ scaling behaviour, while in the other schemes this scaling should be strict. In this brief analysis, we have ignored the logarithmic $Q$ dependence of the leading hadronisation correction~\cite{Dasgupta:2024znl}, which is an interesting line of research, and deserving of further study in the future. 

\FloatBarrier
\section{Calculation at NNLO}
\label{sec:NNLO}
\begin{figure}
\begin{subfigure}{.32\textwidth}
    \centering
    \includegraphics[width=1\linewidth]{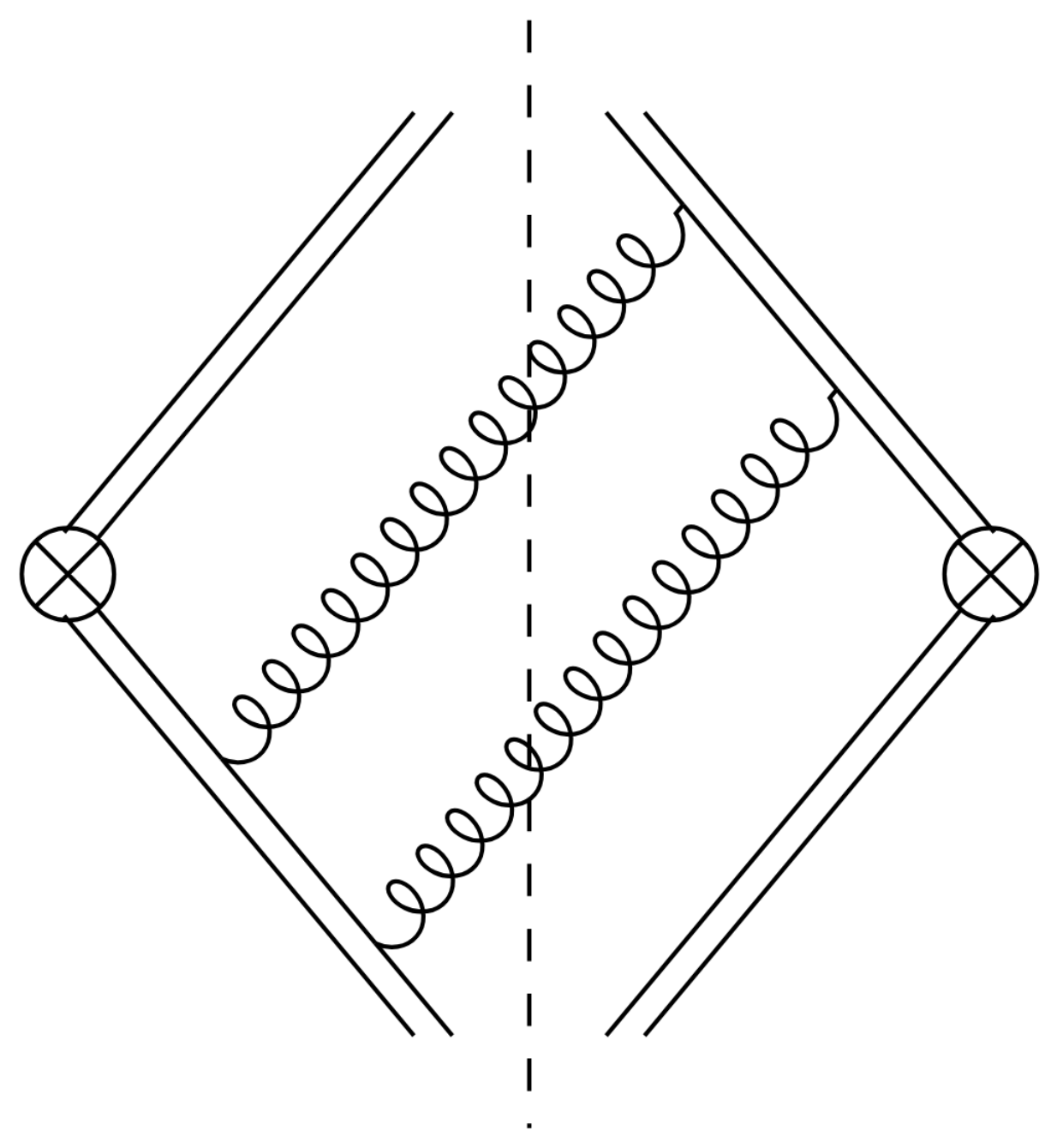}
    \label{fig:CR}
    \caption{}
\end{subfigure}
\begin{subfigure}{.32\textwidth}
    \centering
    \includegraphics[width=1\linewidth]{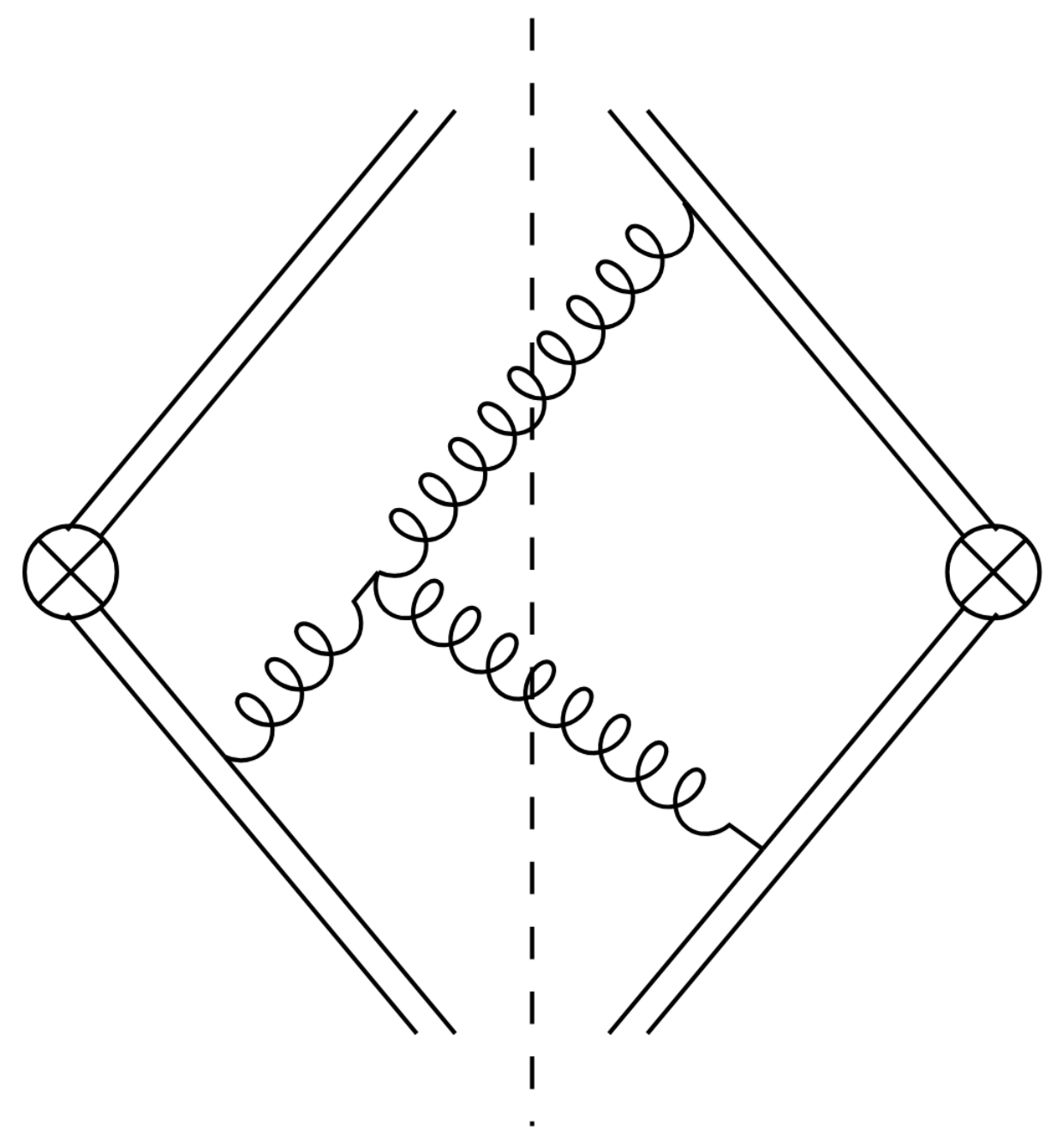}
    \label{fig:CA}
    \caption{}
\end{subfigure}
\begin{subfigure}{.32\textwidth}
    \centering
\includegraphics[width=1\linewidth]{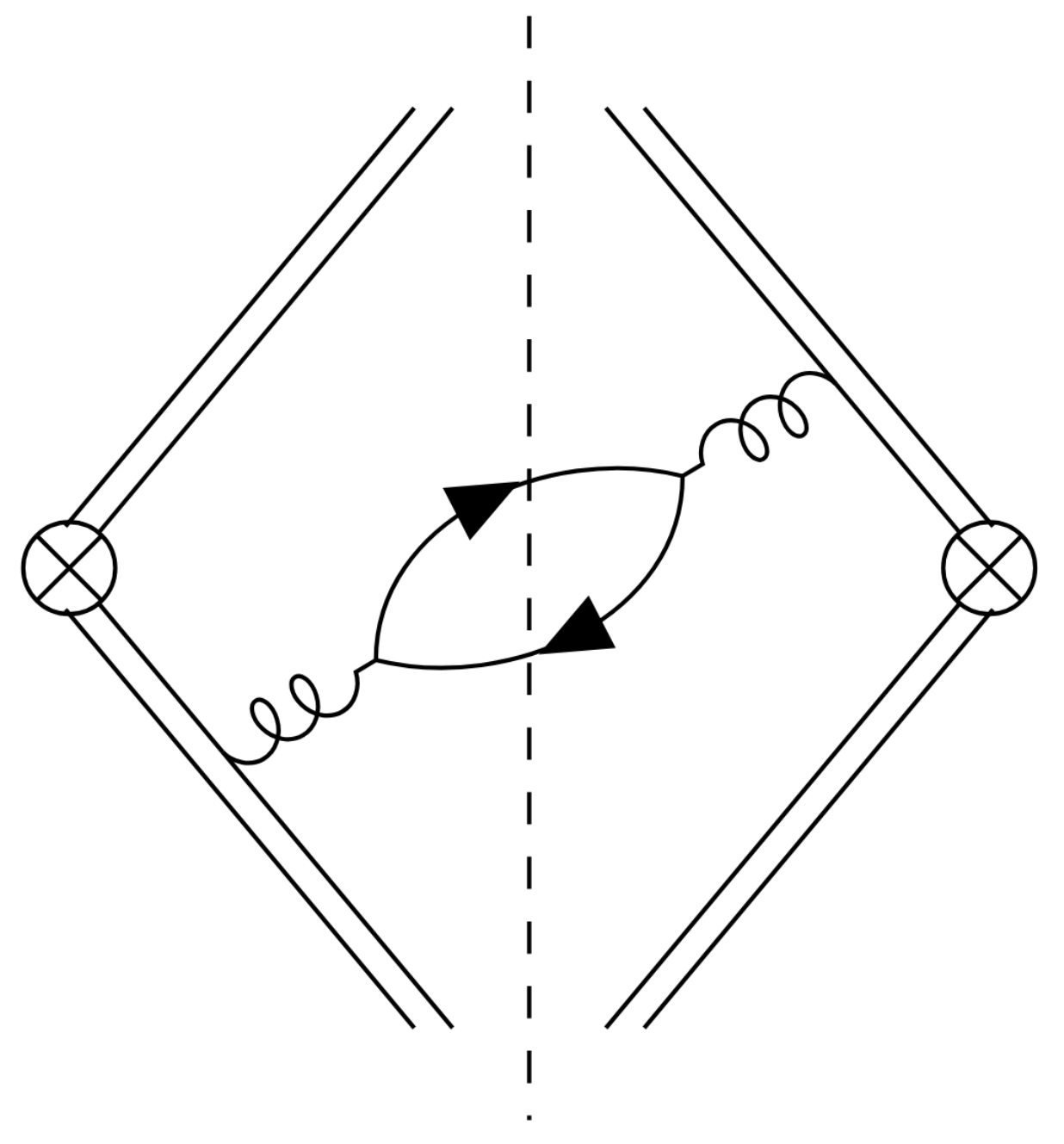}
    \label{fig:nfTF}
    \caption{}
\end{subfigure}
\caption{Example two-loop soft factor diagrams containing the colour factors (a) $C_R^2$, (b) $C_R C_A$ and (c) $C_R n_f T_F$.}
\label{fig:twoloopdiagrams}
\end{figure}

In this section we derive the NNLO soft function for any event shape of the generalised angularity class. It will be convenient to decompose the soft function into its constituent colour structures:
\begin{align}
    S_e^{(1)}&= C_R \, S_e^{(1, C_R)}\,, \\
   S_e^{(2)}&=C_R^2 \, S_e^{(2,C_R^2)}+ C_R C_A  \, S_e^{(2,C_R C_A)} + C_R n_f T_F  \, S_e^{(2,C_R n_f T_F)}\,.
   \label{eq:colour_decomp}
\end{align}
Example diagrams which give rise to the three colour factors in \cref{eq:colour_decomp} are depicted in \cref{fig:twoloopdiagrams}.
For the $C_R^2$ term, we utilise non-abelian exponentiation~\cite{Gatheral:1983cz,Frenkel:1984pz} to write the two-loop soft function as a convolution of the one-loop result,
\begin{equation}
S^{(2,C_R^2)}_{e}(\mathcal{T}) = 
\frac{1}{2}S^{(1,C_R)}_e (\mathcal{T})  \otimes_{\mathcal{T}}S^{(1,C_R)}_e(\mathcal{T}) \,.
\end{equation}
The necessary convolutions of plus distributions are given in, e.g. ref.~\cite{Ligeti:2008ac}. The result agrees with the structure predicted by \cref{eq:SRGE2},
\begin{align}
    \begin{split}
    C_R^2 \, S_e^{(2,C_R^2)}&=\frac{1}{2}(\Gamma^0_e)^2 \, \cL_3(\cT,\mu) 
    +
    \frac{3}{2} \hat{\gamma}^0_e \Gamma^0_e \, \cL_2(\cT,\mu)\\
    &+
    \left[
    (\hat{\gamma}^0_e)^2
    - S_{e,\delta}^{(1)} \, \Gamma^0_e 
    - (\Gamma^0_e)^2 \zeta_2
    \right] \, \cL_1(\cT,\mu) \\
    &+
    \left[
    \Gamma^0_e(-\hat{\gamma}^0_e \zeta_2 +\Gamma^0_e \zeta_3)
- S_{e,\delta}^{(1)} \, \hat{\gamma}^0_e\ 
     \right] \, \cL_0(\cT,\mu)
     +C_R^2 \, S_{e,\delta}^{(2,C_R^2)} \,  \delta(\cT)\,,
     \label{eq:SRGE_CR2}
    \end{split}
\end{align}
where the two-loop $C_R^2$ boundary coefficient is fully determined by the one-loop constants \cref{eq:Gamma0e,eq:gamma0e,eq:Sdelta1}:
\begin{equation}
    C_R^2 \, S^{(2,C_R^2)}_{e,\delta}=\frac{1}{2}\left( (S_{e,\delta}^{(1)} )^2- \zeta_2(\hat{\gamma}^0_e)^2 -\frac{\zeta_4}{4}(\Gamma^0_e)^2\right)+ \zeta_3 \hat{\gamma}^0_e \Gamma^0_e\,.
    \label{eq:CR2}
\end{equation}
Subtracting the $C_R^2$ result from \cref{eq:SRGE2}, we see that remaining `correlated' colour channels have a much simpler structure:
\begin{align}
\begin{split}
  S_e^{(2,C_R \, c)}  &= \beta_0^{(c)}\Gamma^{(0,C_R)}  \, \cL_2(\cT,\mu)+
    \left[
     2 \beta_0^{(c)} \hat{\gamma}^{(0,C_R)}_e 
    - \Gamma^{(1,C_R \, c)}_e 
    \right] \, \cL_1(\cT,\mu) \\
    &+
    \left[
    -\hat{\gamma}^{(1,C_R \, c)}_e
-  2 \beta_0^{(c)} S_{e,\delta}^{(1,C_R)} \ 
     \right] \, \cL_0(\cT,\mu)
     +S_{e,\delta}^{(2, C_R \,c)} \,  \delta(\cT)\,,
     \label{eq:S2corr-structure}
    \end{split}
    \end{align}
with $c \in \{C_A, n_f T_F\}$, and where we have applied the colour decomposition of \cref{eq:colour_decomp} to $\beta_0$ and the other anomalous dimensions, for example,
\begin{equation}
    \beta_0 = C_A \, \beta_0^{(C_A)} + n_f T_F  \, \beta_0^{(n_f T_F)} \,.
\end{equation}
To compute the soft function for the correlated colour channels, we follow the approach of refs.~\cite{Bauer:2020npd, Abreu:2022sdc, Abreu:2022zgo, Bennett:2025jli, Buonocore:2026yai}\footnote{Note that a similar decomposition into inclusive and correction terms has also been used in the context of studies of non-perturbative effects in event shapes, see refs.~\cite{Dokshitzer:1997iz,Dokshitzer:1998pt,Dasgupta:1999mb}.} and first compute the soft function for the simpler \emph{inclusive} event shape, which acts as a measurement on the total emitted radiation,
\begin{equation}
   \cM_{e_I}(\cT) = \delta\left(\cT - \cT_{e}\big(\sum_i k_i\big)\right)\,.
   \label{eq:Minc}
\end{equation}
The full soft function is then obtained by computing a correction to the inclusive soft function,
\begin{equation}
    S_e^{(2,C_R\,c)} (\cT) = S_{e_I}^{(2,C_R\,c)} (\cT)+ \Delta S_{e}^{(2,C_R\,c)}  (\cT)\,.
    \label{eq:inclusive_split}
\end{equation}
For zero or one real emission the inclusive measurement coincides with the full measurement. For two real emissions, the two measurements coincide when either parton is soft, or when both emitted particles are collinear to each other. The inclusive soft function therefore captures much of the divergence structure of the full two-loop correlated soft function. The only source of divergence is when both emitted partons become collinear to the same jet direction, where the measurements do not coincide. This means $\Delta S_e^{(2, C_R\,c )}$ can have at worst a single pole in $\epsilon$, which comes from integrating over the rapidity of the whole system.

In the next subsection we provide details on the calculation of the inclusive soft function $S_{e_I}^{(2,c)}$, while in the following subsection we provide details on the calculation of the correction $\Delta S_e^{(2,c)}$.

\subsection{Inclusive soft function}

Dimensional analysis and symmetry of the measurement in $\eta \leftrightarrow -\eta$ constrains the inclusive soft function to have the form~\cite{Bennett:2025jli,Buonocore:2026yai}
\begin{equation}  
S_{e_I}^{\mathrm{bare}}(\mathcal{T}) = \sum_{\ell=0} \left(\frac{\alpha_s^\bare}{4 \pi} \right)^\ell F_T^{(\ell)}(\epsilon, N_c, n_f)
\int \frac{\mathrm{d} q^+ \mathrm{d} q^- }{q^+  q^-}
\left(\frac{\mu^2}{q^+  q^-}\right)^{\ell \epsilon} \delta(\mathcal{T}_{e_I}-\mathcal{T})\,,
\end{equation}
where $F_T^{(\ell)}$ is some function of $\epsilon$ and colour factors, which is to be determined. 
We note that the remaining integral is identical to the one-loop integral \cref{eq:SNLOint}, but with $\epsilon \mapsto \epsilon \ell$. The dependence of the inclusive soft function on the integrated rapidity weight, $F_e$, is therefore given by
\begin{equation}
    S_{e_I}^{\bare}(\mathcal{T}) =
    \sum_{\ell=0} \left(\frac{\alpha_s^\bare}{4 \pi} \right)^\ell 
    F_T^{(\ell)}(\epsilon, N_c, n_f) F_e( \ell \epsilon) \frac{1}{\mu}\left(\frac{\mathcal{T}}{\mu} \right)^{-1-2 \ell\epsilon} \,.
    \label{eq:Sell}
\end{equation}
It remains to determine $F_T$ to two loops. Setting $\ell=1$ and comparing \cref{eq:Sell} to \cref{eq:SNLO}, we can immediately read off the one-loop result,
\begin{equation}
    F_T^{(1)}(\epsilon, N_c, n_f) = \frac{8 C_R e^{\epsilon \gamma_E}}{\Gamma(1-\epsilon)}\,.
\end{equation}
In ref.~\cite{Bennett:2025jli}, $F_T^{(2)}$ was obtained from the endpoint limit $x\to 1$ of the NNLO virtuality-dependent beam function computed in ref.~\cite{Gaunt:2014cfa,Gaunt:2014xga,Boughezal:2017tdd, Baranowski:2020xlp}. As noted in ref.~\cite{Buonocore:2026yai}, a convenient alternative way to determine $F_T^{(\ell)}$ is by recognising that the inclusive soft function can be written as an integral of the double-differential threshold soft function \cite{Anastasiou:2014vaa,Li:2014afw,Billis:2019vxg}
\begin{equation} S_{e}^{\mathrm{bare}(l)}(\mathcal{T}) 
= 
\int \mathrm{d} q^+ \mathrm{d} q^- 
S^{\mathrm{bare}(l)}(q^+,q^-)
\delta(\mathcal{T}_{e_I}-\mathcal{T})\,.
\end{equation}
Then, by comparing with \cref{eq:Sell}, the double-differential soft function must have the form
\begin{equation}   
\sum_{n=0} \left(\frac{\alpha_s(\mu)}{4 \pi} \right)^n  S_{e}^{\mathrm{bare}(n)}(q^+, q^-) 
= 
\sum_{\ell=0} \left(\frac{\alpha_s^\bare}{4 \pi} \right)^\ell 
F_T^{(l)}(\epsilon, N_c, n_f)
\frac{1}{\mu^2}
\left(\frac{\mu^2}{q^+ q^-}\right)^{-1+l \epsilon}\,.
\end{equation}
Expanding in $\epsilon$ and comparing with, e.g.~ref.~\cite{Billis:2019vxg}, we can then read off $F_T^{(2)}$ to $\cO(\epsilon^0)$. Going one step further, we can determine $F_T^{(2)}$ to all orders in $\epsilon$
by noting that the threshold soft function can be written as an integral of the fully differential soft function of refs.~\cite{Mantry:2010mk,Li:2011zp} over the transverse momentum of the final state,
\begin{equation}
 S_{e}^{\mathrm{bare}(n)}(q^+, q^-) = \int \mathrm{d}^2 \vec{q}_\perp   \, S_{e}^{\mathrm{bare}(n)}(q^+, q^-,\vec{q}_\perp)\,.
\end{equation}
Then we can identify $F_T$ as the integral over the transverse degrees of freedom of the fully-differential soft function. We find
\begin{align}
F_T^{(2,C_R^2)}(\epsilon) &=-
\frac{128 \, e^{2 \epsilon \gamma_E} }{\epsilon}
\frac{\Gamma (1-\epsilon )^2 \Gamma (-2 \epsilon )}{\Gamma (1-2 \epsilon )^3}\,,
\\
F_T^{(2,C_R n_f T_F)}(\epsilon) &= \frac{32 \, e^{2\epsilon \gamma_E}(1-\epsilon)\Gamma(1-\epsilon)\Gamma(-\epsilon)}{(3-2\epsilon)(1-2\epsilon)\Gamma(1-2\epsilon)^2}  \,,
\\
\begin{split}
F_T^{(2,C_R C_A)}(\epsilon) &= \frac{16 \, e^{2\epsilon \gamma_E}}{\epsilon}\Bigg[
\frac{2 \Gamma (-2 \epsilon ) \Gamma (1-\epsilon )^2}{\Gamma (1-2 \epsilon )^3}-\frac{\pi  \epsilon  \Gamma (\epsilon )^2 \Gamma (1-\epsilon )^2}{\Gamma (1-2 \epsilon ) \Gamma \left(\frac{1}{2}-\epsilon \right) \Gamma \left(\epsilon +\frac{1}{2}\right)}\\
&\hspace{-4 em}
-\frac{\Gamma (-\epsilon ) \, _3F_2(-\epsilon ,-\epsilon ,-\epsilon ;1-3 \epsilon ,1-\epsilon ;1)}{\Gamma (1-3 \epsilon )}
-\frac{(3-\epsilon) (2-\epsilon ) \Gamma (-\epsilon )^2}{4 (3 -2 \epsilon ) (1-2 \epsilon ) \Gamma (1-2 \epsilon ) \Gamma (-2 \epsilon )}
\Bigg]\,.
\end{split}
\end{align}
At two loops, after performing the coupling renormalisation, we can write the bare inclusive soft function in terms of $F_T$, and $F_e$,
\begin{equation}
    S_{e_I}^{\mathrm{bare}(2)}(\mathcal{T}) 
= F_T^{(2)}(\epsilon) F_e(2\epsilon) \left[\frac{1}{\mu}\left(\frac{\cT}{\mu}\right)^{-1-4 \epsilon}\right]
-
\frac{\beta_0}{\epsilon}
F_T^{(1)}(\epsilon) F_e(\epsilon) \left[\frac{1}{\mu}\left(\frac{\cT}{\mu}\right)^{-1-2 \epsilon}\right]\,.
\end{equation}
Expanding the renormalisation equation \cref{eq:ren} to NNLO, 
\begin{equation}
    S_{e}^{\bare(2)}(\mathcal{T}) = Z_{e}^{(2)}(\mathcal{T},\mu) + S_{e}^{(2)}(\mathcal{T},\mu)+Z_{e}^{(1)}(\mathcal{T},\mu)\otimes_\cT S_{e}^{(1)}(\mathcal{T},\mu) \,,
    \label{eq:ren2}
\end{equation}
we see that the UV renormalisation of the correlated channels $c \in \{ C_R C_A, C_R n_f T_F\}$,
\begin{equation}
    S_{e}^{\bare(2,c)}(\mathcal{T}) = Z_{e}^{(2,c)}(\mathcal{T},\mu) + S_{e}^{(2,c)}(\mathcal{T},\mu)
    \label{eq:ren2_corr}
\end{equation}
has the same structure as the NLO renormalisation condition \cref{eq:ren1}, and that one can simply subtract the IR divergent terms from the bare soft function to obtain the renormalised soft function. Following ref.~\cite{Bennett:2025jli}, we define the inclusive analogue of \cref{eq:ren2_corr},
\begin{equation}
     S_{e_I}^{\bare(2,c)}(\mathcal{T}) = Z_{e_I}^{(2,c)}(\mathcal{T},\mu) + S_{e_I}^{(2,c)}(\mathcal{T},\mu)\,.
    \label{eq:ren2_corr_inc}
\end{equation}
Finally, decomposing the anomalous dimension along similar lines to \cref{eq:inclusive_split},
\begin{equation}
    \Gamma_e =  \Gamma_{e_I} + \Delta \Gamma_e\,, \qquad \qquad
    \hat{\gamma}_e =  \hat{\gamma}_{e_I} + \Delta \hat{\gamma}_e\,,
\end{equation}
we can determine the inclusive anomalous dimension and boundary condition in terms of $F_e$. For the cusp anomalous dimension, we find
\begin{equation}
\Gamma^1_{e_I} = 4 \, F_e^{[-1]} \Gamma^1_{\mathrm{cusp}}\,,
\label{eq:Gamma1I}
\end{equation}
as in \cref{eq:Gamma0e}, where $\Gamma^1_{\mathrm{cusp}} = 
4C_R[C_A(\tfrac{67}{9}-\tfrac{\pi^2}{3})
-\tfrac{20}{9}T_F n_f] $ is the two-loop coefficient of the usual cusp anomalous dimension. We conjecture that to all orders, the inclusive soft function captures the full cusp contribution:
\begin{equation}
\Gamma^n_{e_I} = \Gamma^n_{e} = 4 \, F_e^{[-1]} \Gamma^n_{\mathrm{cusp}}\,.
\label{eq:Gamma1e_inc}
\end{equation}
For the non-cusp coefficient, we find that, just as in \cref{eq:Gamma0e}, the first two terms of the Laurent expansion of $F_e$ are sufficient,
\begin{subequations}
\label{eq:gammae1_inc}
\begin{align}
\hat{\gamma}^{(1,\,C_R n_f T_F)}_{e_I} &=
\left(
\frac{448}{27}-\frac{16}{3}\zeta_2
\right)
F_e^{[-1]} 
+
\frac{160}{9}F_e^{[0]}\,,\\
\hat{\gamma}^{(1,\, C_R C_A)}_{e_I} &=
\left(
-\frac{1616}{27}+\frac{44}{3}\zeta_2+56 \zeta_3
\right)
F_e^{[-1]} 
+
\left(-
\frac{536}{9}
+16 \zeta_2
\right)F_e^{[0]}\,.
\end{align}
\end{subequations}
We recall that these first two terms are the same regardless of the behaviour in the central soft region --- we always have $F_e^{[-1]} = 1/(1-a)$ and $F_e^{[0]}=0$\,. Thus, as one might expect, these anomalous dimensions are independent of the shape of $f_e$ in the central region. On the other hand, the inclusive boundary conditions do depend on this shape, through a dependence on $F_e^{[1]}$ and $F_e^{[2]}$:
\begin{subequations}
\label{eq:Sdelta2}
\begin{align}
\begin{split}
S^{(2,\, C_R n_f T_F)}_{e_I,\delta} &=
\left(
\frac{656}{81}
-\frac{20}{3}\zeta_2
-\frac{40}{9}\zeta_3
\right)
F_e^{[-1]} 
\\
&+
\left(
\frac{224}{27}
-
\frac{16}{3}\zeta_2
\right)
F_e^{[0]}
+\frac{80}{9}
F_e^{[1]}
+
\frac{16}{3}
F_e^{[2]}\,,
\end{split}
\\
\begin{split}
S^{(2,\, C_R C_A)}_{e_I,\delta} &=
\left(
-\frac{2428}{81}
+\frac{67}{3}\zeta_2
+\frac{110}{9}\zeta_3
+10 \zeta_4
\right)
F_e^{[-1]} \\
&+
\left(
-
\frac{808}{27}
+
\frac{44}{3}\zeta_2
+28
\zeta_3
\right)
F_e^{[0]}
+
\left(
-
\frac{268}{9}+8\zeta_2
\right)
F_e^{[1]}
-
\frac{44}{3}
F_e^{[2]}\,.
\end{split}
\end{align}
\end{subequations}
With \cref{eq:Gamma1e_inc,eq:gammae1_inc,eq:Sdelta2} we can immediately write down the inclusive soft function for any generalised angularity event shape by computing \cref{eq:Fe} to $\cO(\epsilon^2)$\,. For the families of angularity we introduced in \cref{sec:setup}, the requisite coefficients of $\epsilon^2$ were listed in \cref{eq:F2list}.

In the next subsection we describe the calculation of the correction term needed to obtain the full soft function, i.e. $\Delta S_e^{(2)}$ of \cref{eq:inclusive_split}. 

\subsection{Correction to inclusive soft function}

Before computing the correction from the inclusive soft function to the full soft function, we can already use the solution of the RGE, that is,  \cref{eq:S2corr-structure}, to predict the structure of this term. Noting that the inclusive and full soft functions coincide at NLO,
\begin{equation}
    S_e^{(1)}(\cT)=S_{e_I}^{(1)}(\cT) \quad \implies
    \Delta S_e^{(1)}(\cT)=0\,,
\end{equation}
and using the fact that the inclusive soft function captures the full cusp anomalous dimension at NNLO (see \cref{eq:Gamma1I}), we see that \cref{eq:S2corr-structure} simplifies dramatically to
\begin{align}
\begin{split}
  \Delta S_e^{(2, C_R \, c)}  &= 
    -\Delta \hat{\gamma}^{(1,C_R \, c)}_e
 \cL_0(\cT,\mu)
     +\Delta S_{e,\delta}^{(2, C_R \,c)} \,  \delta(\cT)\,.
     \label{eq:DS_RGE}
    \end{split}
\end{align}
This is the structure observed in ref.~\cite{Bennett:2025jli}, for the calculation of this correction for the specific case of C-angularity, $\Delta S^{(2,C_R \, c)}_{C,a}$, and we follow the method of that paper closely. In particular, we compute the cumulant of the correction,
\begin{align}
    \Delta S_e^{(2,C_R \, c)}(\cTcut)  &= \int_0^{\cTcut} \mathrm{d} \cT \, \Delta S_e^{(2,C_R \, c)}(\cT)
    \\
    &=
    -\Delta \hat{\gamma}^{(1,C_R \, c)}_e
 \ln\left(\frac{\cTcut}{\mu}\right)
     +\Delta S_{e,\delta}^{(2, C_R \,c)} \,,
\end{align}
as an integral over double-real-emission soft matrix elements on the support of the difference of measurement functions
\begin{align}
    \Delta \cM^{RR}_e(\cTcut) &=  \cM^{RR}_e(\cTcut)-\cM^{RR}_{e_I}(\cTcut) \\
    &=\theta
    \Big(\cT(k_1)+\cT(k_2)<\cTcut
    \Big)-
    \theta
    \Big(\cT(k_1+k_2)<\cTcut
    \Big)\,.
\end{align}
Specifically, the correction is given by 
\begin{align}
\begin{split}
\Delta S_e^{\bare(2,C_R \, c)}(\cTcut)=
   \int 
   \bigg(
   \prod_{i \in \{1,2 \}}
   \!
   \left[\mathrm{d} k_i
    \right]
    \bigg)
    \Delta \cM^{RR}_e(\cTcut) \cA_{RR}^{(2,C_R \, c)}\,,
    \label{eq:DeltaS2}
\end{split}
\end{align}
with phase-space measure
\begin{equation}
    [\mathrm{d}k_i]=\frac{\mathrm{d}^d k_i}{(2\pi)^d}
    (2\pi)\delta(k_i^2)\theta(k_i^0)\,,
\end{equation}
and where $\cA_{RR}^{(2,C_R \, c)}$ are the amplitudes for the emission of two soft partons, squared, and summed over final state helicities and colours, which are given in ref.~\cite{Hornig:2011iu}. The superscript indicates this function is the coefficient of $\alpha_s(\mu)^2/(4 \pi)^2$ and colour factor $ c \in \{ C_A, n_f T_F\} $.

Following the method of ref.~\cite{Bennett:2025jli} we split \cref{eq:DeltaS2} into two pieces:
\begin{subequations}
\label{eq:ytsplit}
\begin{align}
\left(\frac{\mu}{\cTcut} \right)^{4\epsilon}
    \frac{1}{2(1-a)\epsilon} I^c_{e,\mathrm{div.}}
    &=
   \int 
   \left[\mathrm{d} k_1\right]
   \left[\mathrm{d} k_2\right]
   2\theta(y_t>0)
   \left[
    \lim_{y_t \to \infty }\Delta \cM^{RR}_e \right]\cA_{RR}^{(2,C_R \, c)}\,,
    \label{eq:Idiv-def}
    \\  
    2\left(\frac{\mu}{\cTcut} \right)^{4\epsilon}
    I^c_{e,\mathrm{reg.}}
   &=
   \int 
   \left[\mathrm{d} k_1\right]
   \left[\mathrm{d} k_2\right]
    2\theta(y_t>0)
   \left\{
   \Delta \cM^{RR}_e
   -
   \left[
    \lim_{y_t \to \infty }\Delta \cM^{RR}_e \right]
    \right\}
    \cA_{RR}^{(2,C_R \, c)}\,.
    \label{eq:Ireg-def}
\end{align}
\end{subequations}
The factors of $(\mu/\cTcut)^{4\epsilon}$ on the LHS are fixed by dimensional analysis. We further factor out $(2(1-a)\epsilon)^{-1}$ in \cref{eq:Idiv-def}, which is the pole generated upon integrating over the average rapidity of the two particles,
\begin{equation}
    y_t =\frac{1}{2}( y_1 + y_2 )= \frac{1}{4}\ln\left(
    \frac{k_1^- k_2^-}{k_1^+ k_2^+}
    \right)\,.
\end{equation}
The decomposition \cref{eq:ytsplit} is thus similar in spirit to our discussion of $F_e$, with
$$\int \mathrm{d} y_t \, y_t^{-4 \epsilon} \Delta \cM_e^{RR} $$
playing the role of \cref{eq:Fe}.
Indeed, just as in \cref{eq:Fem1}, $I^c_{e,\mathrm{div.}}$ is common to all event shapes of the class $\tau_{e,a}$, and as we will see shortly, also gives rise to a universal angularity anomalous dimension. 

The two integrals $I^{c}_{e,\mathrm{div.}}$ and $I^{c}_{e,\mathrm{reg.}}$, implicitly defined  by \cref{eq:ytsplit}, are finite in $\epsilon$. In order to perform these integrals (either analytically or numerically), we expand in $\epsilon$:
\begin{align}
    I^{c}_{e,\mathrm{div.}}=
    \sum_{n=0}^{\infty}
    \epsilon^n  I^{c[n]}_{e,\mathrm{div.}}\,,
    \qquad \qquad
    I^{c}_{e,\mathrm{reg.}}=
    \sum_{n=0}^{\infty}
    \epsilon^n  I^{c[n]}_{e,\mathrm{reg.}}\,.
\end{align}
Expanding the correction to the soft function in $\epsilon$, and performing renormalisation, we find
\begin{align}
    \Delta S_e^{(2,C_R \, c)}(\cTcut,\mu)=
    -\frac{2
    I^{c[0]}_{e,\mathrm{div.}}
    }{(1-a)}\ln \left(\frac{\cTcut}{\mu}
    \right)
    +
    \frac{
    I^{c[1]}_{e,\mathrm{div.}}
    }{2(1-a)}+
    2I^{c[0]}_{e,\mathrm{div.}}
    +\cO(\epsilon)\,.
    \label{eq:DeltaSren}
\end{align}
Comparing to \cref{eq:DS_RGE} we can write the anomalous dimension and boundary coefficients in terms of three integrals ($I^{c[0]}_{e,\mathrm{div.}}$, $I^{c[1]}_{e,\mathrm{div.}}$ and $I^{c[0]}_{e,\mathrm{reg.}}$), which are each finite in four dimensions:
\begin{align}
    \Delta \Gamma_e^{(1,C_R \, c)}&=0\,,\\
    \Delta \hat{\gamma}_e^{(1,C_R \, c)}&=\frac{2}{(1-a)}
    I^{c[0]}_{e,\mathrm{div.}}\,,\\
    \Delta S_{e,\delta}^{(2,C_R \, c)}&=\frac{1}{2(1-a)}
    I^{c[1]}_{e,\mathrm{div.}}+2 I^{c[0]}_{e,\mathrm{reg.}}\,.
    \label{eq:DSdelta}
\end{align}
As mentioned above, two of these integrals ($I^{c[0]}_{e,\mathrm{div.}}$ and $I^{c[1]}_{e,\mathrm{div.}}$) are common to all angularity dijet event shapes, and we can therefore use the results of ref.~\cite{Bennett:2025jli}, which computed these integrals analytically as an expansion in $a$, 
\begin{align}
    I^{c}_{e,\mathrm{div.}}=
    \sum_{m=1}^{\infty}
    \sum_{n=0}^{\infty}
    \epsilon^n a^m I^{c[n,m]}_{e,\mathrm{div.}}\,,
    \qquad \qquad
    I^{c}_{e,\mathrm{reg.}}=
    \sum_{m=0}^{\infty}
    \sum_{n=0}^{\infty}
    \epsilon^n a^m I^{c[n,m]}_{e,\mathrm{reg.}}\,.
    \label{eq:Ia}
\end{align}
up to order $a^3$, which we collect in \cref{tab:Icnm} for the convenience of the reader. 
\begin{table}[htb]
\centering
$
\begin{array}{|c|c|c|}
\hline
I^{c[n,m]}
& c= n_f T_F &  c= C_A
\\ \hhline{|=|=|=|}
\vphantom{\Big[}
\Idiv^{c[0,1]}
& 
\frac{10}{3}
&     
-\frac{41}{3}+\frac{4}{3}\pi^2+8 \zeta_3
\\ \hline
\vphantom{\Big[}
\Idiv^{c[0,2]}
& 
\frac{56}{45}
-
\frac{12}{5}\zeta_3
&   
-\frac{118}{45}
+\frac{1}{3}\pi^2
-\frac{24}{5}\zeta_3
\\ \hline
\vphantom{\Big[}
\Idiv^{c[0,3]} 
&    
\frac{11}{54}
-
\frac{2}{9}\zeta_3
& 
-\frac{65}{108}
+\frac{1}{9}\pi^2
-\frac{8}{9}\zeta_3
+\frac{1}{90}\pi^4
\\ \hline
\vphantom{\Big[}
\Idiv^{c[0,4]} 
&    
\frac{13}{210}
-
\frac{2}{15}\zeta_3
+
\frac{1}{7}
\zeta_5
& 
-\frac{19}{105}
+\frac{1}{24}\pi^2
-\frac{13}{30}\zeta_3
+\frac{1}{120}\pi^4
-\frac{4}{7}\zeta_5
\\ \hhline{|=|=|=|}
\vphantom{\Big[}
\Idiv^{c[1,1]} 
&  
\frac{239}{9}
-\frac{32}{3}\zeta_3
& 
-\frac{1793}{18}
+\frac{16}{3}\pi^2
+\frac{160}{3}\zeta_3
+\frac{4}{9}\pi^4
\\ \hline
\vphantom{\Big[}
\Idiv^{c[1,2]}
&  
\frac{2162}{225}
-\frac{2}{9}\pi^2
-\frac{212}{75}\zeta_3
-\frac{1}{9}\pi^4
& 
-\frac{2741}{225}
+\frac{4}{9}\pi^2
+\frac{2116}{75}\zeta_3
-\frac{26}{45}\pi^4
\\ \hline
\vphantom{\Big[}
\Idiv^{c[1,3]} 
& 
\frac{467}{1620}
-\frac{1}{27}\pi^2
+\frac{202}{135}\zeta_3
-\frac{7}{405}\pi^4
&    
-\frac{4187}{3240}
+\frac{559}{135}\zeta_3
-\frac{31}{405}\pi^4
+\frac{2}{9}\pi^2\zeta_3
+\zeta_5
\\ \hline
\end{array}
$
\caption{The coefficients of $\epsilon^n a^m$ and colour structure $c$ in the expansion \cref{eq:Ia} of the integrals in \cref{eq:Idiv-def,eq:Ireg-def}. These values are taken from ref.~\cite{Bennett:2025jli}.}
\label{tab:Icnm}
\end{table}

This leaves a single integral ($I^{c[0]}_{e,\mathrm{reg.}}$) which is sensitive to the rapidity modulation, $f_e$, of the observable in question. We provide in the supplemental material a \texttt{Mathematica} \cite{Mathematica} file which computes $I^{c[0]}_{e,\mathrm{reg.}}$ as a three-fold numerical integral, which takes an arbitrary rapidity modulation function, $f_e$, as a user input. 

\begin{figure}
\centering
\begin{subfigure}{\textwidth}
    \centering
    \hspace{0.5 em}
    \includegraphics[width=0.7\linewidth]{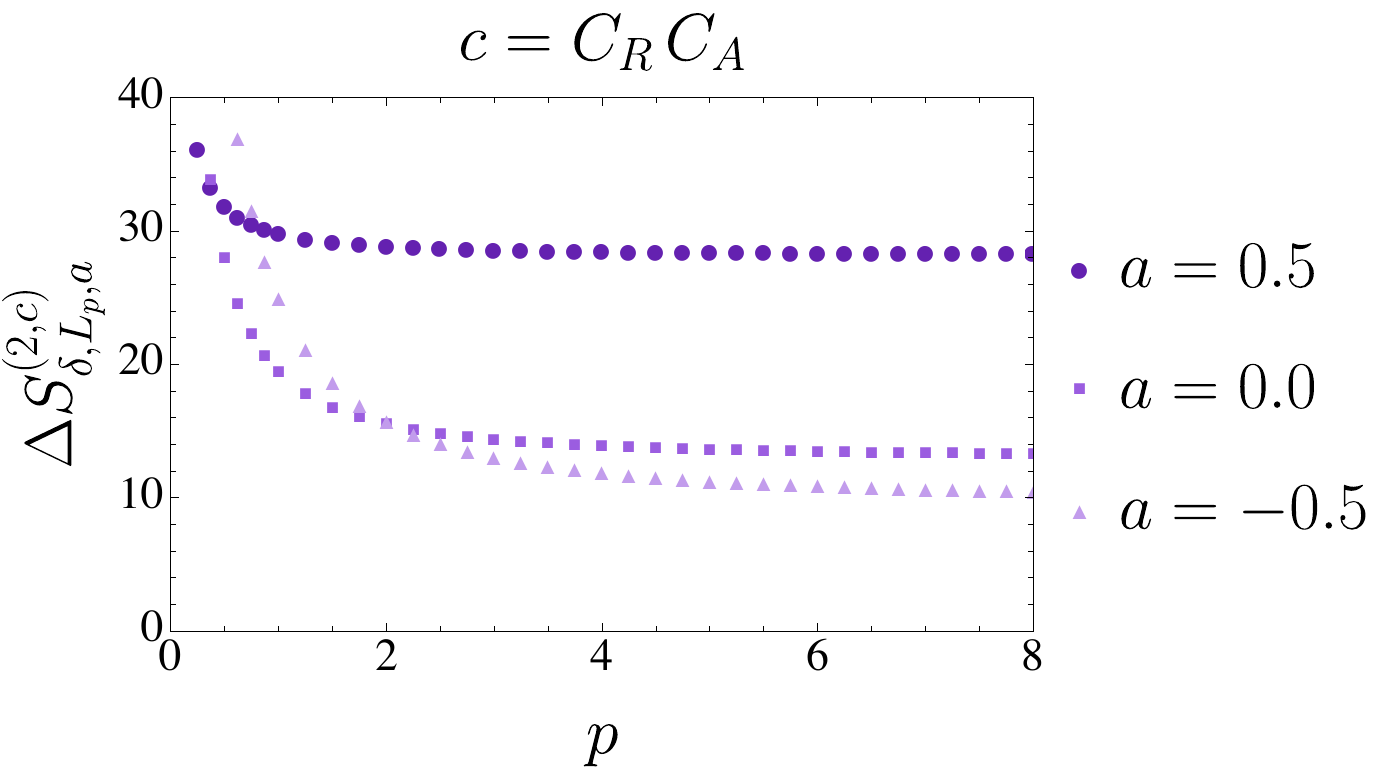}
    \label{fig:DS_Lp_CA}
    \vspace{-0.5 em}
    \caption{}
\end{subfigure}
\\
\begin{subfigure}{\textwidth}
    \centering
    \vspace{1.5 em}
    \hspace{0.5 em}
    \includegraphics[width=0.7\linewidth]{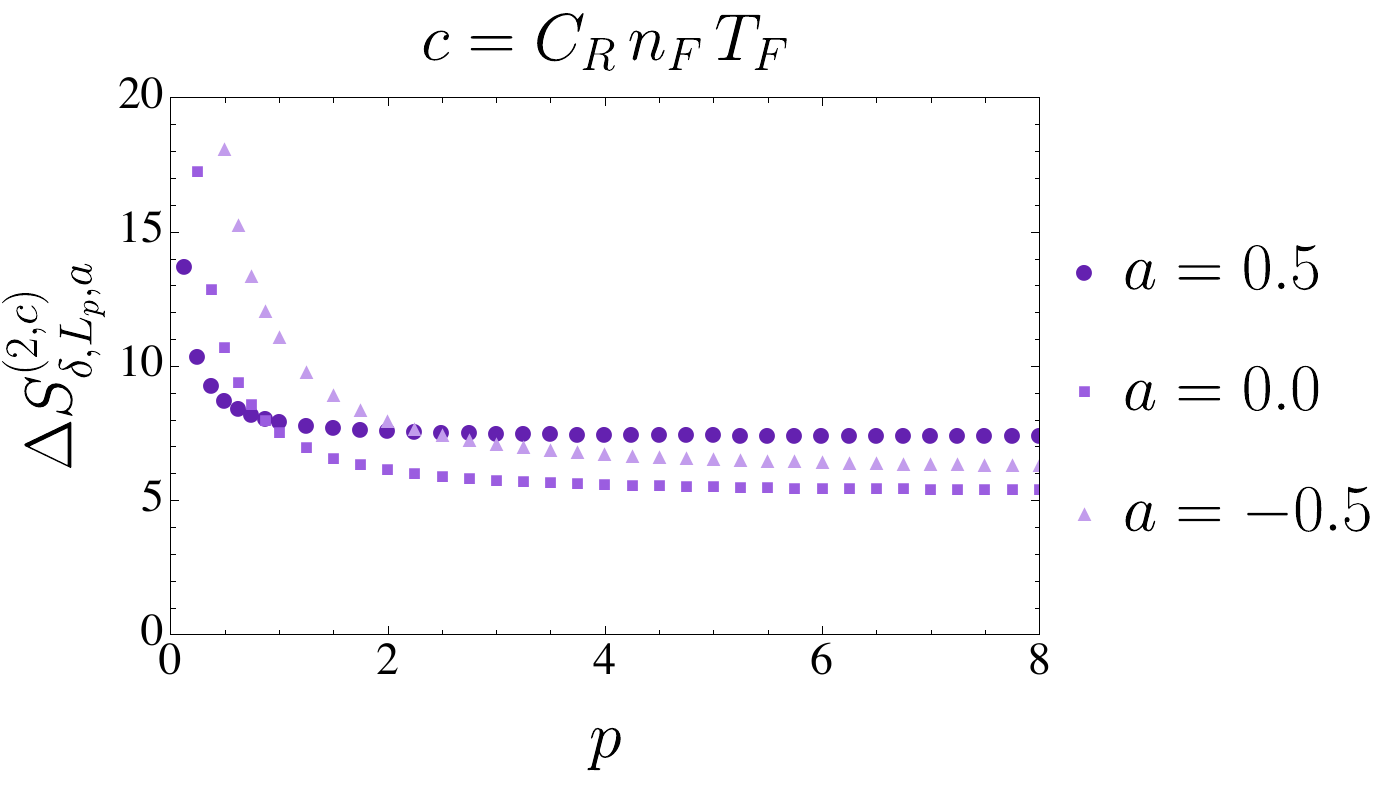}
    \vspace{-0.5 em}
    \label{fig:DS_Lp_nfTF}
    \caption{}
\end{subfigure}
 \\
\begin{subfigure}{\textwidth}
    \centering
    \vspace{1.5 em}
    \hspace{-1.7 em}
    \includegraphics[width=0.67\linewidth]{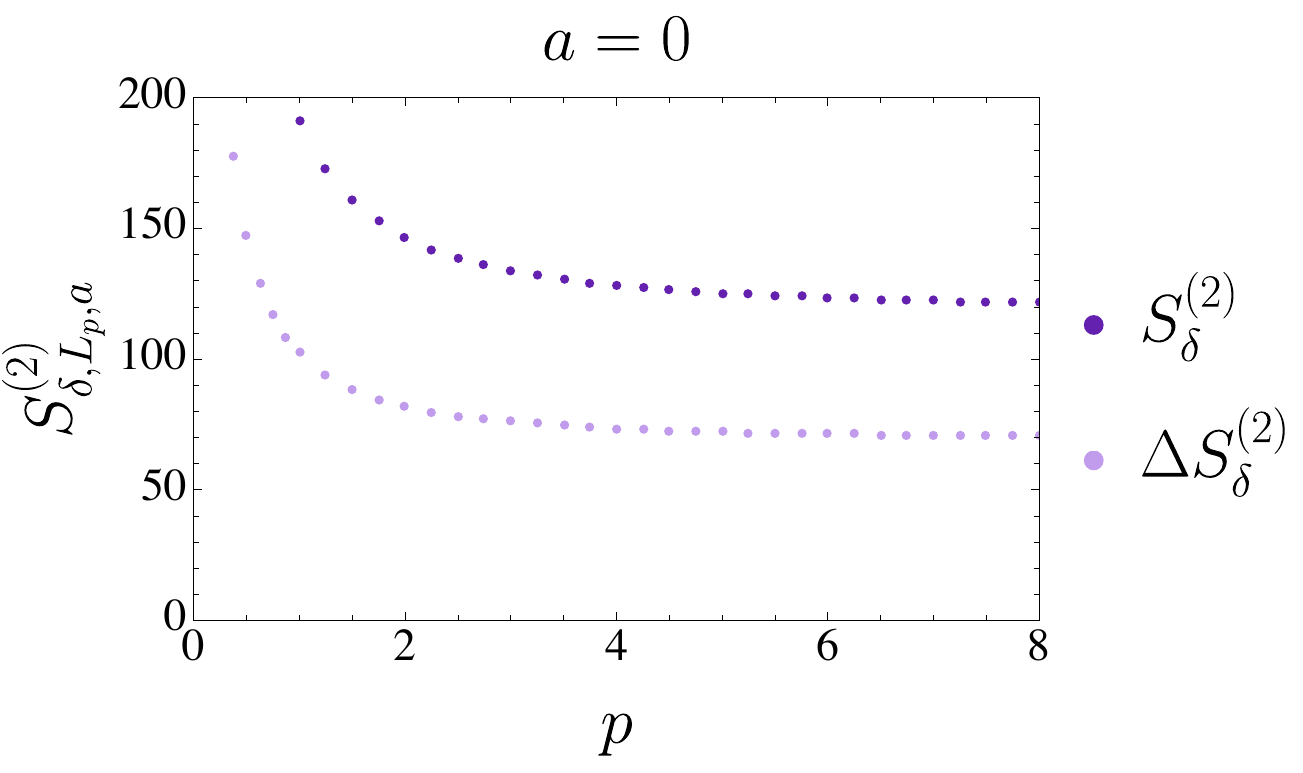}
    \vspace{-0.5 em}
    \label{fig:S_Lp}
    \caption{}
\end{subfigure}
\caption{Numerical results for the two-loop $L_p$ angularity boundary coefficient. Panels (a) and (b) plot $\Delta S^{(2,c)}_{e,\delta}$ for $c=C_R C_A$ and $c=C_R n_f T_F$ respectively, which have been computed numerically. Panel (c) compares the full colour-dressed boundary coefficient, $S^{(2)}_{e,\delta}$, for $a=0$ with the piece that is computed numerically.
\label{fig:Sdelta_Lp}}
\end{figure}

\begin{figure}
\centering
\begin{subfigure}{.8\textwidth}
    \centering
    \includegraphics[width=1\linewidth]{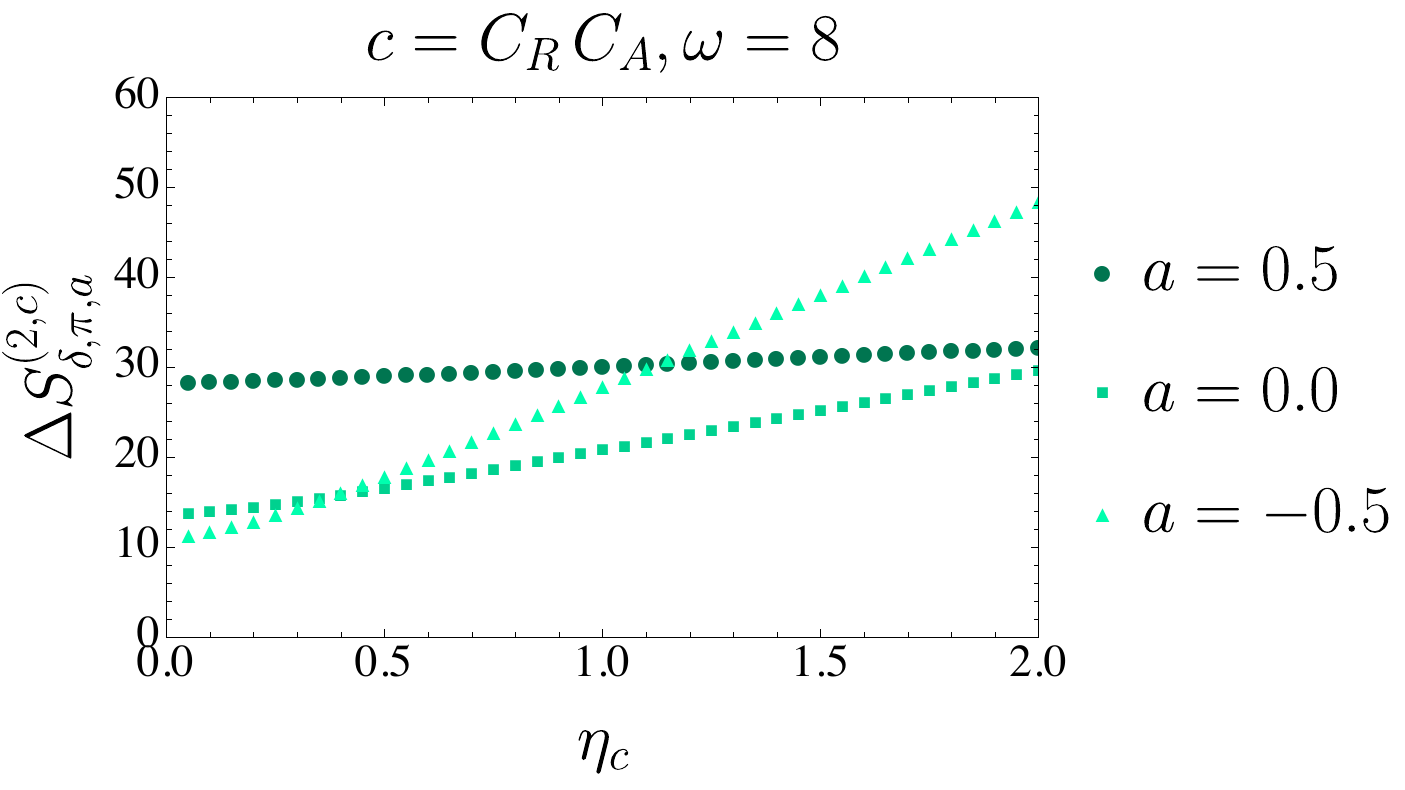}
    \vspace{-1.5 em}
    \label{fig:DS_mp_CA}
    \caption{}
\end{subfigure}
\\
\begin{subfigure}{.8\textwidth}
    \centering
    \vspace{1 em}
    \includegraphics[width=1\linewidth]{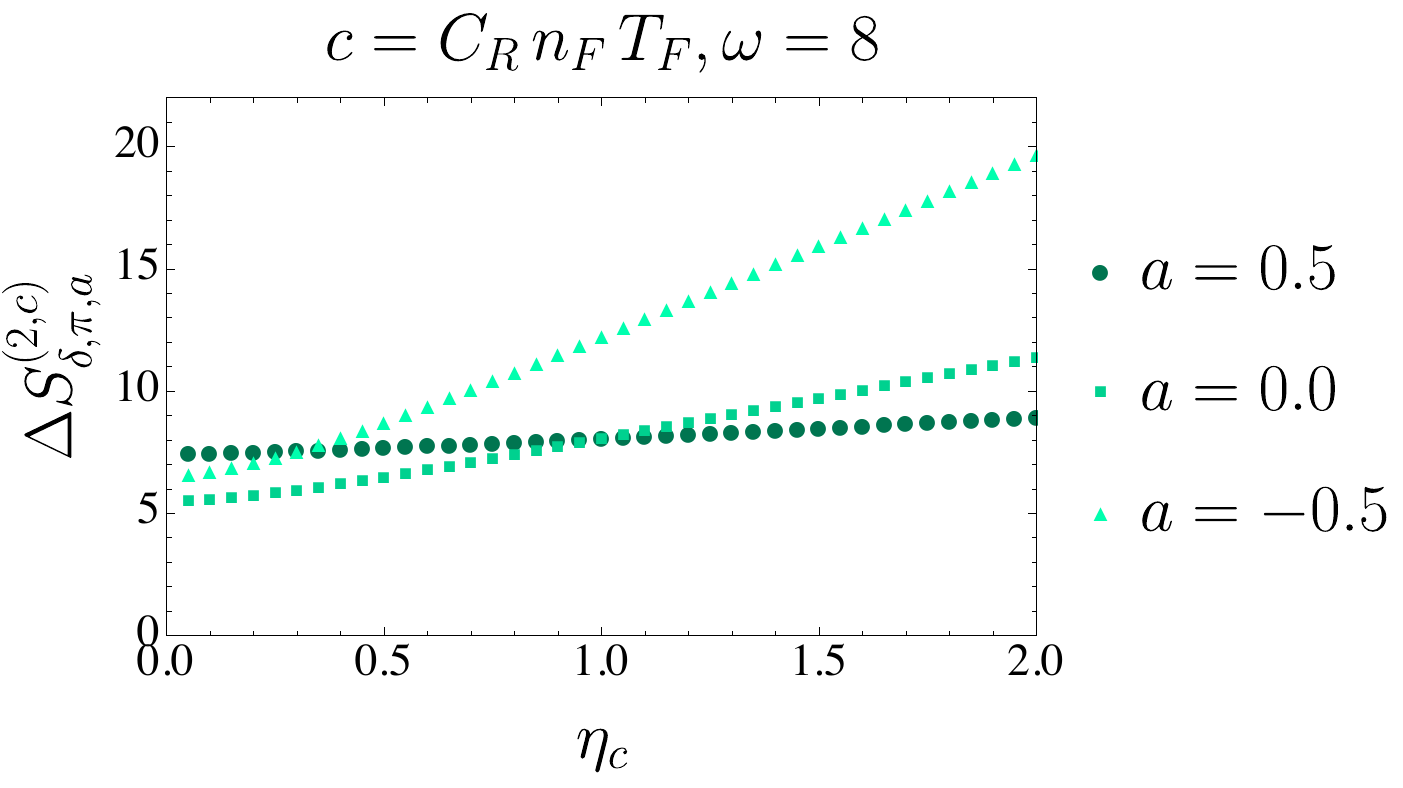}
    \vspace{-1.5 em}
    \label{fig:DS_mp_nfTF}
    \caption{}
\end{subfigure}
 \\
\begin{subfigure}{.77\textwidth}
    \centering
    \vspace{1 em}
    \hspace{-2.4 em}
    \includegraphics[width=1\linewidth]{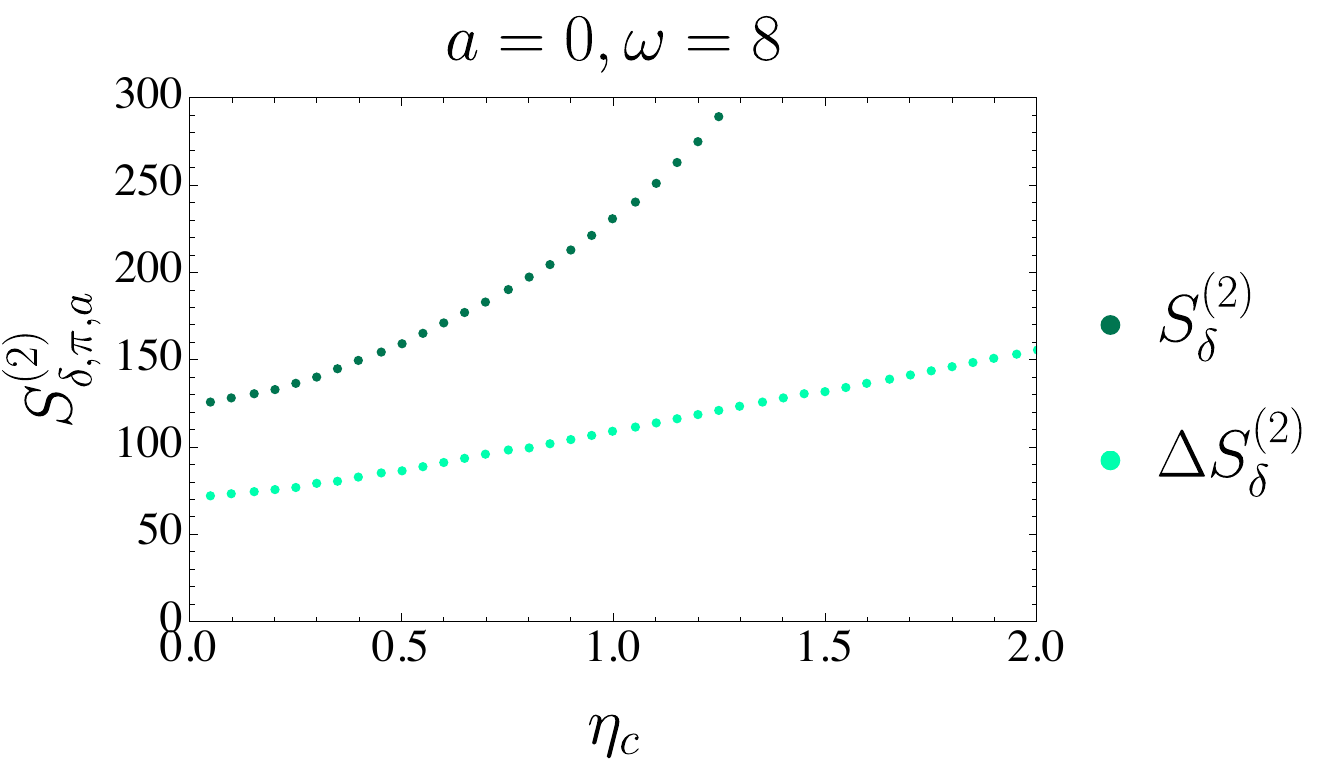}
    \label{fig:S_mp}
    \caption{}
\end{subfigure}
\caption{Similar plots to \cref{fig:Sdelta_Lp} but for the multiplicative plateau angularity, $\pi,a$.}
\label{fig:Sdelta_mp}
\end{figure}

\begin{figure}
\centering
\begin{subfigure}{.8\textwidth}
    \centering
    \includegraphics[width=1\linewidth]{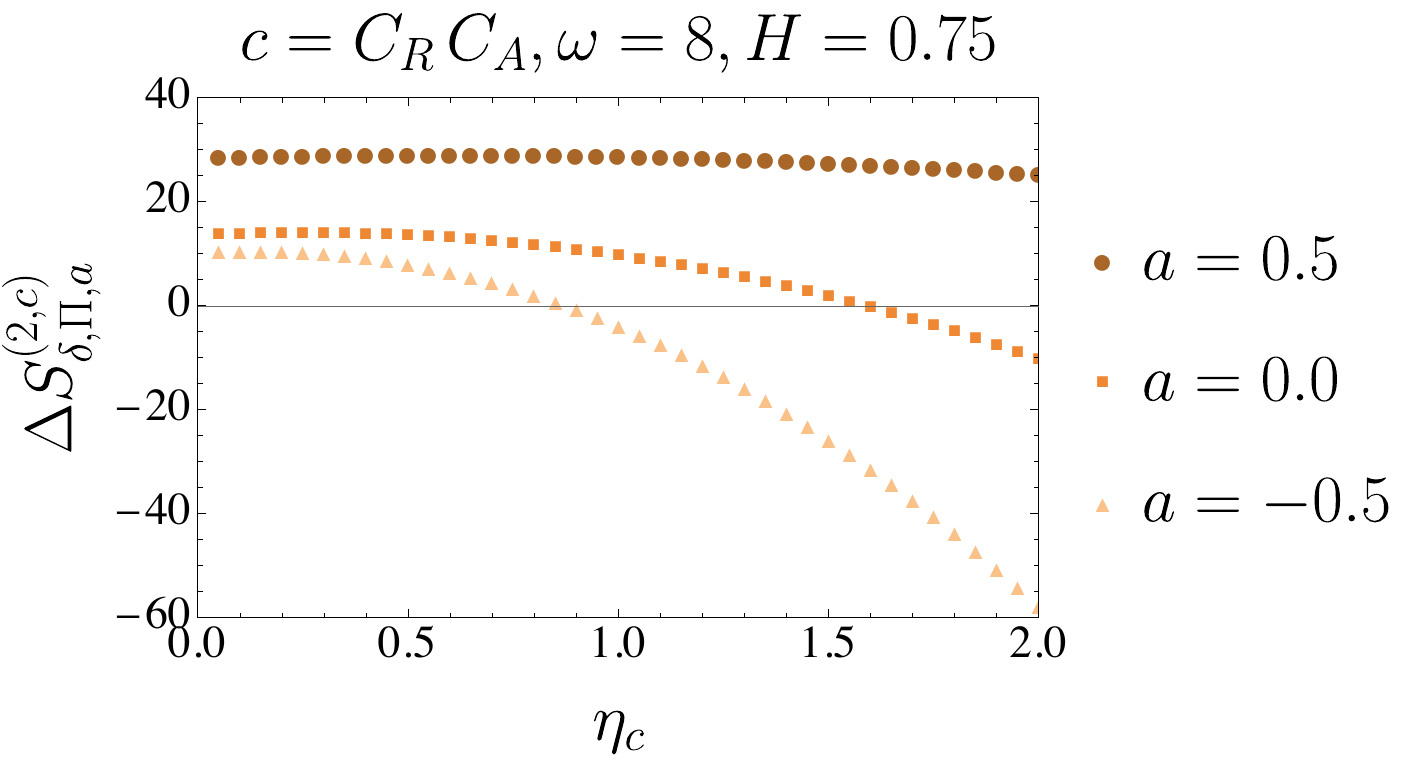}
    \vspace{-1.5 em}
    \label{fig:DS_ap_CA}
    \caption{}
\end{subfigure}
\\
\begin{subfigure}{.8\textwidth}
    \centering
    \vspace{1 em}
    \includegraphics[width=1\linewidth]{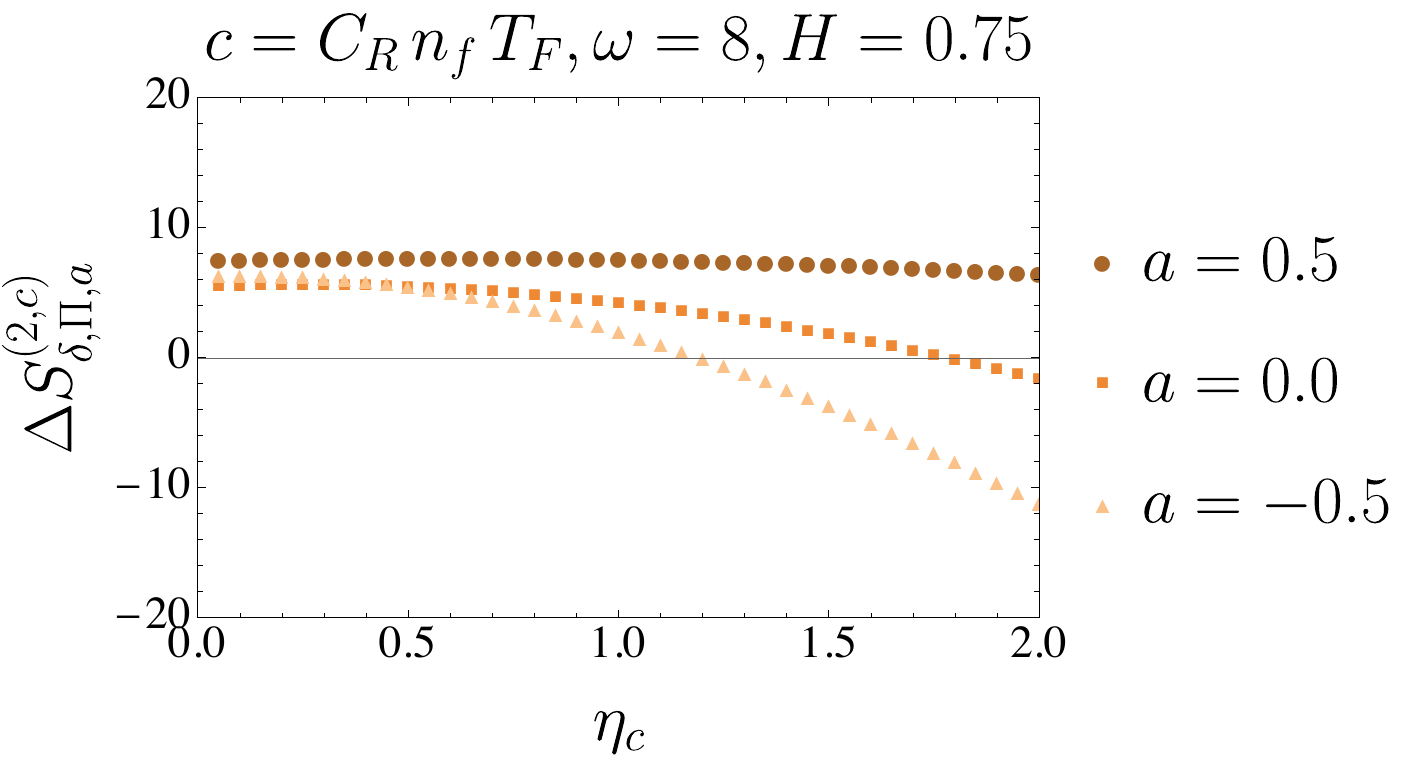}
    \vspace{-1.5 em}
    \label{fig:DS_ap_nfTF}
    \caption{}
\end{subfigure}
 \\
\begin{subfigure}{.77\textwidth}
    \centering
    \vspace{1 em}
    \hspace{-2.4 em}
    \includegraphics[width=1\linewidth]{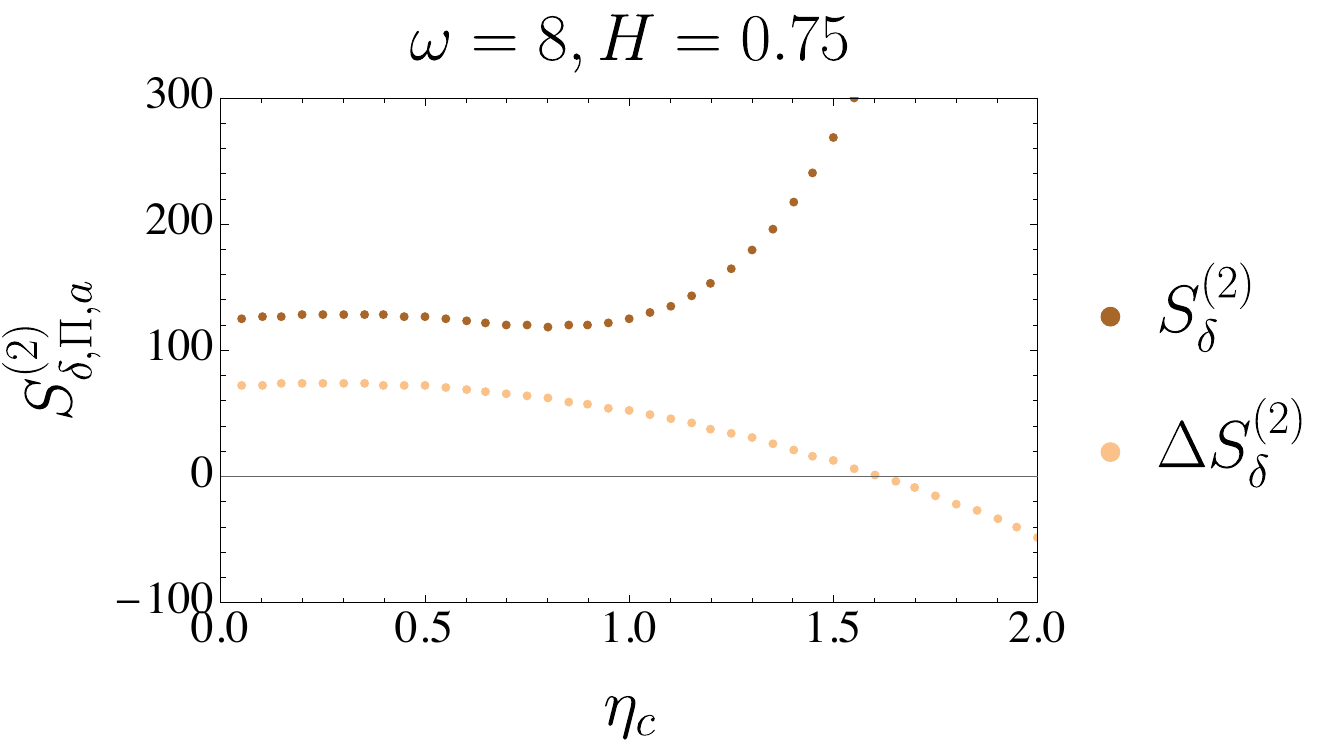}
    \label{fig:S_ap}
    \caption{}
\end{subfigure}
\caption{Similar plots to \cref{fig:Sdelta_Lp} but for the additive plateau angularity, $\Pi,a$.}
\label{fig:Sdelta_ap}
\end{figure}

In \cref{fig:Sdelta_Lp,fig:Sdelta_mp,fig:Sdelta_ap} we plot $\Delta S^{(2)}_{e,\delta}$ for the three families of generalised angularity introduced in \cref{sec:setup}. The fixed parameters and plotted range were chosen to match that of \cref{fig:Lp,fig:mp,fig:AddPlateau} respectively. In panel $(c)$ of \cref{fig:Sdelta_Lp,fig:Sdelta_mp,fig:Sdelta_ap} we compare the total boundary term $S^{(2)}_{e, \delta}$ with the numerical contribution from \cref{eq:DSdelta}. As a rough rule, we see that approximately half of this boundary term is under analytic control, while half is obtained from a three-fold numerical integral. Similar to the one-loop boundary terms, we see the two-loop boundary term diverge at low $p$ and large $\eta_c$, where the SCET$_\mathrm{I}$ factorisation breaks down. 

As a point of comparison for panels $(a)$ and $(b)$ of these plots, we quote analytic results for $\Delta S^{(2)}_{e,\delta}$ for thrust~\cite{Monni:2011gb}
\begin{align}
\Delta S_{\tau,\delta}^{(2,C_R n_f T_F)}&=
-\frac{64}{9}+\frac{208}{9}\zeta_2-\frac{64}{3}\zeta_3
\,,\\
\Delta S_{\tau,\delta}^{(2,C_R C_A)}&=
\frac{32}{9}
-\frac{536}{9}\zeta_2
+\frac{176}{3}\zeta_3
+34\zeta_4\,,
\end{align}
and for C-parameter~\cite{Bell:2018oqa}
\begin{align}
\Delta S_{C,\delta}^{(2,C_R n_f T_F)}&=
-\frac{16}{3}+\frac{32}{3}\zeta_3
\,,\\
\Delta S_{C,\delta}^{(2,C_R C_A)}&=
\frac{8}{3}
-\frac{88}{3}\zeta_3
+48\zeta_4
\,.
\end{align}

\begin{table}[htb]
\centering
$
\begin{array}{|c|c|c|}
\hline
\Delta S^{(2,c)}_{\tau,a,\delta}
& c=C_R n_f T_F &  c= C_R C_A
\\ \hhline{|=|=|=|}
\vphantom{\Big[}
a=-0.5
&    
6.169(63\pm 06)
& 
9.875(73\pm 16)
\\ \hline
\vphantom{\Big[}
a=0.0
& 
5.261262
&   
12.910255
\\ \hline
\vphantom{\Big[}
a=0.5
& 
7.411(00\pm09)
&     
28.264(79\pm 27)
\\ \hline
\end{array}
\quad
\begin{array}{|c|c|c|}
\hline
\Delta S^{(2,c)}_{C,a,\delta}
& c=C_R n_f T_F &  c= C_R C_A
\\ \hhline{|=|=|=|}
\vphantom{\Big[}
a=-0.5
& 
11.124(45\pm05)
&     
24.952(58\pm14)
\\ \hline
\vphantom{\Big[}
a=0.0
& 
7.488607
&   
19.357846
\\ \hline
\vphantom{\Big[}
a=0.5
&    
7.968(88\pm09)
& 
29.835(46\pm27)
\\ \hline
\end{array}
$
\caption{Some values of $\Delta S^{(2)}_{e,\delta}$ for angularity and C-angularity.}
\label{tab:DS}
\end{table}
We tabulate the corresponding numerical values for C-angularity and conventional angularity in \cref{tab:DS} for $a\in \{-0.5,0,0.5\}$. 
As a point of reference for panel $(c)$ in \cref{fig:Sdelta_Lp,fig:Sdelta_mp,fig:Sdelta_ap}, we quote the full analytic boundary coefficient, given by \cref{eq:CR2} and \cref{eq:Sdelta2}, for thrust
\begin{align}
\begin{split}
    S_{\tau,\delta}^{(2)}&= C_R\Bigg[
    C_R\left(-27 \zeta_4\right)
    +C_A\left(
    -\frac{2428}{81}
    +\frac{67 }{3}\zeta_2
    +\frac{110 }{9}\zeta_3
    +10 \zeta_4 
    \right) \\
    &\qquad \quad
    +n_f T_F
    \left( 
    \frac{656}{81}
    -\frac{20 }{3}\zeta_2
    -\frac{40 }{9}\zeta_3
    \right)
    \Bigg]+\Delta S^{(2)}_{\tau,\delta}\,,
    \\
    &=118.96695\quad(n_f=5)\,,
\end{split}
\end{align}
which agrees with ref.~\cite{Monni:2011gb}, and for C-parameter 
\begin{align}
\begin{split}
    S_{C,\delta}^{(2)}&= C_R\Bigg[
    C_R\left(13 \zeta_4\right)
    +C_A\left(
    -\frac{2428}{81}
    +\frac{469 }{9}\zeta_2
    -\frac{154 }{9}\zeta_3
    -10 \zeta_4 
    \right) \\
    &\qquad \quad
    +n_f T_F
    \left( 
    \frac{656}{81}
    -\frac{140 }{9}\zeta_2
    +\frac{56 }{9}\zeta_3
    \right)
    \Bigg]+\Delta S^{(2)}_{C,\delta}\,,
    \\
    &=191.4507446 \quad (n_f=5)\,,
\end{split}
\end{align}
which agrees with ref.~\cite{Bell:2018oqa}. In the above equations we have included a numerical evaluation for $n_f=5$ and $T_F=\tfrac{1}{2}$. As expected, in \cref{fig:Sdelta_Lp} (c) we see that the $L_p$ event shape agrees with the C-parameter value at $p=1$ and tends towards the thrust value as $p$ increases. One reaches per-mille agreement with thrust at $p$=42.

\section{Conclusion}
\label{sec:conc}
We have introduced a class of additive dijet event shapes known as generalised angularities, where the contribution from a particle reduces to that of angularity in the forward limit, but the behaviour of the event shape in the central region can be rather general. The former property means that in the resummation of these event shapes in $e^+e^-$ or DIS collisions, the collinear functions (i.e. beam and/or jet functions) are the same as that for angularity, which are already known to two loop order. Thus, for these event shapes the only missing ingredient required to enable NNLL$'$ predictions is the soft function up to two loop order.

We have introduced a fast and efficient method to compute the NNLO soft function for any event shape in this generalised angularity class, which involves splitting the `correlated' $C_R C_A$ and $C_R n_f T_F$ pieces into an `inclusive' piece and a correction term. The $C_R^2$ and inclusive pieces of the soft function are written in terms of two one-dimensional integrals which can be performed analytically in many cases. The correction can be written as a three-dimensional integral, and we provide a numerical implementation to compute this for any rapidity weighting. 

We proposed three concrete families of event shape in the generalised angularity class, which between them cover a wide variety of possible shapes in the central soft region. One of these, the $L_p$-angularity, is a natural interpolation between standard angularity and the C-angularity shape that we studied in a previous paper. To illustrate our general algorithm, we computed the NNLO soft function for event shapes in these three families with various reasonable values of the event shape parameters, along with the shape-dependent hadronization correction coefficients $c_e$ appropriate to the dijet tail region of the event shape.

Modifying the behaviour of the event shape only in the central region changes the contribution of non-perturbative hadronisation corrections whilst preserving the perturbative resummation structure. Thus, studying a range of generalised angularity event shapes in the context of $e^+e^-$ collisions or DIS would be an interesting new avenue to study and constrain hadronisation corrections in the dijet limit, and to perform a global fit of the hadronisation corrections together with the fundamental strong coupling constant. 

\section*{Acknowledgments}

The diagrams in this paper were produced with the aid of {\tt FeynCraft} \cite{Gaunt:2025peq}, {\tt quiver} \cite{Quiver}, and the {\tt TikZ-Feynman} package \cite{Ellis:2016jkw}. The authors acknowledge the use of Google Gemini AI, in brainstorming the initial ideas for the project, proposing observable shapes, and cross-checking many of the results in this paper. The authors take full responsibility for the content of this paper.

\appendix
\section{Kinematic conventions}
\label{sec:kin}
We decompose an arbitrary four-vector $k^\mu$ as
\begin{equation}
    k^\mu = \frac{1}{2}\left( k^+ n^\mu +k^- \bar{n}^\mu\right)+k_\perp^\mu\,,
\end{equation}
where $n^\mu$ and $\bar{n}^\mu$ are light-like reference vectors which satisfy
\begin{align}
    n^2=\bar{n}^2=0\,, \qquad n \cdot \bar{n}=2\,.
\end{align}
We take these reference vectors to lie on the thrust axis, and use a coordinate system in which the $z$ axis is aligned with this thrust axis:
\begin{align}
    n=(1,0,0,1)\,, \qquad  \bar{n}=(1,0,0,-1)
\end{align}
such that
\begin{align}
    k^\pm=k^0 \pm k^z\,.
\end{align}
We denote by $\vec{k}_\perp$ the Euclidean two-vector comprising the transverse components of the Lorentzian four-momentum $k_\perp$, and we denote by $k_T$ the scalar magnitude of $k_\perp$. These objects are related via
\begin{equation}
   -k_\perp \cdot k_\perp = \vec{k}_\perp \cdot \vec{k}_\perp 
   =|\vec{k}_\perp|^2 =k_T^2 \ge 0\,,
\end{equation}
where we use the symbol ``$\cdot$" to denote either the Lorentzian or Euclidean inner product depending on the context. 

\bibliography{refs.bib}
\end{document}